\documentclass[pdftex,twocolumn,epjc3]{svjour3}          

\RequirePackage[T1]{fontenc}

\smartqed  

\RequirePackage{graphicx}
\RequirePackage{mathptmx}      
\RequirePackage{flushend}
\RequirePackage[numbers,sort&compress]{natbib}
\RequirePackage[colorlinks,citecolor=blue,urlcolor=blue,linkcolor=blue]{hyperref}
\RequirePackage{amsmath} 
\RequirePackage{physics} 
\RequirePackage{booktabs}
\RequirePackage{longtable}
\RequirePackage{lipsum}
\RequirePackage{widetext}

\journalname{Eur. Phys. J. C}

\begin{document}

\title{Double-mixing CP violation in $B$ decays}

\author{Wen-Jie Song\thanksref{e1,addr1} \and Yin-Fa Shen\thanksref{e2,addr1} \and Qin Qin\thanksref{e3,addr1} 
}

\thankstext{e1}{e-mail: jaysong@hust.edu.cn}
\thankstext{e2}{e-mail: syf70280@hust.edu.cn}
\thankstext{e3}{e-mail: qqin@hust.edu.cn}

\institute{School of Physics, Huazhong University of Science and Technology, Wuhan 430074, China\label{addr1}}

\date{Received: 21 July 2024 / Accepted: 5 September 2024}

\maketitle

\begin{abstract}
We study a long-overlooked CP violation observable, termed double-mixing CP violation, which arises from the interference between two neutral meson oscillating paths involved in a decay chain. The double-mixing CP violation is beneficial for the precise test of the Standard Model CKM mechanism, as it offers the potential to extract weak phases without pollution from strong dynamics. To provide a comprehensive understanding of the double-mixing CP violation, we perform phenomenological analyses on the cascade decays of the $B^0_d$ and $B^0_s$ mesons. Our results show that the double-mixing CP violation can be very significant in certain decay channels, such as $B^0_d \to J/\psi K \to J/\psi(\pi^+\ell^-\bar{\nu}_\ell)$, where the total CP violation arises almost entirely from the double-mixing CP violation. In addition, we employ the decay process $B^0_d \to D^0 K \to (K^-\pi^+)(\pi\ell\nu)$ to demonstrate that the involved strong and weak phases can be directly determined from the experimental data without any theoretical inputs. Specifically, the overall weak phase of this process is $2\beta+\gamma$, which is of significant importance for extracting the CKM phase angles. We also use $B^0_s \to \rho^0 K \to \rho^0(\pi^-\ell^+\nu_\ell)$ to illustrate and discussion the significant importance of $K^0_S - K^0_L$ interference.
\end{abstract}

\section{Introduction}

CP violation plays a crucial role in flavor physics. It is one of the fundamental criteria for explaining the matter-antimatter asymmetry of the universe~\cite{Sakharov:1967dj}. However, the standard model (SM) does not account for a sufficiently large CP violation to explain the observed asymmetry~\cite{Bernreuther:2002uj, Canetti:2012zc}. This suggests the presence of additional sources of CP violation beyond the standard Cabibbo-Kobayashi-Maskawa (CKM) mechanism~\cite{Cabibbo:1963yz, Kobayashi:1973fv}. Consequently, the quest for non-standard CP violation is a vital undertaking in the quark flavor sector. One systematic approach involves testing the unitarity of the CKM matrix, which relies on measurements of CP violation in quark flavor processes to determine the values of the CKM phase angles. Therefore, it is imperative to conduct comprehensive investigations into CP violation in flavor physics, to precisely test and explore potential sources of CP violation beyond the framework of the CKM mechanism.

Since the unexpected discovery of indirect CP violation in the neutral $K$ meson system in 1964~\cite{Christenson:1964fg}, significant theoretical and experimental advancements have been achieved ~\cite{NA31:1988eyf, NA48:1999szy, KTeV:1999kad, BaBar:2001pki, Belle:2001zzw, Belle:2004nch, BaBar:2004gyj, BaBar:2010hvw, Belle:2010xyn, LHCb:2011aa, LHCb:2012xkq, LHCb:2013syl, LHCb:2014kcb, LHCb:2019hro}, including the detection of CP violation in the $B$ and $D$ meson decays~\cite{BaBar:2001pki, Belle:2001zzw, LHCb:2019hro}. These studies have successfully observed three types of CP violation under the CKM mechanism in meson systems, along with measurements of CP-violating phases and testing of the unitarity of the CKM matrix. As no evidence of non-standard CP violation has been observed yet, collaborative efforts in both theoretical and experimental domains are essential to conduct more precise tests. In general, experimental precision of channels takes advantage of powerful flavor experiments such as BESIII~\cite{BESIII:2020nme}, Belle II~\cite{Belle-II:2018jsg}, and LHCb~\cite{LHCb:2012myk, LHCb:2018roe, LHCb:2023hlw}. The primary theoretical challenge in improving precision lies in disentangling the effects of strong interactions within the SM and non-standard contributions. Despite numerous studies dedicated to strong interactions (see {\it e.g.}~\cite{Beneke:1999br, Beneke:2000ry, Keum:2000wi, Lu:2000em, Bauer:2000yr, Bauer:2001yt, Li:2012cfa, Qin:2013tje, Qin:2014xta, Qin:2021tve}), the theoretical predictions arising from strong interactions still struggle to align well with the current experimental precision. Therefore, there is an anticipation to provide CP violation observables that are less affected or unaffected by strong interactions.

Of the three types of observed CP violation---direct CP violation, CP violation in mixing, and CP violation in interference between a decay without mixing $M^0 \to f$ and a decay with mixing $M^0 \to \bar{M}^0 \to f$---the last one has the potential to avoid contamination from strong dynamics. The visualization of CP violation induced by the CKM mechanism typically involves interference between amplitudes with distinct CKM phases and CP-conserving phases, often including strong phases. However, the $M^0 \to f$ and $M^0 \to \bar{M}^0 \to f$ amplitudes have a relative CP-conserving phase induced by different time evolution, making strong phases not a necessity. In fact, the first significant CP violation effect observed by $B$ factories in $B^0_d\to J/\psi K^0_S$~\cite{BaBar:2001pki, Belle:2001zzw} serves as a representative example of this type. The CP violation in the interference between $i\to M^0$ and $i\to \bar{M}^0\to M^0$~\cite{Yu:2017oky,Meca:1998ee,Amorim:1998pi} also capitalizes on the same advantage provided by neutral meson mixing. It motivates us to develop more CP violation observables that involve mixing. 

In our recent study~\cite{Shen:2023nuw}, we introduced a long-overlooked form of CP violation arising from the interference between two neutral meson oscillating paths involved in a decay chain. Such a decay, involving two neutral mesons $M^0_{1,2}$, typically occurs via two amplitudes: $M^0_1 \to M^0_2 \to \bar{M}^0_2 \to f$ and $M^0_1 \to \bar{M}^0_1 \to \bar{M}^0_2 \to f$, as depicted in Fig.~\ref{fig:2path}, where the particles associated with $M_2$ are hidden for convenience. We refer to this type of CP violation as the double-mixing CP violation. This introduces a new experimental approach, allowing for measuring the dependence of CP violation on two time variables, specifically the evolution periods of $M_{1,2}$. Our previous research~\cite{Shen:2023nuw} revealed that the existence of double-mixing CP violation does not necessarily require a nonzero strong phase, offering the potential to extract weak phases without the influence of strong dynamics. In some cases where a strong phase is involved, it is also feasible to directly extract the strong phase from experimental data, thereby avoiding significant theoretical uncertainties. To provide a more comprehensive understanding, we will explore the double-mixing CP violation in typical $B^0_d$ and $B_s^0$ meson cascade decays in this study. In our previous studies~\cite{Shen:2023nuw}, we have conducted a phenomenological analysis of $B^0_s \to \rho^0 K \to \rho^0 (\pi^-\ell^+\nu_\ell )$ and $B^0_d \to D K \to (K^-\pi^+)(\pi^+\ell^-\bar{\nu}_\ell)$. For the sake of completeness in our discussion, we are still carrying out the corresponding analysis of these two processes.

The rest of the paper is organized as follows. In Sect.~\ref{sec:formulae}, we provide the general formulas for the double-mixing CP violation in the process $M^0_1(t_1) \to M_2(t_2) \to f$, where $M_2$ is either $M^0_2$ or $\bar{M}^0_2$. In Sect.~\ref{Section3}, we introduce a numerical analysis strategy to calculate the double-mixing CP violation in cascade decays of $M^0_1$, where $M^0_1$ can be either $B^0_d$ or $B^0_s$. Different initial and final states exhibit markedly distinct patterns. With frequently oscillating $B^0_{(s)}$ and $K^0$ involved, the double-mixing CP violation effects are generally more promising. In the case of semi-leptonic final states, such as $B^0_d \to J/\psi K \to J/\psi(\pi\ell\nu)$, the total CP violation predominantly stems from contributions of the double-mixing CP violation. Conversely, the contribution of the double-mixing CP violation is exceedingly small when $D^0$ is involved, primarily due to the negligible mixing of $D^0$. We employ certain decay processes, such as $B^0_s \to \rho^0 K \to \rho^0 (\pi^\pm\ell^\mp\nu_\ell)$, to reveal that the double-mixing CP violation is possibly significantly large, exceeding 50\%. These decay modes could serve as the initial experimental attempts to search for the double-mixing CP violation. Of greater significance, the double-mixing CP violation in certain decay channels under investigation can be utilized for CKM phase extraction, such as in the case of $B^0_d \to D^0 K \to (K^-\pi^+)(\pi \ell \nu)$. Despite involving both strong and weak phases simultaneously, both phases can be extracted through experimental measurements of the double-mixing CP violation without relying on any theoretical inputs. In Sect.~\ref{Conclusion}, we summarize our results and present a conclusion.

\section{Formulae}\label{sec:formulae}

We adopt the conventional formulae that the mass eigenstates $M_{H, L}$ of the flavored neutral mesons are superpositions of their flavor eigenstates $M^0$ and $\bar{M}^0$~\cite{Workman:2022ynf}, 
\begin{eqnarray}
\ket{M_\mathrm{H,L} } = p \ket{ M^0 } \mp q \ket{ \bar{M}^0 } \; ,
\end{eqnarray}
where $q,p$ are complex coefficients. Then, the time evolution starting from a flavor eigenstate in the case of CPT conservation is given by,
\begin{align}\label{eq2}
    \ket{M^0(t)} &= g_+(t)\ket{M^0} - \frac{q}{p}g_-(t)\ket{\bar{M}^0}, \nonumber \\
    \ket{\bar{M}^0(t)} &= g_+(t)\ket{\bar{M}^0} - \frac{p}{q}g_-(t)\ket{M^0},
\end{align}
where the functions $g_{\pm}(t)$ of time are given by
\begin{align}\label{eq3}
    g_{\pm}(t)=\frac{1}{2}\left[\mathrm{exp}\left(-im_\mathrm{H}t-\frac{1}{2}\Gamma_\mathrm{H}t\right)\pm \mathrm{exp}\left(-im_\mathrm{L}t-\frac{1}{2}\Gamma_\mathrm{L}t\right)\right],
\end{align}
with $m_\mathrm{H(L)}$ and $\Gamma_\mathrm{H(L)}$ being the mass and decay width of the `Heavy(Light)' mass eigenstate $M_\mathrm{H(L)}$, respectively.

\begin{figure}[htbp]
    \centering
\includegraphics[keepaspectratio,width=8cm]{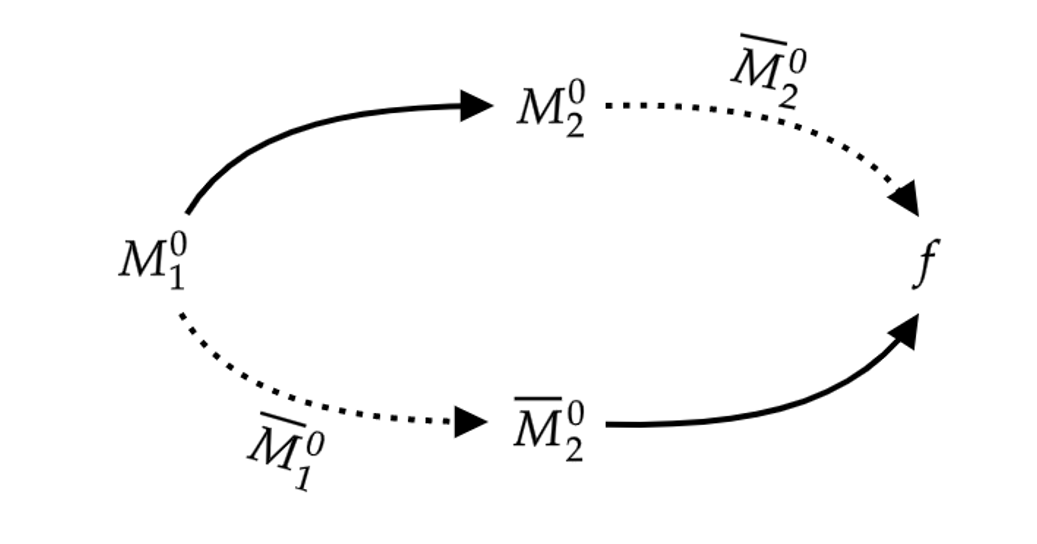}
\caption{Two oscillation paths in the cascade decay $M_1^0\to M_2\to f$ with only $M_1^0\to M_2^0$, $\bar M_2^0\to f$ and their CP conjugate processes being allowed. The decay products associated with $M_2$ are not displayed.}\label{fig:2path}
\end{figure}

The double-mixing CP violation will occur in the cascade decay containing at least two different neutral mesons. The process illustrated in  Fig.~\ref{fig:2path} serves as a clear example. During the time interval $t_1$, a primary neutral meson $M^0_1$ either remains in its original state or oscillates into its antiparticle $\bar{M}^0_1$, which subsequently decays into a secondary neutral meson $\bar{M}^0_2(M^0_2)$. Other accompanying particles are not shown in the figure for simplicity. Similarly, the intermediate state $M_2(t_2)$, representing a time-evolved neutral meson, $M^0_2(t_2)$ or $\bar{M}^0_2(t_2)$, decays into a final state $f$. Here, $t_2$ denotes the time for the oscillation of $M^0_2 (\bar{M}^0_2)$ in the rest frame of $M^0_2 (\bar{M}^0_2)$. A solid line and a dotted line represent the direct decay and flavor oscillation of the neutral meson before decay, respectively. The time-dependent CP asymmetry in the chain $M^0_1(t_1) \to M_2(t_2) \to f$ can be defined as 
\begin{align}
    A_{\rm CP}(t_1,t_2)
    \equiv&\ \frac{\abs{\mathcal{M}}^2(t_1,t_2) - \abs{\bar{\mathcal{M}}}^2(t_1, t_2)}{\abs{\mathcal{M}}^2(t_1,t_2) + \abs{\bar{\mathcal{M}}}^2(t_1, t_2)}=\frac{N(t_1,t_2)}{D(t_1,t_2)},\label{eq4} \\ 
    N(t_1,t_2)=&\ e^{-\Gamma_1 t_1}\Big[C_h(t_2)\cosh{\frac{\Delta \Gamma_1 t_1}{2}}+C_n(t_2)\cos{\Delta m_1 t_1}\nonumber\\
    &+S_n(t_2)\sin{\Delta m_1  t_1}+S_h(t_2)\sinh{\frac{\Delta\Gamma_1 t_1}{2}}\Big],\nonumber\\
    D(t_1,t_2)=&\ e^{-\Gamma_1 t_1}\Big[C_h^{\prime}(t_2)\cosh{\frac{\Delta \Gamma_1 t_1}{2}}+C_n^{\prime}(t_2)\cos{\Delta m_1 t_1}\nonumber\\
    &+S_n^{\prime}(t_2)\sin{\Delta m_1 t_1}+S_h^{\prime}(t_2)\sinh{\frac{\Delta\Gamma_1 t_1}{2}}\Big],\nonumber
\end{align}
where the amplitude $\mathcal{M}(t_1, t_2)$ is the sum of amplitudes
of all possible paths, while $\bar{\mathcal{M}}(t_1, t_2)$ represents the CP conjugate of $\mathcal{M}(t_1, t_2)$. The differences in widths and masses are defined as $\Delta\Gamma \equiv \Gamma_\mathrm{H}-\Gamma_\mathrm{L}$ and $ \Delta m \equiv m_\mathrm{H}-m_\mathrm{L}$, respectively, and the average of widths is $\Gamma \equiv (\Gamma_\mathrm{H}+\Gamma_\mathrm{L})/2$. Here, the subscripts 1 and 2 denote the primary meson $M_1$ and the secondary meson $M_2$, respectively. Evidently, according to the $t_1$ dependence, $N(t_1, t_2)$ and $D(t_1, t_2)$ can be written in four terms. Alternatively, a different $t_1$ function basis can be chosen, for example, rewriting the two terms in $N(t_1,t_2)$ as 
\begin{align}
    &e^{-\Gamma_1 t_1}\Big[C_h(t_2)\cosh{\frac{\Delta \Gamma_1 t_1}{2}} + C_n(t_2)\cos{\Delta m_1 t_1}\Big]  \nonumber\\
    \to \quad &C_+(t_2)\abs{g_{+,1}(t_1)}^2 + C_-(t_2)\abs{g_{-,1}(t_1)}^2,
\end{align}
and the same substitution also applies to $D(t_1,t_2)$.

We will provide the explicit expressions of the terms in~\eqref{eq4} for different cases. We categorize all cases into three groups based on the number of possible oscillation and decay paths. Besides the two-path scenario shown in Fig.~\ref{fig:2path}, there are also four- and eight-path ones, which will be discussed in more detail below.

\noindent \textbf{Category 1}

In the first category, we consider decay processes that occur via two possible paths, as exemplified in~Fig.~\ref{fig:2path}. In this study, we are only interested in cases where direct CP violation is negligible, so we have $\braket{\bar{M}^0_2}{\bar{M}^0_1} = \braket{M^0_2}{M^0_1}e^{i\omega}$ with where $\omega$ being a pure weak phase, and $\braket{f}{\bar{M}^0_2} = \braket{\bar{f}}{M^0_2}$. In order to avoid confusion, we used $2\omega$ in our previous research~\cite{Shen:2023nuw}, whereas here we are directly employing a more general phase representation $\omega$. The mixing parameters of $M_{1,2}$ are written as $(q/p)_{1,2} = \abs{(q/p)}_{1,2}e^{-i\phi_{1,2}}$, with $\phi_{1,2}$ complex phases.

The amplitude $\mathcal{M}(t_1, t_2)$ of the process $M_1^0(t_1)\to M_2(t_2)\to f$ and its CP conjugate $\bar{\mathcal{M}}(t_1, t_2)$ are given by
\begin{align}\label{eq6}
    \mathcal{M}(t_1, t_2) =& \braket{f}{M^0_2(t_2)}\braket{M^0_2}{M^0_1(t_1)}\nonumber\\
    &+\braket{f}{\bar{M}^0_2(t_2)}\braket{\bar{M}^0_2}{M^0_1(t_1)},\nonumber \\
    \bar{\mathcal{M}}(t_1, t_2)=&\braket{\bar{f}}{M^0_2(t_2)}\braket{M^0_2}{\bar{M}^0_1(t_1)}\nonumber\\
    &+\braket{\bar{f}}{\bar{M}^0_2(t_2)}\braket{\bar{M}^0_2}{\bar{M}^0_1(t_1)}. 
\end{align}
Considering the time evolution of the neutral mesons~\eqref{eq2} -~\eqref{eq3}, we can make the following substitutions 
\begin{align}\label{eq7}
    &\braket{M^0_2}{M^0_1(t_1)} = g_{+,1}(t_1)\braket{M^0_2}{M^0_1},\qquad \nonumber\\
    &\braket{\bar{M}^0_2}{M^0_1(t_1)} = -\frac{q_1}{p_1}g_{-,1}(t_1)\braket{\bar{M}^0_2}{\bar{M}^0_1}, \nonumber\\
    &\braket{\bar{M}^0_2}{\bar{M}^0_1(t_1)}  = g_{+,1}(t_1)\braket{\bar{M}^0_2}{\bar{M}^0_1},\qquad \nonumber\\
    &\braket{M^0_2}{\bar{M}^0_1(t_1)}=-\frac{p_1}{q_1}g_{-,1}(t_1)\braket{M^0_2}{M^0_1}, \nonumber\\
    &\braket{f}{M^0_2(t_2)}=-\frac{q_2}{p_2}g_{-,2}(t_2)\braket{f}{\bar{M}^0_2},\qquad  \nonumber\\
    &\braket{\bar{f}}{M^0_2(t_2)} = g_{+,2}(t_2)\braket{\bar{f}}{M^0_2}, \nonumber\\
    &\braket{\bar{f}}{\bar{M}^0_2(t_2)}=-\frac{p_2}{q_2}g_{-,2}(t_2)\braket{\bar{f}}{M^0_2},\qquad \nonumber\\
    &\braket{f}{\bar{M}^0_2(t_2)} = g_{+,2}(t_2)\braket{f}{\bar{M}^0_2}.
\end{align}
Then, it arrives that the total time-dependent CP asymmetry~\eqref{eq4}. 

The corresponding double-mixing CP asymmetry is reflected by the $S_h$ and $S_n$ terms, which are induced by the interference between the upper and lower paths in~Fig.~\ref{fig:2path}. The results are given by
\begin{align}\label{eq9}
    S_h(t_2) = &\ \frac{e^{-\Gamma_2 t_2}}{2}\Big[\sinh\frac{\Delta \Gamma_2 t_2}{2}\cos{(\omega-\phi_1+\phi_2)}\Big(\abs{\frac{q_1}{p_1}}\abs{\frac{q_2}{p_2}}\nonumber\\
    &-\abs{\frac{p_1}{q_1}}\abs{\frac{p_2}{q_2}}\Big)+\sin{\Delta m_2 t_2}\sin{(\omega-\phi_1+\phi_2)}\nonumber\\
    &\times\pqty{\abs{\frac{q_1}{p_1}}\abs{\frac{q_2}{p_2}}+\abs{\frac{p_1}{q_1}}\abs{\frac{p_2}{q_2}}}\Big], \\
    S_n(t_2) = &\ \frac{e^{-\Gamma_2 t_2}}{2}\Big[-\sinh{\frac{\Delta \Gamma_2 t_2}{2}}\sin{(\omega-\phi_1+\phi_2)}\Big(\abs{\frac{p_1}{q_1}}\abs{\frac{p_2}{q_2}}\nonumber\\
    &+\abs{\frac{q_1}{p_1}}\abs{\frac{q_2}{p_2}}\Big)+\sin{\Delta m_2 t_2}\cos{(\omega-\phi_1+\phi_2)}\nonumber\\
    &\times\pqty{\abs{\frac{q_1}{p_1}}\abs{\frac{q_2}{p_2}}-\abs{\frac{p_1}{q_1}}\abs{\frac{p_2}{q_2}}}\Big].
\end{align}
The common factors $\abs{\braket{M^0_2}{M^0_1}}^2$ and $\abs{\braket{f}{\bar{M}^0_2}}^2$ are omitted in the above expression. The remaining terms in the numerator, which are not the focus of our current discussion, are given by
\begin{align}\label{eq11}
    C_+(t_2)&=\pqty{\abs{\frac{q_2}{p_2}}^2-\abs{\frac{p_2}{q_2}}^2}\abs{g_{-,2}(t_2)}^2,\nonumber\\
    C_-(t_2) &= \pqty{\abs{\frac{q_1}{p_1}}^2-\abs{\frac{p_1}{q_1}}^2}\abs{g_{+,2}(t_2)}^2 \;.
\end{align}
The $C_+(t_2)$ term is contributed by purely the upper path in~Fig.~\ref{fig:2path}, where $M^0_2 \to \bar{M}^0_2$ oscillation is the only CP violation source, while the $C_-(t_2)$ term is contributed purely by the lower path, where CP violation is induced by $M^0_1 \to \bar{M}^0_1$ oscillation. 
The terms contributing to $D(t_1,t_2)$ are given by
\begin{align}
    C_+^{\prime}(t_2)=&\pqty{\abs{\frac{q_2}{p_2}}^2+\abs{\frac{p_2}{q_2}}^2}\abs{g_{-,2}(t_2)}^2,\nonumber\\
    C_-^{\prime}(t_2) =& \pqty{\abs{\frac{q_1}{p_1}}^2+\abs{\frac{p_1}{q_1}}^2}\abs{g_{+,2}(t_2)}^2,\label{eq12}\\
    S_h^{\prime}(t_2) =&\frac{e^{-\Gamma_2 t_2}}{2}\Big[\sinh\frac{\Delta \Gamma_2 t_2}{2}\cos{(\omega-\phi_1+\phi_2)}\Big(\abs{\frac{q_1}{p_1}}\abs{\frac{q_2}{p_2}}\nonumber\\
    &+\abs{\frac{p_1}{q_1}}\abs{\frac{p_2}{q_2}}\Big)+\sin{\Delta m_2 t_2}\sin{(\omega-\phi_1+\phi_2)}\nonumber\\
    &\times\pqty{\abs{\frac{q_1}{p_1}}\abs{\frac{q_2}{p_2}}-\abs{\frac{p_1}{q_1}}\abs{\frac{p_2}{q_2}}}\Big],\label{eq13-1}\\
    S_n^{\prime}(t_2) =&\frac{e^{-\Gamma_2 t_2}}{2}\Big[-\sinh{\frac{\Delta \Gamma_2 t_2}{2}}\sin{(\omega-\phi_1+\phi_2)}\Big(\abs{\frac{q_1}{p_1}}\abs{\frac{q_2}{p_2}}\nonumber\\
    &-\abs{\frac{p_1}{q_1}}\abs{\frac{p_2}{q_2}}\Big)+\sin{\Delta m_2 t_2}\cos{(\omega-\phi_1+\phi_2)}\nonumber\\
    &\times\pqty{\abs{\frac{q_1}{p_1}}\abs{\frac{q_2}{p_2}}+\abs{\frac{p_1}{q_1}}\abs{\frac{p_2}{q_2}}}\Big].\label{eq13-2}
\end{align}

For the convenience of further analysis, we define the two-dimensional time-dependent double-mixing CP asymmetry $A_{dm}(t_1,t_2)$, which is represented as 
\begin{align}
A_{dm}(t_1,t_2) &= \frac{e^{-\Gamma_1 t_1}}{D(t_1, t_2)}\pqty{S_h(t_2)\sinh{\frac{\Delta \Gamma_1 t_1}{2}} + S_n(t_2)\sin{\Delta m_1 t_1}}\label{eq10-1}\\
&= A_{h,dm}(t_1,t_2) + A_{n,dm}(t_1, t_2).\label{eq10-2}
\end{align}
It represents the double-mixing component within the $A_{\rm CP}(t_1,t_2)$ in~\eqref{eq4}. In fact, the remaining part is negligible. 

For decay processes occurring via $M^0_1 \to \bar{M}^0_2 \to M_2^0 \to f$ and $M^0_1 \to \bar{M}^0_1 \to M_2^0 \to f$, the results can be directly obtained by applying the replacement $p_2/q_2\to q_2/p_2$ from~\eqref{eq12} -~\eqref{eq13-2}, {\it i.e.}, $|p_2/q_2|\to |q_2/p_2|$ and $\phi_2\to -\phi_2$. In order to see how this works, we will conduct a detailed analysis of the situation. First, we need to make the following modification to \eqref{eq7}:
\begin{align}
    &\braket{M^0_2}{M^0_1(t_1)} = -\frac{q_1}{p_1}g_{-,1}(t_1)\braket{M^0_2}{\bar{M}^0_1},\nonumber\\ &\braket{\bar{M}^0_2}{M^0_1(t_1)} = g_{+,1}(t_1)\braket{\bar{M}^0_2}{M^0_1}, \nonumber\\
    &\braket{\bar{M}^0_2}{\bar{M}^0_1(t_1)}  = -\frac{p_1}{q_1}g_{-,1}(t_1)\braket{\bar{M}^0_2}{M^0_1},\nonumber\\
    &\braket{M^0_2}{\bar{M}^0_1(t_1)}=g_{+,1}(t_1)\braket{M^0_2}{\bar{M}^0_1},\nonumber\\
    &\braket{f}{M^0_2(t_2)}=g_{+,2}(t_2)\braket{f}{\bar{M}^0_2},\nonumber\\ &\braket{\bar{f}}{M^0_2(t_2)} = -\frac{q_2}{p_2}g_{-,2}(t_2)\braket{\bar{f}}{M^0_2}, \nonumber\\
    &\braket{\bar{f}}{\bar{M}^0_2(t_2)}=g_{+,2}(t_2)\braket{\bar{f}}{M^0_2},\nonumber\\ &\braket{f}{\bar{M}^0_2(t_2)} = -\frac{p_2}{q_2}g_{-,2}(t_2)\braket{f}{\bar{M}^0_2}.
\end{align}
When we substitute all the components into \eqref{eq6}, we notice a disparity. In this scenario, the first term of $\mathcal{M}(t_1,t_2)$ in Equation \eqref{eq6} involves $q_1/p_1$, contributing $-\phi_1$, while the second term involves $p_2/q_2$, contributing $\phi_2$. When these two terms interfere, the sign of the mixed angle in one of the terms is flipped, resulting in a final manifestation of $\phi_1+\phi_2$. However, in the previous scenario, the first term in $\mathcal{M}(t_1,t_2)$ contains $q_2/p_2$, contributing $-\phi_2$, while the second term contains $q_1/p_1$, contributing $-\phi_1$. When these two terms interfere, the final manifestation is $\phi_1-\phi_2$.

\noindent \textbf{Category 2}

\begin{figure}[htbp]
    \centering
\includegraphics[keepaspectratio,width=8cm]{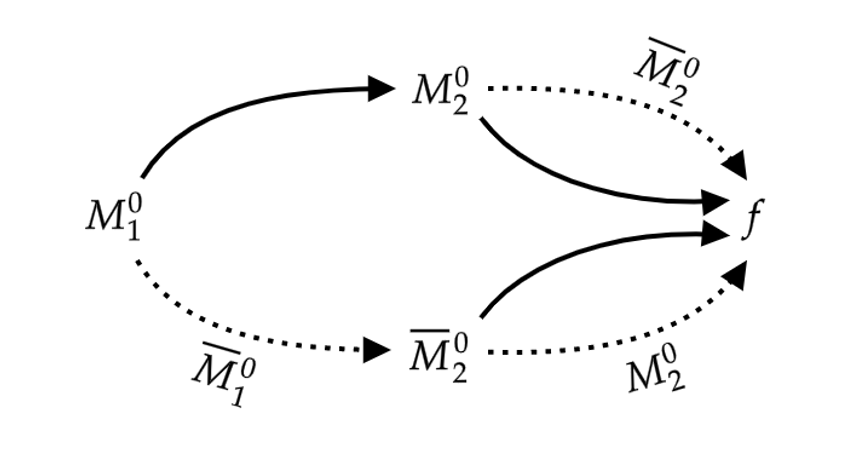}
\caption{Four oscillation paths in the cascade decay $M_1^0\to M_2\to f$ where $f$ is a CP eigenstate, with $M_1^0\to M_2^0$, $\bar M_2^0\to f$ and their CP conjugate processes being allowed.}\label{fig:4path}
\end{figure}

In the second category, we consider that $f$ is a CP eigenstate that can decay from both $M_2^0$ and $\bar M_2^0$. Unlike \textbf{Category 1}, there are four possible paths, as shown in  Fig.~\ref{fig:4path}. Assuming negligible direct CP asymmetries, we can express $\braket{\bar{M}^0_2}{\bar{M}^0_1} = \braket{M^0_2}{M^0_1}e^{i\omega}$, and $\braket{f}{M^0_2} = \braket{f}{\bar{M}^0_2}$ under a suitable convention. It is important to note that in practical terms, $M_2^0$ here must be $K^0$. If it were $D^0$ decaying from $B$ mesons, its antiparticle $\bar D^0$ could also decay from the same $B$ meson, rendering this category inapplicable. This justifies the assumption $\braket{f}{M^0_2} = \braket{f}{\bar{M}^0_2}$ as CP violation is tiny in strange decays. The result for the CP asymmetry~\eqref{eq4} is listed below.

To clarify the physical meaning, we separate the $S_h$ and $S_n$ terms in~\eqref{eq4} into two parts, which are given by 
\begin{align}
    S_{n,1}(t_2)=&\ -\frac{e^{-\Gamma_2 t_2}}{2} \Big[\Big(\abs{\frac{q_2}{p_2}}+\abs{\frac{p_2}{q_2}}\Big)\sin{\Phi}\sinh{\frac{\Delta \Gamma_2 t_2}{2}}\nonumber\\
    &-\pqty{\abs{\frac{q_2}{p_2}}-\abs{\frac{p_2}{q_2}}}\cos{\Phi}\sin{\Delta m_2 t_2}\Big]\times\nonumber\\
    &\pqty{\abs{\frac{q_1}{p_1}}+\abs{\frac{p_1}{q_1}}} , \label{eq14-1}\\
    S_{h,1}(t_2)=&\ \frac{e^{-\Gamma_2 t_2}}{2}\Big[\pqty{\abs{\frac{q_2}{p_2}}+\abs{\frac{p_2}{q_2}}}\cos{\Phi}\sinh{\frac{\Delta \Gamma_2 t_2}{2}}\nonumber\\
    &+\pqty{\abs{\frac{q_2}{p_2}}-\abs{\frac{p_2}{q_2}}}\sin{\Phi}\sin{\Delta m_2 t_2}\Big]\times \nonumber\\
    &\pqty{\abs{\frac{q_1}{p_1}}-\abs{\frac{p_1}{q_1}}},\label{eq14-2} \\
    S_{n,2}(t_2)=&\ -\pqty{\abs{\frac{p_1}{q_1}}+\abs{\frac{q_1}{p_1}}}\Big[\sin{\pqty{\Phi-\phi_2}}\abs{g_{+,2}(t_2)}^2\nonumber\\
    &+\sin{\pqty{\Phi+\phi_2}}\abs{g_{-,2}(t_2)}^2\Big],\label{eq14-3}\\
S_{h,2}(t_2)=&\ \pqty{\abs{\frac{q_1}{p_1}}-\abs{\frac{p_1}{q_1}}}\Big[\cos{\pqty{\Phi-\phi_2}}\abs{g_{+,2}(t_2)}^2\nonumber\\
&+\cos{\pqty{\Phi+\phi_2}}\abs{g_{-,2}(t_2)}^2\Big],\label{eq14-4}
\end{align}
where $\Phi$ is a shorthand for $\omega-\phi_1+\phi_2$. The $S_{n,1}(t_2)$ and $S_{h,1}(t_2)$ terms are induced by the interference between two oscillation paths $M^0_1 \to M^0_2 \to \bar{M}^0_2$ and $M^0_1 \to \bar{M}^0_1 \to \bar{M}^0_2$, as well as the interference between $M^0_1 \to M^0_2$ and $M^0_1 \to \bar{M}^0_1 \to \bar{M}^0_2 \to M^0_2$. The $S_{n,2}(t_2)$ and $S_{h,2}(t_2)$ terms are induced by the interference between $M^0_1 \to M^0_2$ and $M^0_1 \to \bar M^0_1 \to \bar{M}^0_2$ and also the interference between $M^0_1 \to M^0_2\to \bar M^0_2$ and $M^0_1 \to \bar M^0_1 \to \bar{M}^0_2 \to M_2^0$. The other terms involved in~\eqref{eq4} are 
\begin{align}
C_n(t_2) =&\ \frac{e^{-\Gamma_2 t_2}}{2}\Big\{\sinh{\frac{\Delta \Gamma_2 t_2}{2}}\cos{\phi_2}\Big[\abs{\frac{q_2}{p_2}}\pqty{1+\abs{\frac{p_1}{q_1}}^2}\nonumber \\ 
&-\abs{\frac{p_2}{q_2}}\pqty{1+\abs{\frac{q_1}{p_1}}^2} \Big]+\sin{\Delta m_2 t_2}\sin{\phi_2}\Big[\abs{\frac{q_2}{p_2}}\nonumber \\
&\times\pqty{1+\abs{\frac{p_1}{q_1}}^2}+\abs{\frac{p_2}{q_2}}\pqty{1+\abs{\frac{q_1}{p_1}}^2} \Big]\Big\}\nonumber\\
&+\frac{1}{2}\left[\abs{\frac{q_2}{p_2}}^2\pqty{1+\abs{\frac{p_1}{q_1}}^2}-\abs{\frac{p_2}{q_2}}^2\pqty{1+\abs{\frac{q_1}{p_1}}^2}\right]\nonumber\\
&\times\abs{g_{-,2}(t_2)}^2-\frac{1}{2}\pqty{\abs{\frac{q_1}{p_1}}^2-\abs{\frac{p_1}{q_1}}^2}\abs{g_{+,2}(t_2)}^2,\label{eq14-5}
\end{align}
\begin{align}
C_h(t_2) =&\ \frac{e^{-\Gamma_2 t_2}}{2}\Big\{\sinh{\frac{\Delta \Gamma_2 t_2}{2}}\cos{\phi_2}\Big[\abs{\frac{q_2}{p_2}}\pqty{1-\abs{\frac{p_1}{q_1}}^2}\nonumber \\ 
&-\abs{\frac{p_2}{q_2}}\pqty{1-\abs{\frac{q_1}{p_1}}^2} \Big]+\sin{\Delta m_2 t_2}\sin{\phi_2}\Big[\abs{\frac{q_2}{p_2}}\nonumber \\ 
&\times\pqty{1-\abs{\frac{p_1}{q_1}}^2}+\abs{\frac{p_2}{q_2}}\pqty{1-\abs{\frac{q_1}{p_1}}^2} \Big]\Big\}\nonumber\\
&+\frac{1}{2} \left[\abs{\frac{q_2}{p_2}}^2\pqty{1-\abs{\frac{p_1}{q_1}}^2}-\abs{\frac{p_2}{q_2}}^2\pqty{1-\abs{\frac{q_1}{p_1}}^2}\right]\nonumber\\
&\times\abs{g_{-,2}(t_2)}^2+\frac{1}{2}\pqty{\abs{\frac{q_1}{p_1}}^2-\abs{\frac{p_1}{q_1}}^2}\abs{g_{+,2}(t_2)}^2,\label{eq14-6} \\
S_n^\prime(t_2) =& -\frac{1}{2}\pqty{\abs{\frac{q_1}{p_1}}-\abs{\frac{p_1}{q_1}}}\Big\{2\sin{\pqty{\Phi-\phi_2}}\abs{g_{+,2}(t_2)}^2\nonumber\\
&+2\sin{\pqty{\Phi+\phi_2}}\abs{g_{-,2}(t_2)}^2  +e^{-\Gamma_2 t_2}\times\nonumber\\
&\Big[ \pqty{\abs{\frac{q_2}{p_2}}+\abs{\frac{p_2}{q_2}}}\sin{\Phi}\sinh{\frac{\Delta \Gamma_2 t_2}{2}}\nonumber\\
&-\pqty{\abs{\frac{q_2}{p_2}}-\abs{\frac{p_2}{q_2}}}\cos{\Phi}\sin{\Delta m_2 t_2}\Big]\Big\},\label{eq14-7}\\
S_h^\prime(t_2) =&\ \frac{1}{2}\pqty{\abs{\frac{q_1}{p_1}}+\abs{\frac{p_1}{q_1}}}\Big\{2\cos{\pqty{\Phi-\phi_2}}\abs{g_{+,2}(t_2)}^2\nonumber\\
&+2\cos{\pqty{\Phi+\phi_2}}\abs{g_{-,2}(t_2)}^2 +e^{-\Gamma_2 t_2}\times \nonumber\\
&\Big[\pqty{\abs{\frac{q_2}{p_2}}+\abs{\frac{p_2}{q_2}}}\cos{\Phi}\sinh{\frac{\Delta \Gamma_2 t_2}{2}}\nonumber\\
&+\pqty{\abs{\frac{q_2}{p_2}}-\abs{\frac{p_2}{q_2}}}\sin{\Phi}\sin{\Delta m_2 t_2}\Big]\Big\},\label{eq14-8}\\
C_n^\prime(t_2) =&\ \frac{e^{-\Gamma_2 t_2}}{2}\Big\{\sinh{\frac{\Delta \Gamma_2 t_2}{2}}\cos{\phi_2}\Big[\abs{\frac{q_2}{p_2}}\pqty{1-\abs{\frac{p_1}{q_1}}^2}\nonumber\\
&+\abs{\frac{p_2}{q_2}}\pqty{1-\abs{\frac{q_1}{p_1}}^2} \Big]+\sin{\Delta m_2 t_2}\sin{\phi_2}\Big[\abs{\frac{q_2}{p_2}}\nonumber\\
&\times\pqty{1-\abs{\frac{p_1}{q_1}}^2}-\abs{\frac{p_2}{q_2}}\pqty{1-\abs{\frac{q_1}{p_1}}^2} \Big]\Big\}\nonumber\\
& + \frac{1}{2}\Big[\abs{\frac{q_2}{p_2}}^2\pqty{1-\abs{\frac{p_1}{q_1}}^2}+\abs{\frac{p_2}{q_2}}^2\pqty{1-\abs{\frac{q_1}{p_1}}^2}\Big]\nonumber\\
&\times\abs{g_{-,2}(t_2)}^2-\frac{1}{2}\pqty{\abs{\frac{q_1}{p_1}}^2+\abs{\frac{p_1}{q_1}}^2-2}\abs{g_{+,2}(t_2)}^2,\label{eq14-9}
\end{align}

\begin{align}
C_h^\prime(t_2) =&\ \frac{e^{-\Gamma_2 t_2}}{2}\Big\{\sinh{\frac{\Delta \Gamma_2 t_2}{2}}\cos{\phi_2}\Big[\abs{\frac{q_2}{p_2}}\pqty{1+\abs{\frac{p_1}{q_1}}^2}\nonumber\\
&+\abs{\frac{p_2}{q_2}}\pqty{1+\abs{\frac{q_1}{p_1}}^2} \Big]+\sin{\Delta m_2 t_2}\sin{\phi_2}\Big[\abs{\frac{q_2}{p_2}}\nonumber\\
&\times\pqty{1+\abs{\frac{p_1}{q_1}}^2}-\abs{\frac{p_2}{q_2}}\pqty{1+\abs{\frac{q_1}{p_1}}^2} \Big]\Big\}\nonumber\\
&+  \frac{1}{2}\left[\abs{\frac{q_2}{p_2}}^2\pqty{1+\abs{\frac{p_1}{q_1}}^2}+\abs{\frac{p_2}{q_2}}^2\pqty{1+\abs{\frac{q_1}{p_1}}^2}\right]\nonumber\\
&\times\abs{g_{-,2}(t_2)}^2+\frac{1}{2}\pqty{\abs{\frac{q_1}{p_1}}^2+\abs{\frac{p_1}{q_1}}^2+2}\abs{g_{+,2}(t_2)}^2.\label{eq14-10}
\end{align}

Again, if the decay occurs via $M^0_1 \to \bar{M}^0_2$ instead of $M^0_1 \to M^0_2$, the results can be directly obtained from~\eqref{eq14-1} -~\eqref{eq14-10} by replacing $p_2/q_2$ with $q_2/p_2$.

\noindent \textbf{Category 3}

\begin{figure}[htbp]
    \centering
\includegraphics[keepaspectratio,width=8cm]{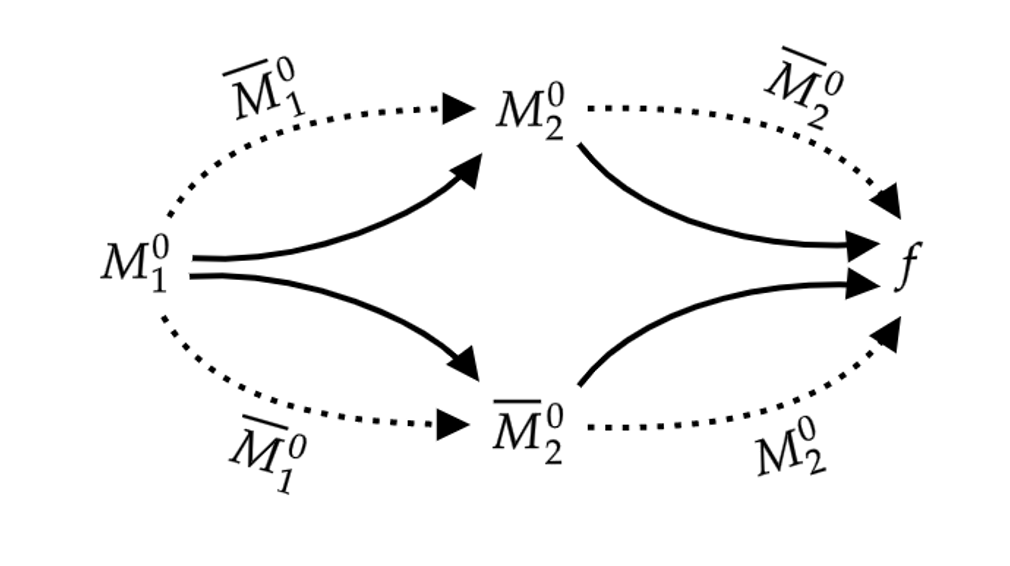}
\caption{Eight oscillation paths in the cascade decay $M_1^0\to M_2\to f$, with $M_1^0\to M_2^0,\bar M_2^0$, $\bar M_2^0\to f,\bar f$ and their CP conjugate processes being allowed.}\label{fig:8path}
\end{figure}

In the third category, $M^0_1$ can decay into both $M^0_2$ and $\bar{M}^0_2$, so the number of possible paths for $M_1^0\to M_2\to f$ is doubled compared with the second category, as shown in Fig.~\ref{fig:8path}. To evaluate the CP asymmetry of such a process, the following parameters need to be introduced, 
\begin{align}\label{eq15}
& \frac{\braket{ M^0_2}{M^0_1}}{\braket{\bar{M}^0_2}{M^0_1}} = r_1 e^{i(\delta_1 + \theta_1)} ,\qquad \frac{\braket{f}{\bar{M}^0_2}}{\braket{f}{M^0_2}} = r_2 e^{i\pqty{\delta_2+\theta_2}}  , \nonumber\\
 &\frac{\braket{M^0_2}{\bar{M}^0_1}}{\braket{\bar{M}^0_2}{M^0_1}} = e^{i\theta_3}, 
\end{align}
where $\delta_{1,2}$ are the relative strong phases, $\theta_{1,2,3}$ are the relative weak phases, and $r_{1,2}$ are the magnitude ratios. All complete results can be found in Appendix 1, where~\eqref{A1-3} and~\eqref{A1-5} correspond to the double-mixing CP violation terms. 
This is the most complex and comprehensive scenario, and by making appropriate approximations for this category, we can derive the complete results of the preceding two categories.

\section{Numerical analysis}\label{Section3}

In this section, we will conduct numerical analyses of the double-mixing CP violation for various channels. As $B_d^0$ and $B_s^0$ have very different oscillation rates, we will discuss their decays separately. The input parameters involved are detailed in Table~\ref{input}.

\begin{table*}
\centering
\caption{The input parameters and their values, with $x_M \equiv \Delta m_M/\Gamma_M$ and $y_M \equiv \Delta\Gamma_M / (2\Gamma_M)$, respectively.}\label{input}
\begin{tabular}{cr|cr}
\hline
 \textbf{Parameter}    &  \textbf{Value} & \textbf{Parameter}    &  \textbf{Value} \\
\hline
  $|q_{B_d}/p_{B_d}|$   &    $1.0010 \pm 0.0008$~\cite{HFLAV:2022pwe} & $x_{B_d}$ & $0.769\pm 0.004$~\cite{Workman:2022ynf}\\
  $|q_{B_s}/p_{B_s}|$   &    $1.0003\pm 0.0014$~\cite{HFLAV:2022pwe}  & $x_{B_s}$ & $27.03\pm 0.09$~\cite{Workman:2022ynf}\\
  $|q_K/p_K|$   &    $0.996774\pm 0.000019$~\cite{Workman:2022ynf} & $x_K$ & $0.946\pm 0.002$~\cite{Workman:2022ynf} \\
  $|q_D/p_D|$   &    $0.995\pm 0.016$~\cite{HFLAV:2022pwe}  & $x_D$ & $\pqty{4.1\pm 0.5}\times 10^{-3}$~\cite{HFLAV:2022pwe} \\
  $\phi_{B_d}$      &    $\pqty{44.4 \pm 1.4}^\circ$~\cite{Workman:2022ynf}  & $y_{B_d}$ & $-0.0005\pm 0.0050$~\cite{Workman:2022ynf} \\
  $\phi_{B_s}$      &    $-\pqty{2.106\pm 0.135}^\circ$~\cite{Workman:2022ynf} & $y_{B_s}$ & $-0.064\pm 0.003$~\cite{Workman:2022ynf}  \\
  $\phi_K$      &    $\pqty{0.176\pm 0.001}^\circ$~\cite{Workman:2022ynf} & $y_K$ & $-0.996506\pm 0.000016$~\cite{Workman:2022ynf} \\
  $\phi_D$      &    $-\pqty{177.5\pm 1.2}^\circ$~\cite{Workman:2022ynf} & $y_D$ & $\pqty{6.2\pm 0.6}\times 10^{-3}$~\cite{HFLAV:2022pwe}\\
  $\omega_{B_d}$      &    $\pqty{0.00372^{+0.00025}_{-0.00023}}^\circ$~\cite{Workman:2022ynf} & $\omega_{B_s}$ & $\pqty{-0.070^{+0.004}_{-0.005}}^\circ$~\cite{Workman:2022ynf}\\
  $\omega_K$      &    $\pqty{43.52\pm 0.05}^\circ$~\cite{Workman:2022ynf} & $\omega_D$ & $\pqty{83.4446^{+32.2995}_{-36.5375}}^\circ$~\cite{Workman:2022ynf}\\
  $\omega_1$      &    $0^\circ$~\cite{Workman:2022ynf} & $\omega_2$ & $\pqty{65.54\pm 1.55}^\circ$~\cite{Workman:2022ynf}\\
  $\omega_3$      &    $0^\circ$~\cite{Workman:2022ynf} & $\omega_4$ & $\pqty{-114.419\pm 1.549}^\circ$~\cite{Workman:2022ynf}\\
  $\omega_5$ &  $\pqty{-131.093\pm 3.094}^\circ$~\cite{Workman:2022ynf} &$r_D$  & $-\pqty{5.857\pm 0.017}\times 10^{-2}$~\cite{HFLAV:2022pwe}\\
    $\delta_D$ & $\pqty{7.2^{+7.9}_{-9.2}}^\circ$ ~\cite{HFLAV:2022pwe} & &\\
\hline
\end{tabular}
\end{table*}
\subsection{\texorpdfstring{$B^0_d$}{} decays}

In this subsection, we discuss the double-mixing CP violation in cascade $B_d^0$ decays. For the $B^0_d$ meson, it is a reasonable approximation to assume that the decay width difference, denoted as $\Delta \Gamma_{B_d}$, is negligible~\cite{HFLAV:2022pwe, Workman:2022ynf}. Consequently, all terms involving $\sinh{\Delta \Gamma_1 t_1/2}$ in~\eqref{eq4} can be disregarded while $\cosh{\Delta \Gamma_1 t_1/2}$ effectively equals 1. Furthermore, the magnitude of ratio $q_{B_d}/p_{B_d}$ is approximated to be 1~\cite{HFLAV:2022pwe, Workman:2022ynf}. Under these approximations, the double-mixing CP violation $A_{dm}$ is simplified to $A_{n, dm}$. Here, $A_{n, dm}$ represents the component of CP violation induced by double-mixing interference and it is proportional to $\sin{\Delta m_1 t_1}$. This simplification aids in a more straightforward analysis of the double-mixing CP violation effect in $B^0_d$ decays.

\subsubsection{\texorpdfstring{$B^0_d \to J/\psi K \to J/\psi (\pi^\pm\ell^\mp \nu)$}{}}

We initially explore the decay channel $B^0_d \to J/\psi K \to J/\psi $
$(\pi^+\ell^-\bar\nu_\ell)$. The decay mode $B^0_d \to J/\psi K^0$, corresponding to the $b \to c\bar{c}s$ transition at the quark level, is frequently employed for the extraction of the CKM angle $\beta$ because of the tiny pollution of strong interaction and the relatively substantial CP violation effect~\cite{Workman:2022ynf}. The cascading mixing effect in this decay channel where the secondary decay is semileptonic has been previously investigated in paper~\cite{Azimov:1997ix}. 

In this decay channel, only the paths $B_d^0 \to J/\psi K^0 \to J/\psi \bar{K}^0 \to J/\psi (\pi^+\ell^-\bar\nu_\ell)$ and $B_d^0 \to \bar{B}_d^0 \to J/\psi \bar{K}^0 \to J/\psi $
$(\pi^+\ell^-\bar\nu_\ell)$ can contribute. This case aligns with the first category defined previously. As a reasonable approximation, we have neglected the direct CP violation in neutral $K$ meson decay, assuming $\braket{\pi^+\ell^-\bar\nu_\ell}{\bar{K}^0}=\braket{\pi^-\ell^+\nu_\ell}{K^0}$. The primary decay amplitudes are related as
\begin{align}
\frac{\braket{J/\psi\bar{K}^0}{\bar{B}^0_d}}{\braket{J/\psi K^0}{B^0_d}} = -e^{i\omega_{B_d}},
\end{align}
where $\omega_{B_d}$ is a pure weak phase, with the latest result given by $\arg(V_{cb}^{} V^\star_{cs}/V^\star_{cb}V_{cs}^{})=\pqty{0.00372^{+0.00025}_{-0.00023}}^\circ$~\cite{Workman:2022ynf}. It is important to note that, the minus sign originates from the convention $CP\ket{J/\psi K^0} = -\ket{J/\psi \bar{K}^0}$. Then, all the components in~\eqref{eq4} are calculated as:
\begin{align}
    A_{n,dm}(t_1,t_2) =& \ \frac{e^{-\pqty{\Gamma_{B_d} t_1 +\Gamma_K t_2}}}{2 D(t_1, t_2)}\Big\{\sinh{\frac{\Delta \Gamma_K t_2}{2}}\sin{\Phi_1}\Big(\abs{\frac{p_K}{q_K}}\nonumber\\
    &+\abs{\frac{q_K}{p_K}}\Big)-\sin{\Delta m_K t_2}\cos{\Phi_1}\Big(\abs{\frac{q_K}{p_K}}\nonumber\\
    &-\abs{\frac{p_K}{q_K}}\Big)\Big\}\sin{\Delta m_{B_d} t_1},\label{eq38-1}\\  
    A_{non-dm}(t_1, t_2) =&\ \frac{1}{D(t_1, t_2)}\pqty{\abs{\frac{q_K}{p_K}}^2-\abs{\frac{p_K}{q_K}}^2}\abs{g_{-,K}(t_2)}^2\nonumber\\
    &\times\abs{g_{+,B_d}(t_1)}^2, \label{eq38-2}
\end{align}
where $\Phi_1$ is defined as $\omega_{B_d} - \phi_{B_d}+\phi_K$, approximated as -2$\beta$. The denominator $D(t_1,t_2)$ is given by~\eqref{eq4} with the $t_2$ functions substituted with \eqref{eq12} - \eqref{eq13-2}. Hence, by measuring the double-mixing CP violation of this channel, the $\beta$ angle can be extracted. The double-mixing CP violation $A_{n, dm}(t_1, t_2)$ is induced by the interference between the decay paths $B_d^0 \to J/\psi K^0 \to J/\psi \bar{K}^0 \to J/\psi(\pi^+\ell^-\bar\nu_\ell)$ and $B_d^0 \to \bar{B}_d^0 \to J/\psi \bar{K}^0$
$ \to J/\psi(\pi^+\ell^-\bar\nu_\ell)$, corresponding to the $S_n(t_2)$ component in~\eqref{eq4}. Meanwhile, $A_{non-dm}(t_1,t_2) $ corresponds to the $C_+(t_2)$ term within $N(t_1,t_2)$ as defined in~\eqref{eq4}.

\begin{figure*}[htbp]
    \centering
    \includegraphics[keepaspectratio,width=5.3cm]{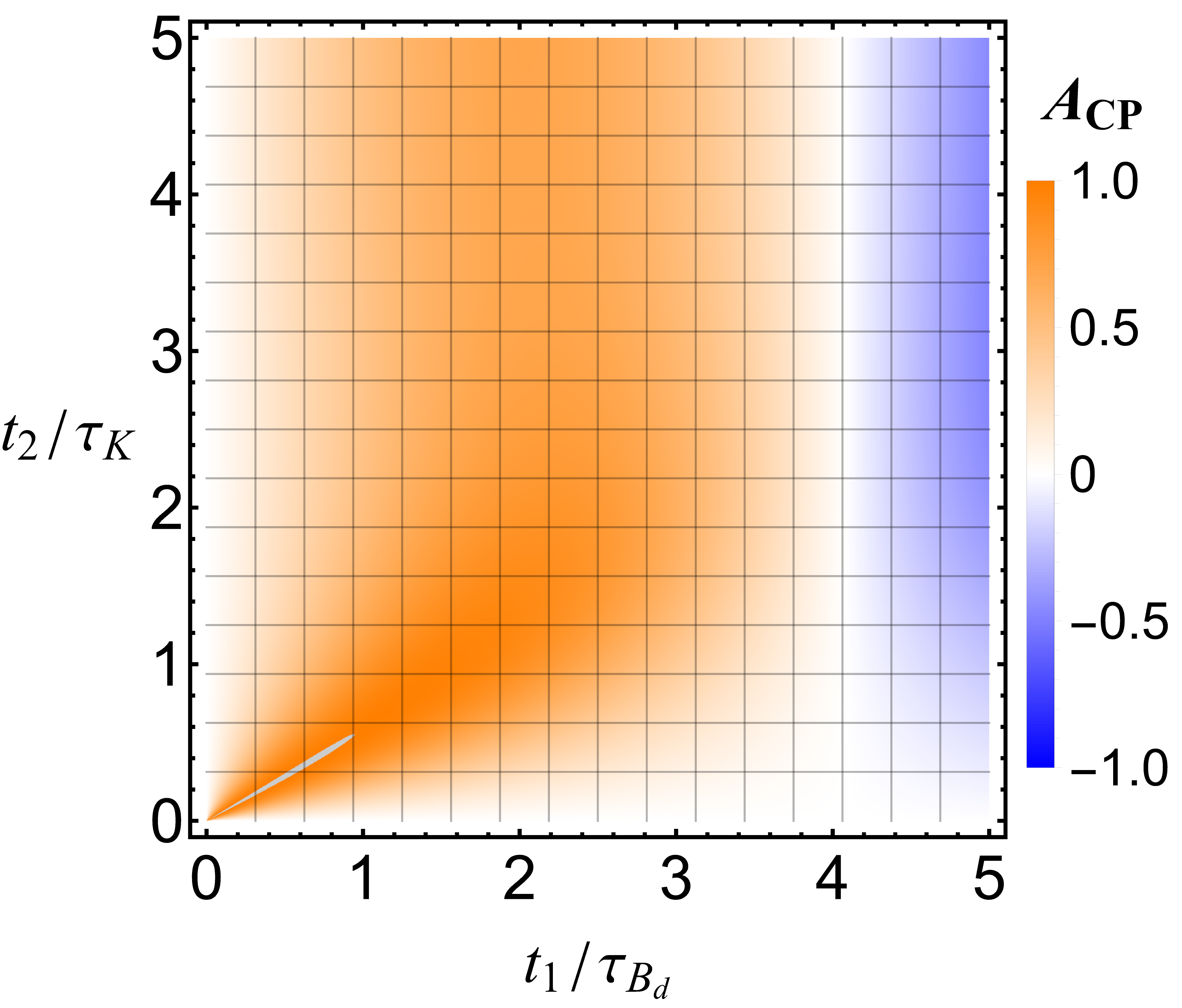}
    \hspace{0.1cm}
    \includegraphics[keepaspectratio,width=4.2cm]{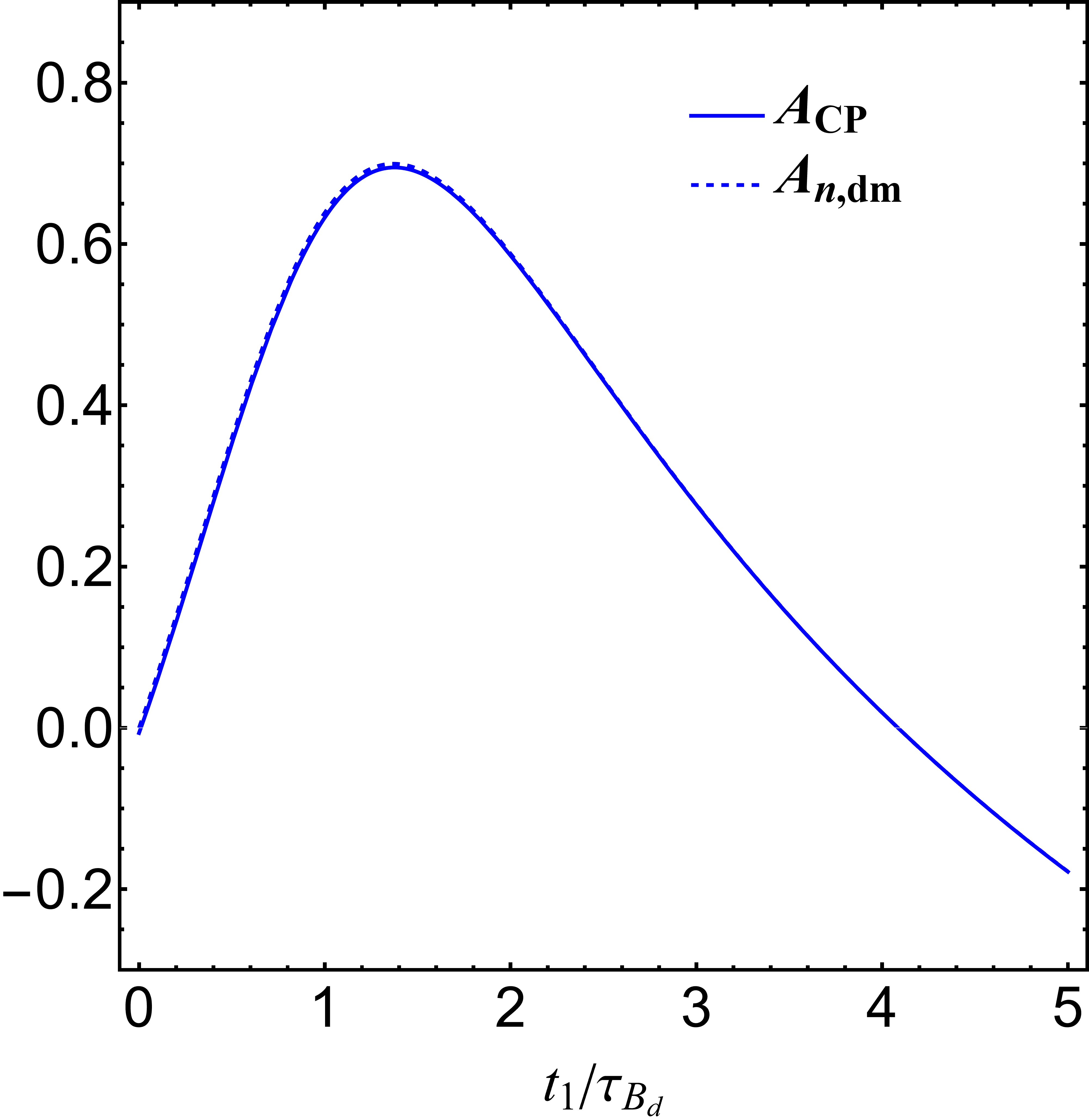}
     \hspace{0.1cm}
    \includegraphics[keepaspectratio,width=4.2cm]{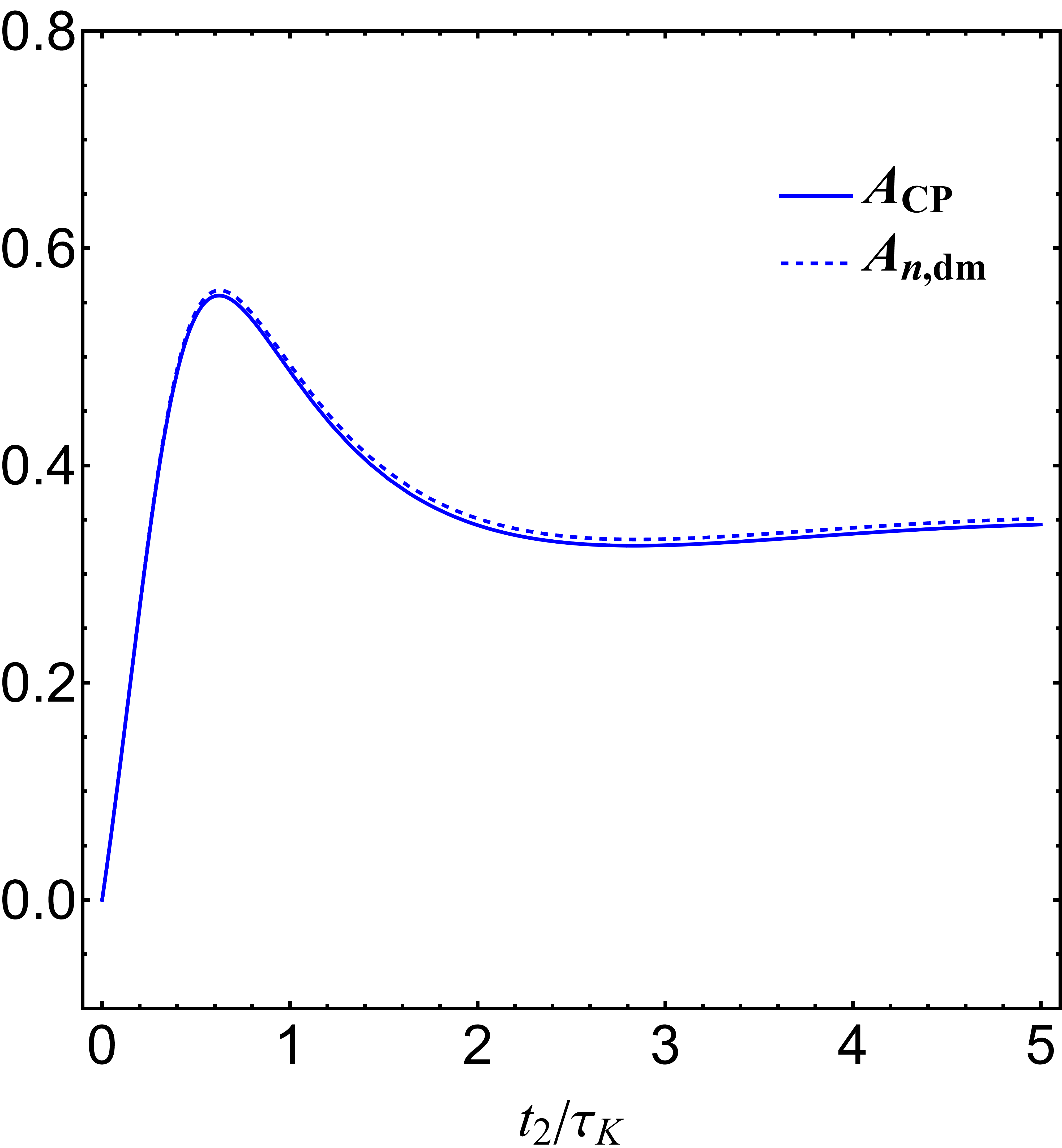}
    \caption{Time dependence of the double-mixing CP asymmetry $A_{\rm CP}$ in $B^0_d(t_1) \to J/\psi K(t_2) \to J/\psi (\pi^+\ell^-\bar\nu_\ell)$. The left panel displays the two-dimensional time dependence. The middle panel and the right panel display the dependence on $t_1$ (with $t_2$ integrated from 0 to $2\tau_{K}$) and $t_2$ (with $t_1$ integrated from $0$ to $3\tau_{B_d}$), respectively.}\label{fig:d1}
\end{figure*}

To investigate the double-mixing CP violation effect, we conducted a numerical analysis of both the total CP asymmetry $A_{\rm CP}$ and the double-mixing CP asymmetry $A_{dm}$. Notably, $A_{dm}$ plays a crucial role in $A_{\rm CP}$, as evidenced by the mathematical formula in~\eqref{eq38-1} and the corresponding numerical results in Fig.~\ref{fig:d1}. The left panel clearly shows that the maximum value of the two-dimensional time-dependent double-mixing CP asymmetry can even exceed 50\%. When integrating $t_2$ from 0 to $2\tau_K$, as depicted in the middle panel, the double-mixing effect almost overlaps with the total CP asymmetry $A_{\rm CP}$, indicating its dominance. Its magnitude initially increases with $t_1$ and then decreases, a pattern attributed to its sine dependence on $t_1$. Within an interval of 4$\tau_{B_d}$, it completes only half an oscillation period, consistent with the small value of $x_{B_d}$ listed in Table \ref{input}. Similar trends are observed when considering the dependence on $t_2$, as shown in the right panel, where $t_1$ is integrated from 0 to $3\tau_{B_d}$. It rapidly increases within a period shorter than $\tau_K$, and then decreases before stabilizing. This phenomenon can be explained by the fact that the numerator is mainly dominated by the first term in~\eqref{eq38-1}, which increases from 0 to $\tau_K$ and then stabilizes. and that the denominator is dominated by $C^\prime_+(t_2)$ and $C^\prime_-(t_2)$, which decreases from 0 to 0.5$\tau_K$ and then steadily increases until around $t_2 \approx 2.5\tau_K$. A comprehensive two-dimensional time integration of $A_{\rm CP}(t_1,t_2)$ yields the results presented in Table $\ref{ACP1}$. For experimental convenience, we perform the integration over different $t_1$ and $t_2$ intervals. The integrated $A_{\rm CP}$ fluctuates around 30-40\% with different configurations.

\begin{table}[h!]
\centering
\caption{The time integrated $A_{\rm CP}$ in $B^0_d \to J/\psi K \to J/\psi (\pi^+\ell^-\bar\nu_\ell)$.}\label{ACP1}
\begin{tabular}{ccc}
\hline
$t_1/\tau_{B_d}$ & $t_2/ \tau_K$ & $A_{\rm CP}$\\
\hline
0$\sim$3 & 0$\sim$ 2 & 40.52\%\\
0$\sim$2 & 0$\sim$ 2 & 39.76\%\\

0$\sim$3 & 0$\sim$ 1 & 42.09\%\\
0$\sim$2 & 0$\sim$ 1 & 48.04\%\\
0$\sim$3 & 0$\sim$ 0.5 & 29.28\% \\
0$\sim$2 & 0$\sim$ 0.5 & 38.76\%\\
\hline
\end{tabular}
\end{table}

The yield of $B^0_d \to J/\psi K^0_S$ has been previously observed at LHCb~\cite{LHCb:2015brj} using data from Run 1 of the LHC at a center-of-mass energy $\sqrt{s}=7\,\rm{TeV}$ with an integrated luminosity 3 $\rm{fb}^{-1}$. This observation involved the final states of $J/\psi \to \mu^+\mu^-$ and $K^0_S \to \pi^+\pi^-$. Based on this study, we estimate the yield of $B^0_d \to J/\psi K \to J/\psi (\pi^+\ell^-\bar\nu_\ell)$ with 300 $\rm{fb}^{-1}$ to be approximately  $3.6 \times 10^4$.

Substituting the semileptonic final state with $\pi^-\ell^+\nu_\ell$ instead of $\pi^+\ell^-\bar{\nu}_{\ell}$ resulting in two viable decay paths, denoted as $B^0_d \to J/\psi K^0 \to J/\psi (\pi^-\ell^+\nu_\ell)$ and $B^0_d \to \bar{B}^0_d \to J/\psi \bar{K}^0 \to J/\psi K^0 \to J/\psi (\pi^-\ell^+\nu_\ell)$. The corresponding CP asymmetries are then calculated to be
\begin{align}
    A_{n,dm}(t_1,t_2) =&\ \frac{e^{-\pqty{\Gamma_{B_d} t_1 +\Gamma_K t_2}}}{2 D(t_1, t_2)}\Big\{\sinh{\frac{\Delta \Gamma_K t_2}{2}}\sin{\Phi_1}\Big(\abs{\frac{p_K}{q_K}}\nonumber\\
    &+\abs{\frac{q_K}{p_K}}\Big)-\sin{\Delta m_K t_2}\cos{\Phi_1}\nonumber \\
    &\times \pqty{\abs{\frac{q_K}{p_K}}-\abs{\frac{p_K}{q_K}}}\Big\}\sin{\Delta m_{B_d} t_1},\label{eq39-1}
\end{align}
\begin{align}
A_{non-dm}(t_1, t_2) =&\ \frac{1}{ D(t_1, t_2)}\pqty{-\abs{\frac{q_K}{p_K}}^2+\abs{\frac{p_K}{q_K}}^2}\abs{g_{-,K}(t_2)}^2\nonumber\\
&\times\abs{g_{-,B_d}(t_1)}^2.\label{eq39-2}
\end{align}
The denominator $D(t_1,t_2)$ here can be obtained by combining~\eqref{eq4} with~\eqref{eq12} - \eqref{eq13-2}, while also performing the substitution $C^\prime_+(t_2)  \leftrightarrow C^\prime_-(t_2)$, and adding an extra negative sign in front of $S^\prime_n(t_2)$.

\begin{figure*}[htbp]
    \centering
    \includegraphics[keepaspectratio,width=5.3cm]{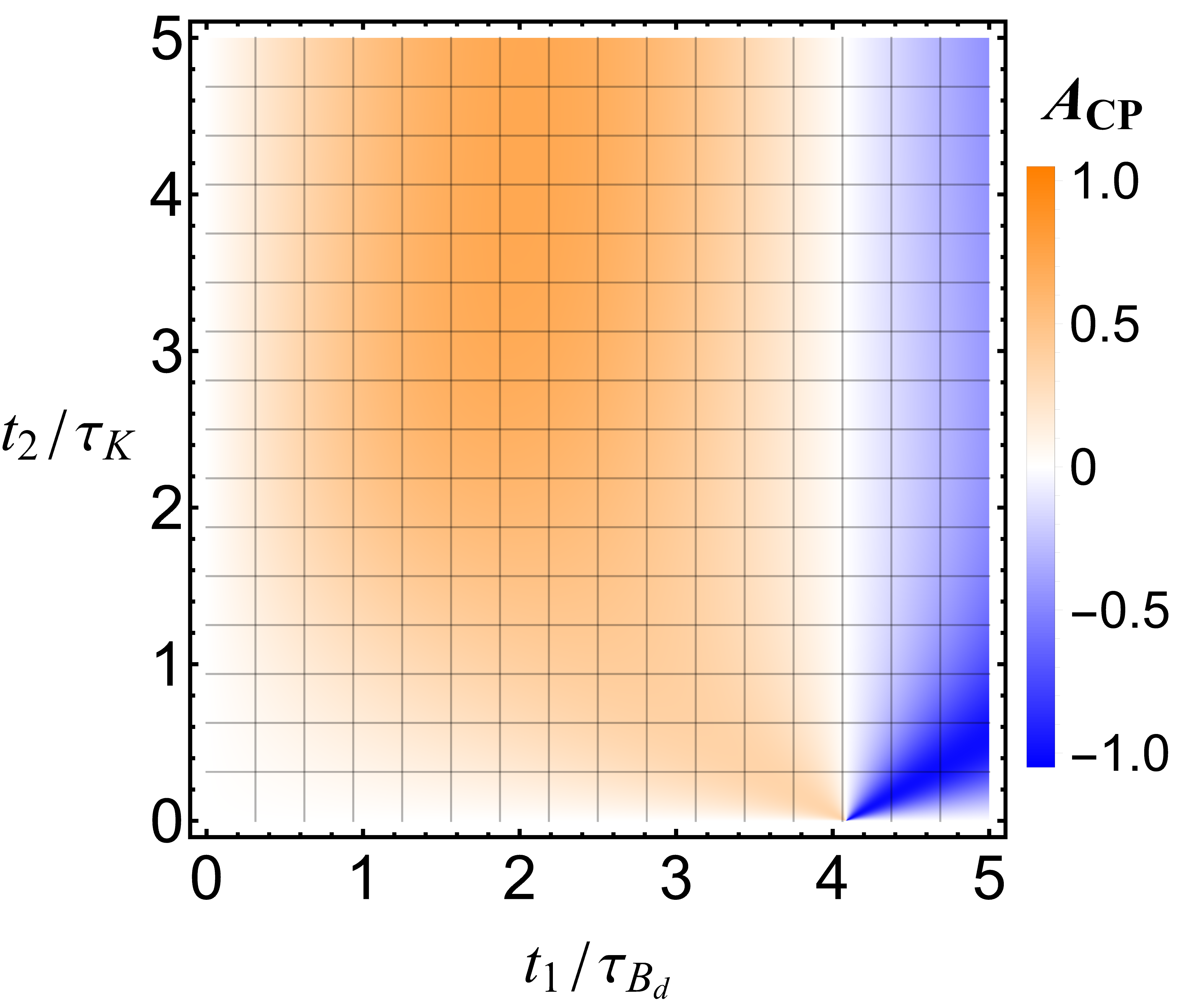}
    \hspace{0.1cm}
    \includegraphics[keepaspectratio,width=4.2cm]{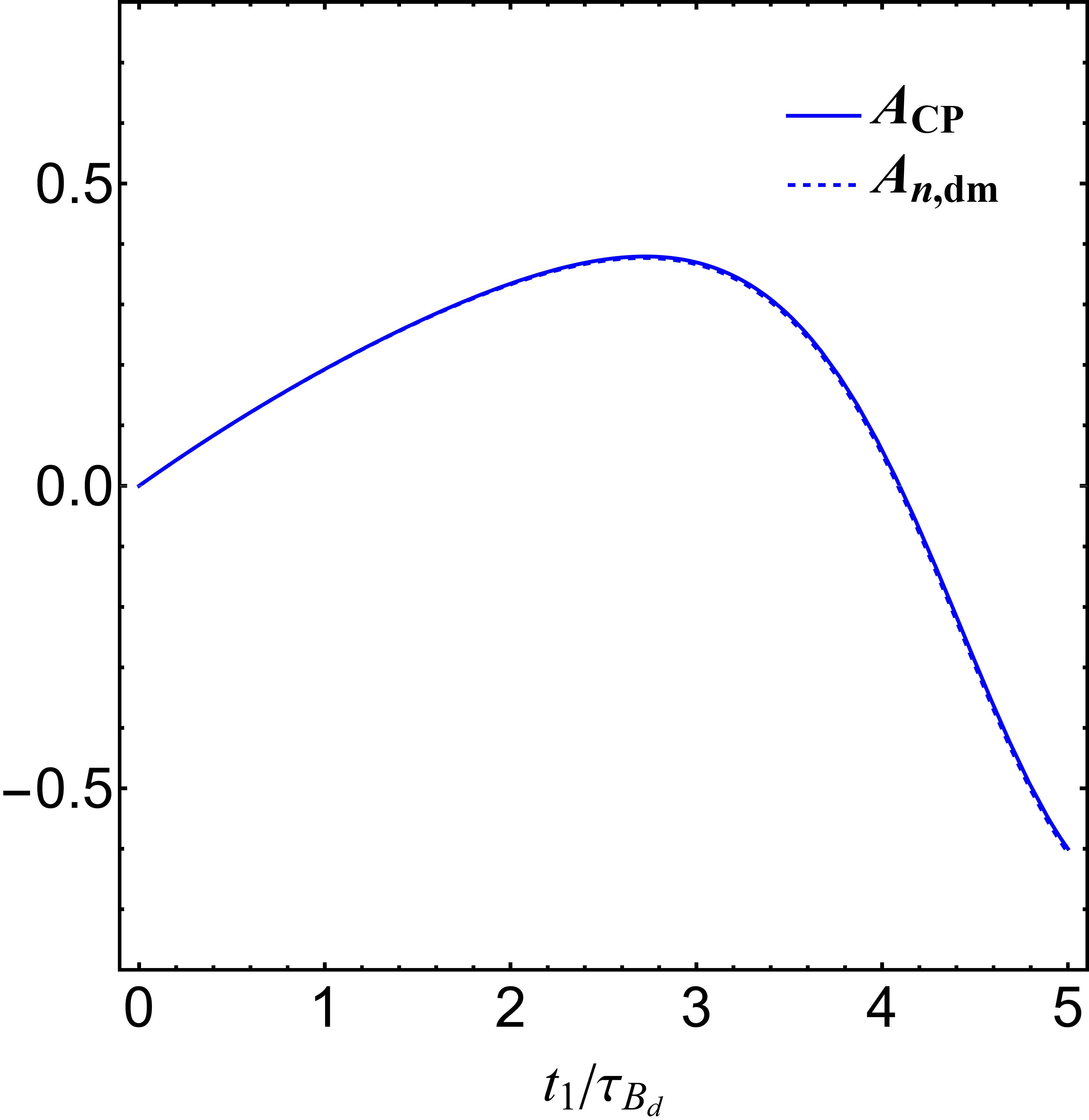}
     \hspace{0.1cm}
    \includegraphics[keepaspectratio,width=4.2cm]{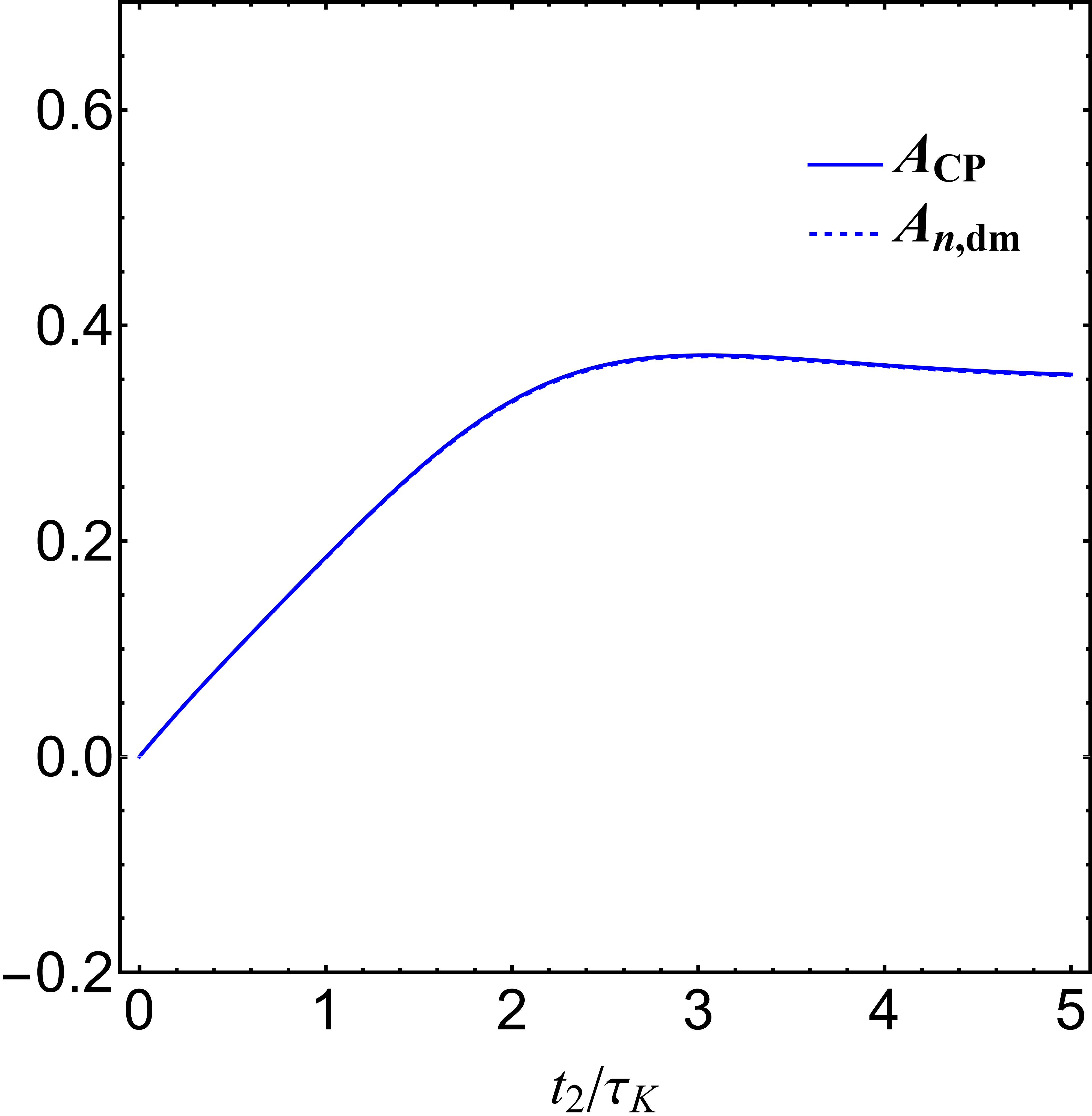}
    \caption{Time dependence of the double-mixing CP asymmetry $A_{\rm CP}$ in $B^0_d(t_1) \to J/\psi K(t_2) \to J/\psi (\pi^-\ell^+\nu_\ell)$. The left panel displays the two-dimensional time dependence. The middle panel and the right panel display the dependence on $t_1$ (with $t_2$ integrated from 0 to $2\tau_{K}$) and $t_2$ (with $t_1$ integrated from $0$ to $3\tau_{B_d}$), respectively.}\label{fig:d1n}
\end{figure*}

Compared to the final state $\pi^+\ell^-\bar{\nu}_{\ell}$, a key difference for the final state $\pi^-\ell^+ \nu_{\ell}$ lies in an additional negative sign in the numerator of the non-double-mixing CP violation term $A_{non-dm}(t_1,t_2)$, as illustrated in Fig.~\ref{fig:d1n}. The double-mixing CP violation remains significant, with the peak occurring later compared to the decay process with the final state $\pi^+\ell^-\bar{\nu}_{\ell}$. For instance, the peak in the middle panel of Fig.~\ref{fig:d1n} occurs at around $t_1 = 3\tau_{B_d}$, in contrast to about $1.5\tau_{B_d}$ for the other final state. This difference is due to an additional negative sign in the term $S^\prime_n(t_2)$ in~\eqref{eq4}. Moreover, the magnitude of CP violation increases significantly over time, reflecting the later occurrence of its peak in both $t_1$ and $t_2$.

\subsubsection{\texorpdfstring{$B^0_d\to J/\psi K \to J/\psi f_+$}{}}

We now analyze the double-mixing CP violation in the decay chain $B^0_d\to J/\psi K \to J/\psi f_+$. In this context, $f_+$ refers to CP-even final states, such as $\pi^+\pi^-$. Notably, the CP-even state under consideration is common to $K^0$ and $\bar{K^0}$. This will lead to four conceivable decay paths: $B^0_d \to K^0 \to f_+$, $B^0_d \to K^0 \to \bar{K}^0 \to f_+$, $B^0_d \to \bar{B}^0_d \to \bar{K}^0 \to f_+$ and $B^0_d \to \bar{B}^0_d \to \bar{K}^0 \to K^0 \to f_+$. Hence, it belongs to the second category in our classification. Considering the negligible direct CP violation in the neutral $K$ meson($\braket{f_+}{K^0} = \braket{f_+}{\bar{K}^0}$) and the suppressed penguin amplitudes in $B^0_d \to J/\psi K^0$ (both Cabbibo-suppressed and loop-suppressed as indicated in~\cite{Workman:2022ynf}), we obtain all components of the two-dimensional time-dependent CP asymmetry $A_{\rm CP}(t_1,t_2)$ in~\eqref{eq4} as follows:
\begin{align}
    A_{n,dm}(t_1,t_2) = &\ \frac{e^{-\pqty{\Gamma_{B_d} t_1 +\Gamma_K t_2}}}{D(t_1, t_2)}\Big\{\sin{\Delta m_K t_2}\Big(\abs{\frac{q_K}{p_K}}-\abs{\frac{p_K}{q_K}}\Big)\nonumber\\
    &\cos{\Phi_1}-\sinh{\frac{\Delta\Gamma_K t_2}{2}}\Big(\abs{\frac{q_K}{p_K}}+\abs{\frac{p_K}{q_K}}\Big)\nonumber\\
    &\times\sin{\Phi_1}\Big\}\sin{\Delta m_{B_d} t_1},\label{eq18-1}\\
    A_{non-dm}(t_1, t_2) =&\ \frac{2e^{-\Gamma_{B_d} t_1 }}{D(t_1,t_2)}\left\{\sin{(\phi_{B_d} - \omega_{B_d})}\abs{g_{+,K}(t_2)}^2 \right.\nonumber\\
    &\left.+ \sin{(\phi_{B_d} -\omega_{B_d} + 2\phi_K)}\abs{g_{-,K}(t_2)}^2\right\}\nonumber\\
    &\times\sin{\Delta m_{B_d} t_1}.\label{eq18-2}
\end{align}

\begin{figure*}[htbp]
    \centering
    \includegraphics[keepaspectratio,width=5.3cm]{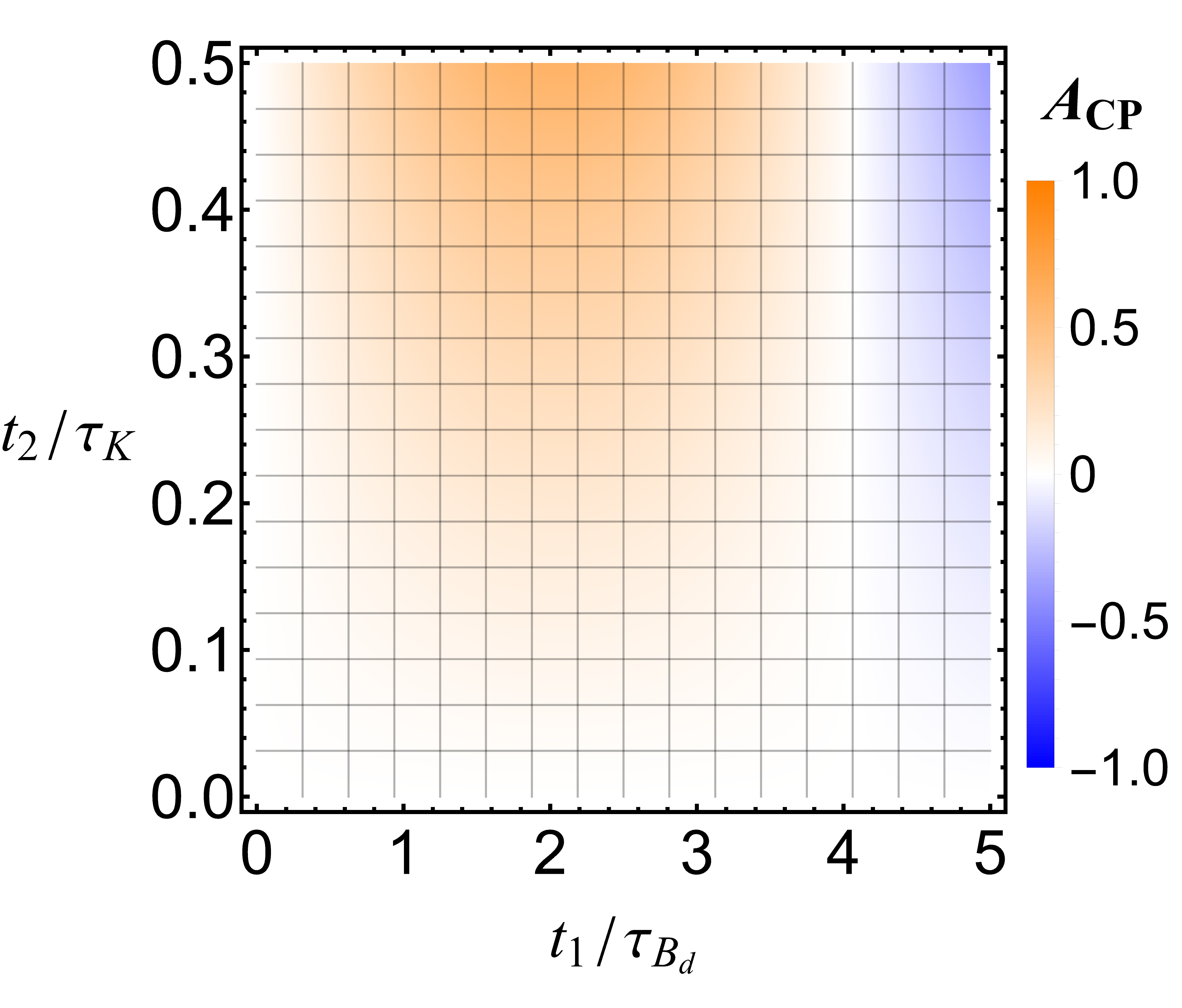}
    \hspace{0.1cm}
    \includegraphics[keepaspectratio,width=4.2cm]{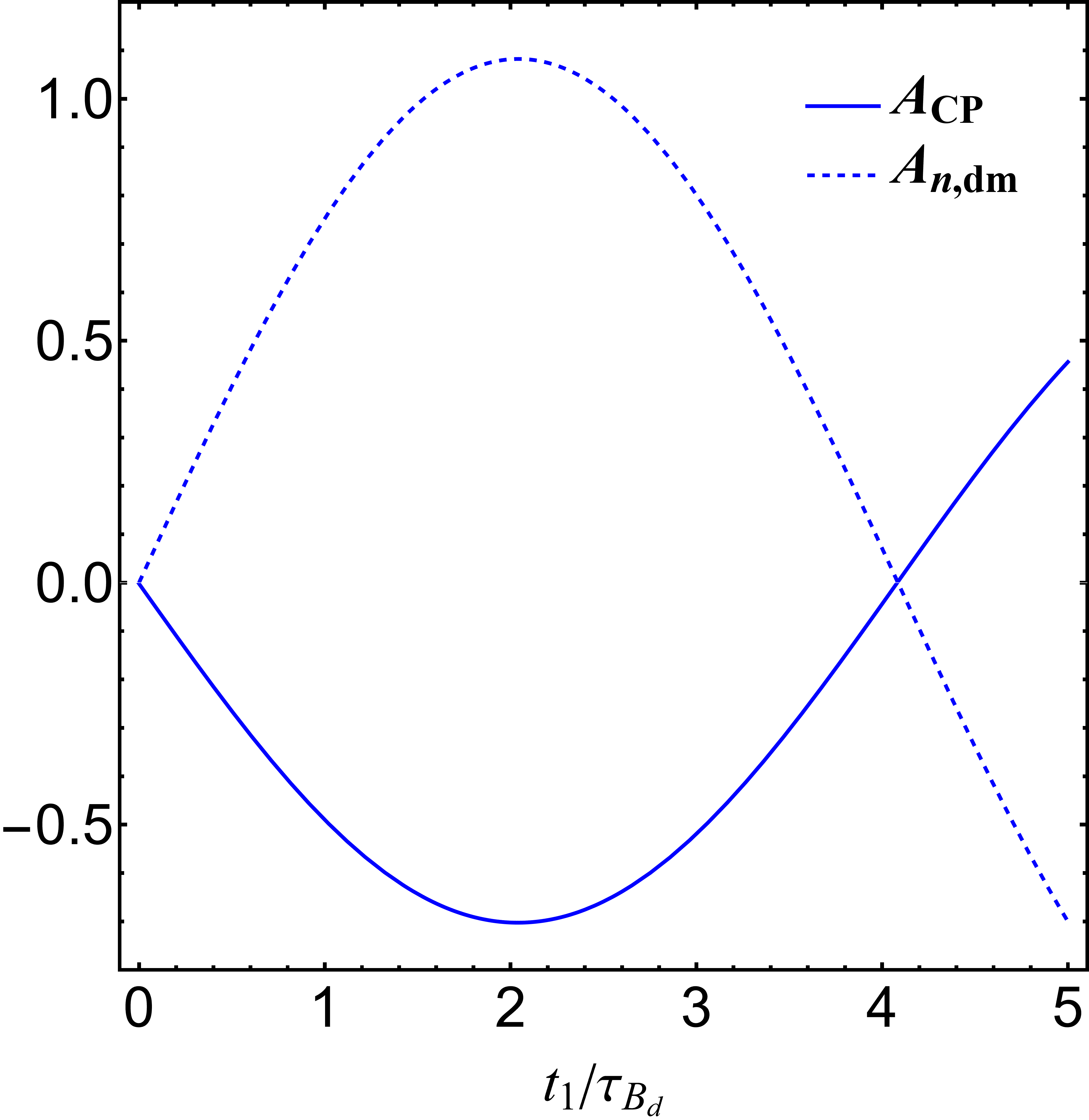}
     \hspace{0.1cm}
    \includegraphics[keepaspectratio,width=4.2cm]{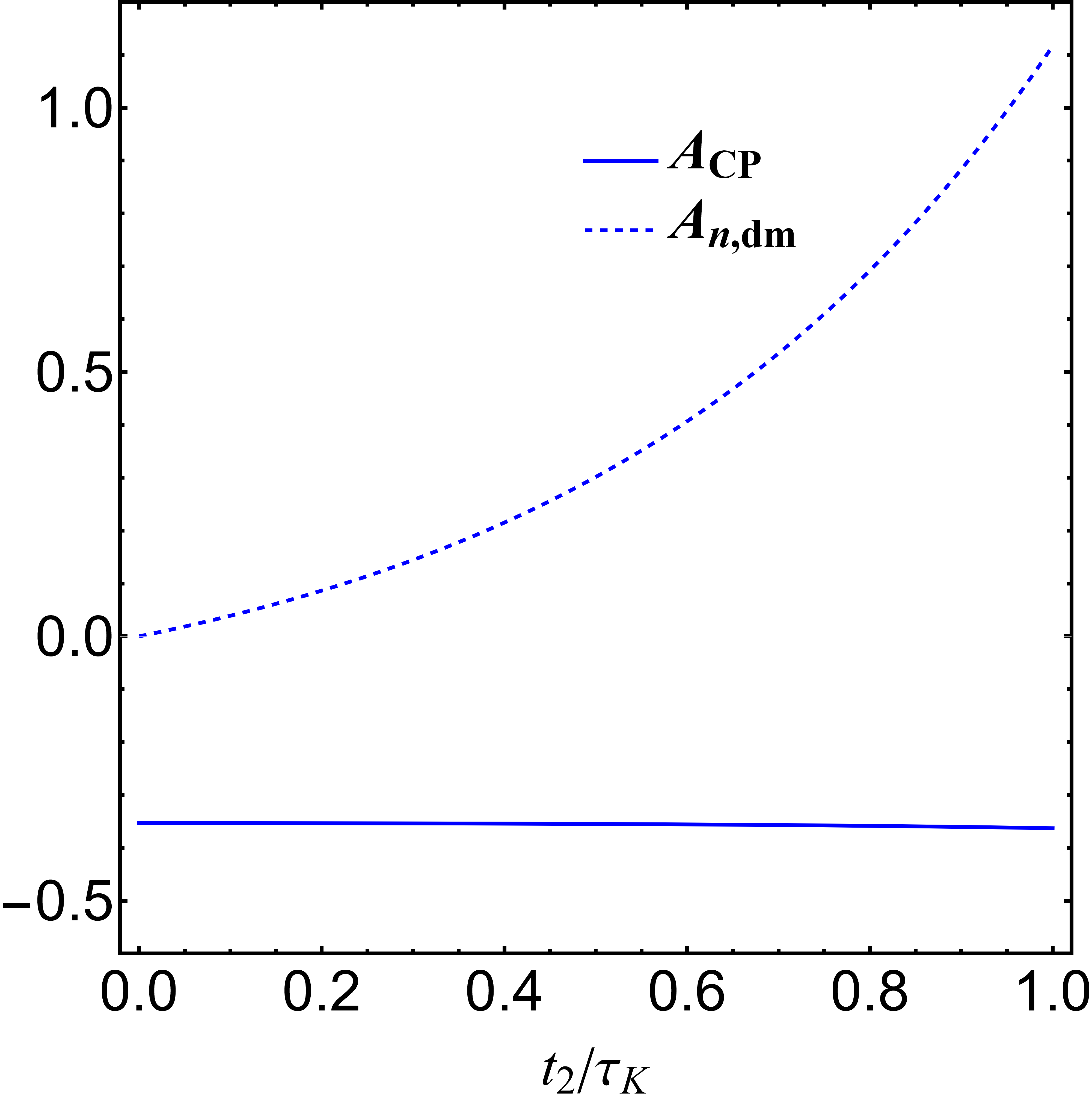}
    \caption{Time dependence of the CP asymmetry $A_{\rm CP}$ in $B^0_d(t_1) \to J/\psi K(t_2) \to J/\psi (\pi^+\pi^-)$. The left panel displays the two-dimensional time dependence. The middle panel and the right panel display the dependence on $t_1$ (with $t_2$ integrated from 0 to $2\tau_{K}$) and $t_2$ (with $t_1$ integrated from $0$ to $3\tau_{B_d}$), respectively.}\label{fig:d2}
\end{figure*}

In this situation, $A_{n, dm}$ and $A_{non-dm}$ constitute $S_n$ in~\eqref{eq4}. Notably, $\Phi_1 \approx -2\beta,\,\phi_{B_d} - \omega_{B_d} \approx 2\beta$ and $\phi_K \approx 0 $. Under the approximation that $\abs{q_K/p_K} \approx 1$, $S_n(t_2)$ simplifies to
\begin{align}
S_n(t_2) = 2\sin{2\beta}e^{-\Gamma_S t_2}.
\end{align}
By taking $q_K/p_K \to 1$, we neglect the indirect CP violation of neutral kaon, so only $K_S^0\to f_+$ contributes to the decay and it explains why only the $e^{-\Gamma_S t_2}$ term survives. As a result, the total CP asymmetry $A_{\rm CP}$ is expressed as
\begin{align}
A_{\rm CP}(t_1,t_2) = \sin{2\beta} \cdot \sin{\Delta m_{B_d} t_1},
\end{align}
which is consistent with the formulas (13.74) and (13.82) in the `` CP Violation in the Quark Sector '' review of the PDG~\cite{Workman:2022ynf}.

\begin{table}
\caption{Time integrated $A_{\rm CP}$ in $B^0_d \to J/\psi K \to J/\psi (\pi^+\pi^-)$.}\label{ACP0}
\begin{center}
{
\begin{tabular}{c c c}
\hline
$t_1/\tau_{B_d}$ & $t_2/ \tau_K$ & $A_{\rm CP}$\\
\hline
0$\sim$3 & 0$\sim$ 2 & -35.36\%\\
0$\sim$2 & 0$\sim$ 2 & -32.35\%\\
0$\sim$3 & 0$\sim$ 1 & -35.30\%\\
0$\sim$2 & 0$\sim$ 1 & -32.28\%\\
0$\sim$3 & 0$\sim$ 0.5 & -35.31\% \\
0$\sim$2 & 0$\sim$ 0.5 & -32.27\%\\
\hline
\end{tabular}
}
\end{center}
\end{table}

We conducted a numerical analysis of $A_{\rm CP}$ and $A_{dm}$ for this process. The two-dimensional time-dependent CP violation result is shown in the left panel of Fig.~\ref{fig:d2}, with its peak appearing around $t_1=2\tau_{B_d}$ and $t_2 > 0.5\tau_K$. By integrating $t_2$ from 0 to $2\tau_K$, the $t_1$-dependent result, displayed in the middle panel, reaches its peak at $t_1=2\tau_{B_d}$. Since both $A_{\rm CP}$ and $A_{n, dm}$ exhibit sine dependence on $t_1$, their oscillations are similar, but with different magnitudes and signs. It can be concluded that the double-mixing contribution and the non-double-mixing contribution have opposite signs, with the former having a smaller magnitude. The right panel of Fig.~\ref{fig:d2} shows that $A_{\rm CP}$ with $t_1$ integrated out is almost independent of $t_2$. On the other hand, the $A_{n, dm}$ part, with its hyperbolic sine dependence on $t_2$, increases with $t_2$ ranging from 0 to $\tau_K$. The two-dimensional time integration of $A_{\rm CP}(t_1,t_2)$ for this channel is given in Table \ref{ACP0}, demonstrating its insensitivity to changes in both $t_1$ and $t_2$, owing to the fact that most of the decays occur earlier.

\subsubsection{\texorpdfstring{$B^0_d \to D K \to (K^-\pi^+) ( \pi^+\ell^-\bar\nu_\ell)$}{}}\label{sec:3.1.3}

The decay $B^0_d \to DK^\star$ is well-known for its precision in measuring the angle $\gamma$. First attempts to constraint $\gamma$ using $B^0_d \to DK^{\star}$ decays have adopted neutral $D$ meson decays into $K^0_S \pi^+\pi^-$~\cite{BaBar:2008mds} and suppressed final states such as $K^-\pi^+$, with relevant branching ratio and parameters measured in several studies~\cite{Belle:2002jgk,BaBar:2005osj,BaBar:2006ftr}. Here, we consider a similar process $B^0_d \to D K $, which was also briefly discussed in our previous work~\cite{Shen:2023nuw}. This channel, comprising two oscillating paths, $B^0_d \to D^0 K^0 \to D^0 \bar{K}^0$ and $B^0_d \to \bar{B}^0_d \to D^0 \bar{K}^0$, falls within the first category. The following relations are employed to evaluate the CP asymmetry, 
\begin{align}\label{eq19}
\frac{A\pqty{\bar{B}^0_d \to D^0 \bar{K}^0}}{A\pqty{B^0_d \to \bar{D}^0 K^0}}&=e^{i\omega_1},\quad \omega_1=\arg{\frac{V_{cb}^{}V^\star_{us}}{V^\star_{cb}V_{us}^{}}}, \\ 
\frac{A\pqty{B^0_d \to D^0 K^0}}{A\pqty{B^0_d \to \bar{D}^0 K^0}}&=r_B e^{i\pqty{\delta_B+\omega_2}},\quad \omega_2=\arg{\frac{V_{ub}^\star V^{}_{cs}}{V^\star_{cb}V_{us}^{}}},
\end{align}
where $r_B$ and $\delta_B$ are the magnitudes and the strong phase of the ratio, respectively. The parameters for the decay channel $B^0_d \to DK^\star $ have been determined by the LHCb experiment~\cite{LHCb:2016aix,LHCb:2016bxi}, and we take similar values for $B^0_d \to DK$: $r_B = 0.366$ and $\delta_B = 164^\circ$. We neglect the direct CP violation in neutral $K$ meson decay, {\it i.e.,} $\braket{f}{\bar K^0}=\braket{\bar{f}}{K^0}$, as well as the amplitude of Cabibbo-suppressed process $\bar D^0 \to K^-\pi^+$ and the mixing of the neutral $D$ meson. The double-mixing term is
\begin{align}
        A_{n,dm}(t_1,t_2)=&\ \frac{e^{-\Gamma_{B_d} t_1}}{D(t_1,t_2)}S_n(t_2)\sin{\Delta m_{B_d} t_1},\label{eq20-1}\\
        S_n(t_2)=&\ e^{-\Gamma_K t_2}r_B\Big\{\abs{\frac{q_K}{p_K}}[\sin{\pqty{\delta_B+\delta_\omega}}\sinh{\frac{1}{2}\Delta\Gamma_K t_2}\nonumber\\        &+\cos{\pqty{\delta_B+\delta_\omega}}\sin{\Delta m_K t_2}]-\abs{\frac{p_K}{q_K}}\nonumber\\
        &\times[\sin{\pqty{\delta_B-\delta_\omega}}\sinh{\frac{1}{2}\Delta\Gamma_K t_2}+\cos{\pqty{\delta_B-\delta_\omega}}\nonumber\\
        &\times\sin{\Delta m_K t_2}]\Big\},\label{eq20-2}
\end{align}
where $\delta_\omega = \omega_2-\omega_1+\phi_{B_d}-\phi_K$, approximating $2\beta + \gamma$. Applying~\eqref{eq12} - \eqref{eq13-2} yields the result of $D(t_1,t_2)$, with $C^\prime_+(t_2)$ has an additional factor of $r^2_B$, and $S_n^\prime(t_2)$ can be obtained by changing the negative sign in front of $\abs{p_K/q_K}$ in~\eqref{eq20-2} to a positive sign. 

\begin{table}
\caption{Time integrated $A_{\rm CP}$ in $B^0_d \to D K \to (K^-\pi^+) ( \pi^+\ell^-\bar\nu_\ell)$.}\label{ACP3}
\begin{center}
{
\begin{tabular}{c c c}
\hline
$t_1/\tau_{B_d}$ & $t_2/ \tau_K$ & $A_{\rm CP}$\\
\hline
0$\sim$3 & 0$\sim$ 2 & 26.36\%\\
0$\sim$2 & 0$\sim$ 2 & 31.28\%\\
0$\sim$3 & 0$\sim$ 1 & 15.68\%\\
0$\sim$2 & 0$\sim$ 1 & 19.79\%\\
0$\sim$3 & 0$\sim$ 0.5 & 8.78\% \\
0$\sim$2 & 0$\sim$ 0.5 & 11.52\%\\
\hline
\end{tabular}
}
\end{center}
\end{table}

\begin{figure*}[htbp]
    \centering
    \includegraphics[keepaspectratio,width=5.3cm]{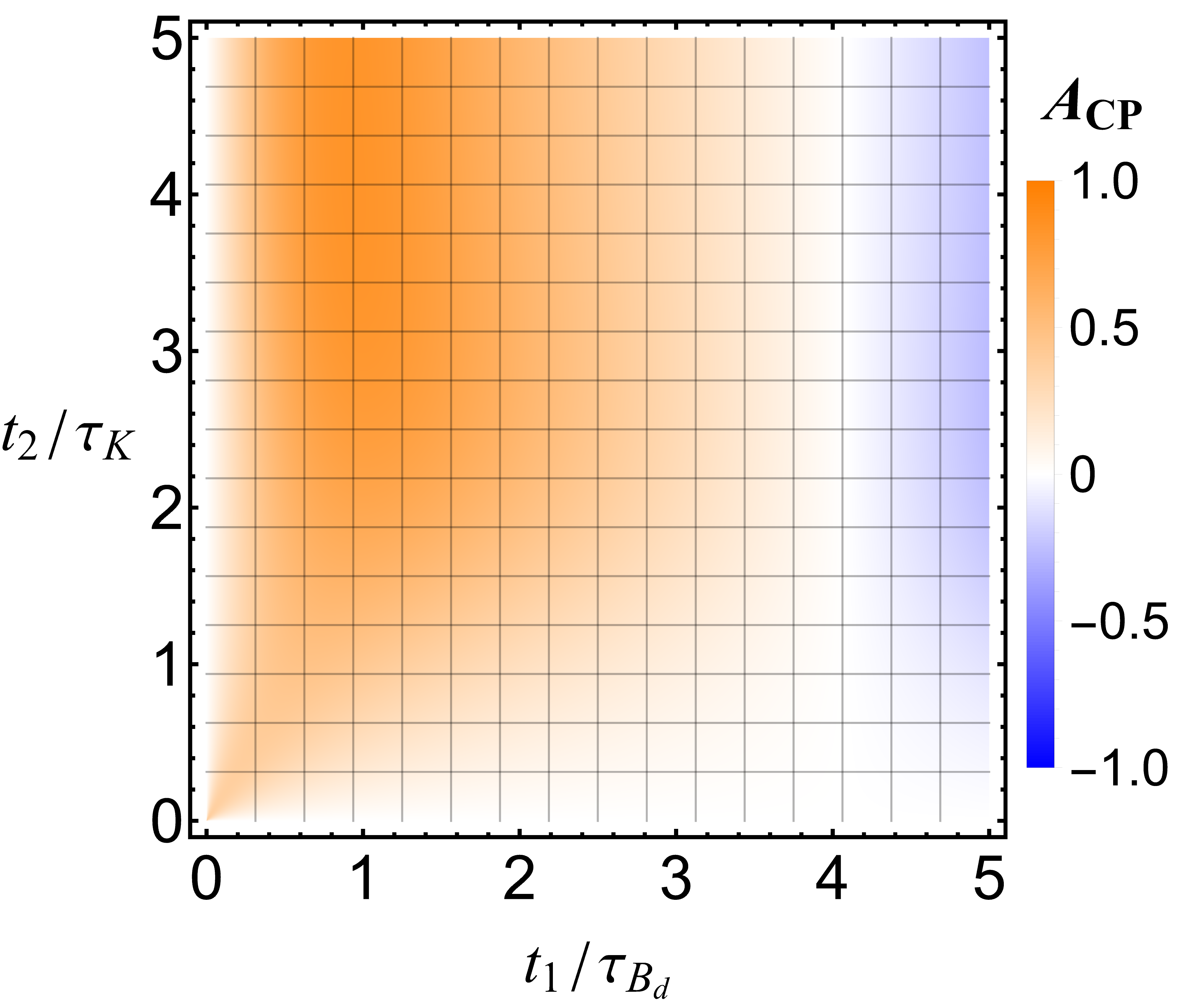}
    \hspace{0.1cm}
    \includegraphics[keepaspectratio,width=4.2cm]{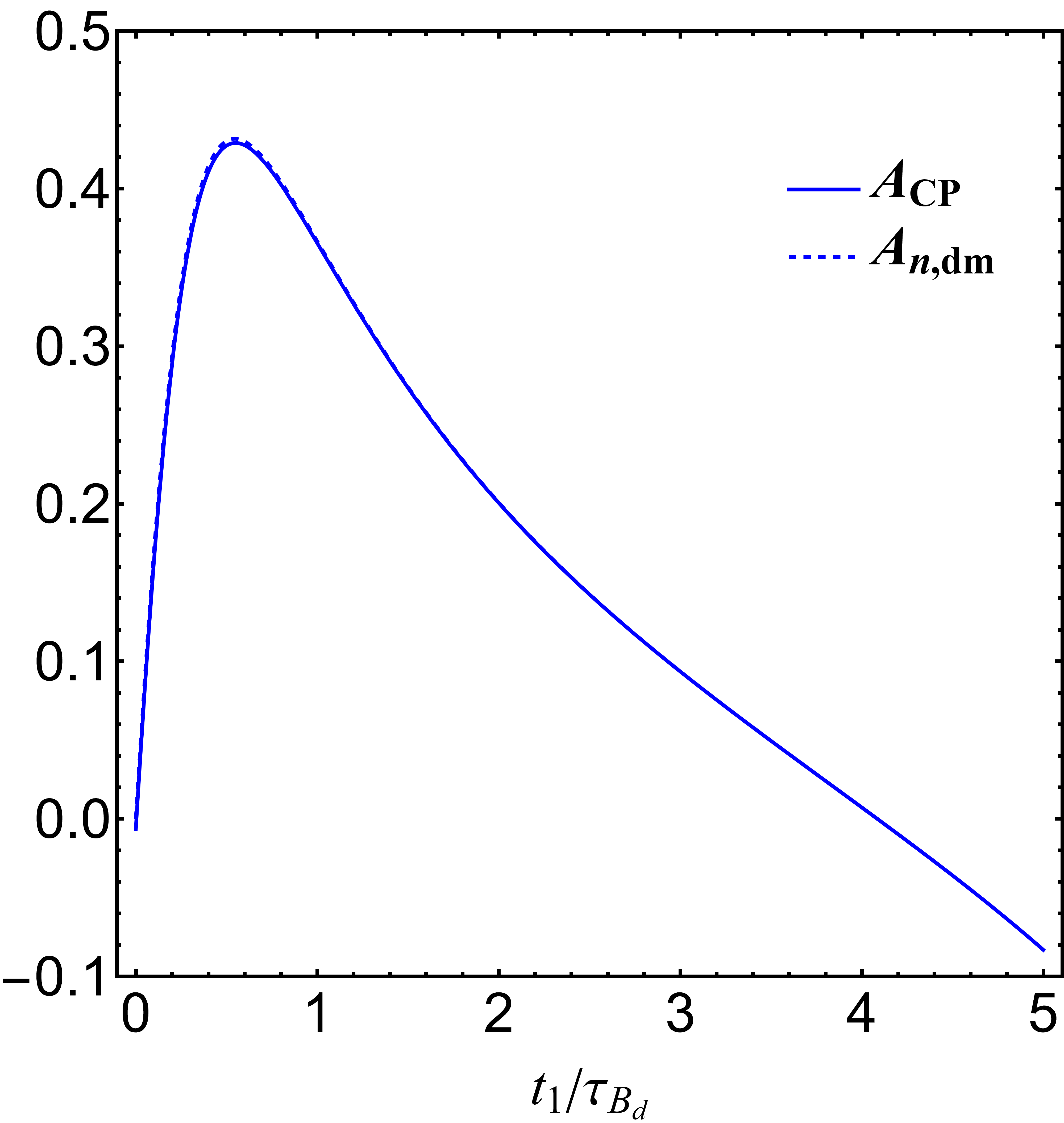}
     \hspace{0.1cm}
    \includegraphics[keepaspectratio,width=4.2cm]{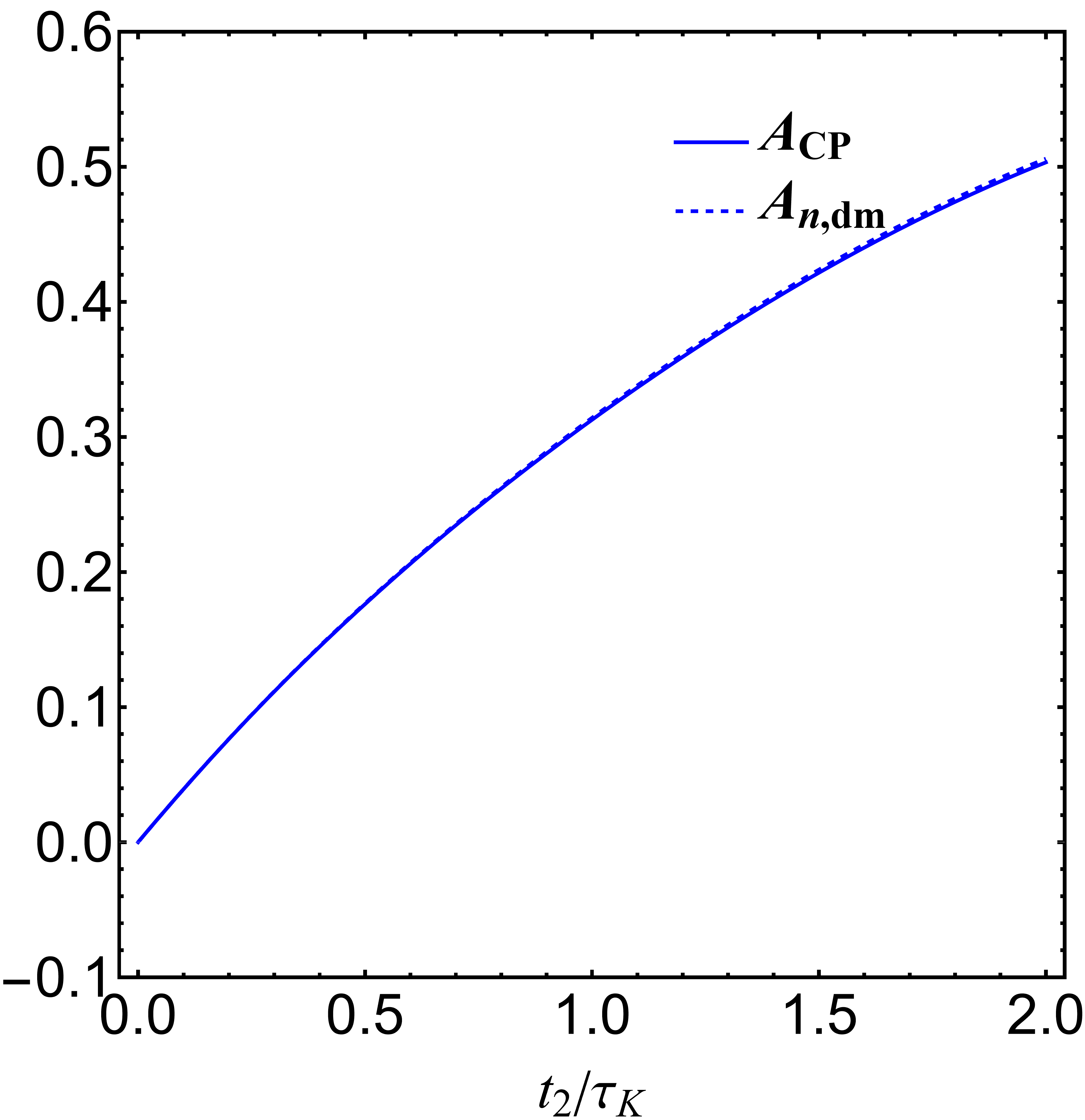}
    \caption{Time dependence of the double-mixing CP asymmetry $A_{\rm CP}$ in $B^0_d(t_1) \to D K(t_2) \to (K^-\pi^+) ( \pi^+\ell^-\bar\nu_\ell)$. The left panel displays the two-dimensional time dependence. The middle panel and the right panel display the dependence on $t_1$ (with $t_2$ integrated from 0 to $2\tau_{K}$) and $t_2$ (with $t_1$ integrated from $0$ to $3\tau_{B_d}$), respectively.}\label{fig:d3}
\end{figure*}

Our numerical analysis addresses both the total CP asymmetry $A_{\rm CP}$ and the double-mixing CP asymmetry $A_{dm}$. As illustrated in the left panel of Fig.~\ref{fig:d3}, the maximum value of two-dimensional time-dependent double-mixing results can surpass 50\% when $t_1$ is small. From the $t_1$ and $t_2$ dependence displayed in the middle and right panels, respectively, it can be seen that the double-mixing effect dominates. This trend resembles that observed in $B^0_d\to J/\psi K \to J/\psi (\pi^+\ell^-\bar\nu_\ell)$, albeit with differences attributed to the suppression factor $r_B$ and a negative sign in $\delta_\omega$. With integrating $t_1$ from 0 to $3\tau_{B_d}$, a steady growth of $A_{\rm CP}$ with respect to $t_2$ is observed. This is supported by the time-integrated $A_{\rm CP}$ presented in Table~\ref{ACP3}, which shows sensitivity to changes in the upper limit of $t_2$.

Given that the weak phase involved in this decay $\delta_\omega \approx 2\beta+\gamma$, its extraction holds significant importance. It should be emphasized that the weak phase can be directly extracted from experimental data, as discussed in our previous study~\cite{Shen:2023nuw}. The research~\cite{Kayser:1999bu} conducted by Kayser and London can also be subjected to similar extraction.

\subsubsection{\texorpdfstring{$B^0_d\to D K^0_S\to f_+ K^0_S$}{}}

Distinct from $B^0_d \to D K \to (K^-\pi^+) (\pi^+\ell^-\bar\nu_\ell)$ discussed previously in Sect.~\ref{sec:3.1.3}, when considering the CP-even final state like $f_+ = \pi^+\pi^-$ of $D$ in $B^0_d\to D K^0_S\to f_+ K^0_S$, both intermediate $D^0$ and $\bar D^0$ contributions are comparable. With primary $b\to u$ and $b \to c$ transitions, both $B^0_d$ and $\bar{B}^0_d$ can decay into states involving either $D^0$ or $\bar{D}^0$. These make this channel a particularly intricate case, categorizing it within the third classification.

The terms related to the double-mixing CP violation are expressed as below:
\begin{align}
        A_{n,dm,1}(t_1,t_2) = &\ f(t_1,t_2)e^{-\Gamma_D t_2}\sinh{\frac{1}{2}\Delta \Gamma_D t_2}\Big\{-\Big(\abs{\frac{p_D}{q_D}}\nonumber\\
        &+\abs{\frac{q_D}{p_D}}\Big)\sin{(\omega_1-\phi_{B_d}-\phi_D)}-r_B   \nonumber\\
        &\times \left[\abs{\frac{p_D}{q_D}}\sin{\pqty{\delta_B+\Phi_2}}-\abs{\frac{q_D}{p_D}}\sin{\pqty{\delta_B-\Phi_2}}\right.\nonumber\\
        &\left. +\abs{\frac{p_D}{q_D}}\sin{\pqty{\delta_B+\Phi_3}}-\abs{\frac{q_D}{p_D}}\sin{\pqty{\delta_B-\Phi_3}}\right] \nonumber\\
        & -r_B^2\pqty{\abs{\frac{p_D}{q_D}}+\abs{\frac{q_D}{p_D}}} \sin{(\Phi_2-\omega_2)} \Big\}, \\ 
        A_{n,dm,2}(t_1,t_2) = &\ f(t_1,t_2)e^{-\Gamma_D t_2}\sin{\Delta m_D t_2}\Big\{\pqty{\abs{\frac{p_D}{q_D}}-\abs{\frac{q_D}{p_D}}}\nonumber\\
        &\cos{(\omega_1-\phi_{B_d}-\phi_D)}  
         -r_B \Big[\abs{\frac{p_D}{q_D}}\nonumber\\
         &\times\cos{\pqty{\delta_B+\Phi_2}}-\abs{\frac{q_D}{p_D}}\cos{\pqty{\delta_B-\Phi_2}}-\nonumber\\     &\abs{\frac{p_D}{q_D}}\cos{\pqty{\delta_B+\Phi_3}}+\abs{\frac{q_D}{p_D}}\cos{\pqty{\delta_B-\Phi_3}}\Big]\nonumber\\
        & -r_B^2\pqty{\abs{\frac{p_D}{q_D}}-\abs{\frac{q_D}{p_D}}} \cos{(\Phi_2-\omega_2)} \Big\}, \label{eq22}
\end{align}
where $\Phi_2 = \omega_1 - \omega_2-\phi_{B_d}+\phi_D$, $\Phi_3 = \omega_1 - \omega_2-\phi_{B_d}-\phi_D$ and $f(t_1,t_2)=e^{-\Gamma_{B_d} t_1}\sin{\Delta m_{B_d} t_1}$ $/D(t_1,t_2)$, and the double-mixing CP violation is the sum of these two contributions. Combining~\eqref{eq4} and~\eqref{eq14-7} - \eqref{eq14-10} yields the denominator $D(t_1,t_2)$. The parameters $r_B$ and $\delta_B$ have been defined in~\eqref{eq19}.

\begin{figure*}[htbp]
    \centering
    \includegraphics[keepaspectratio,width=5.3cm]{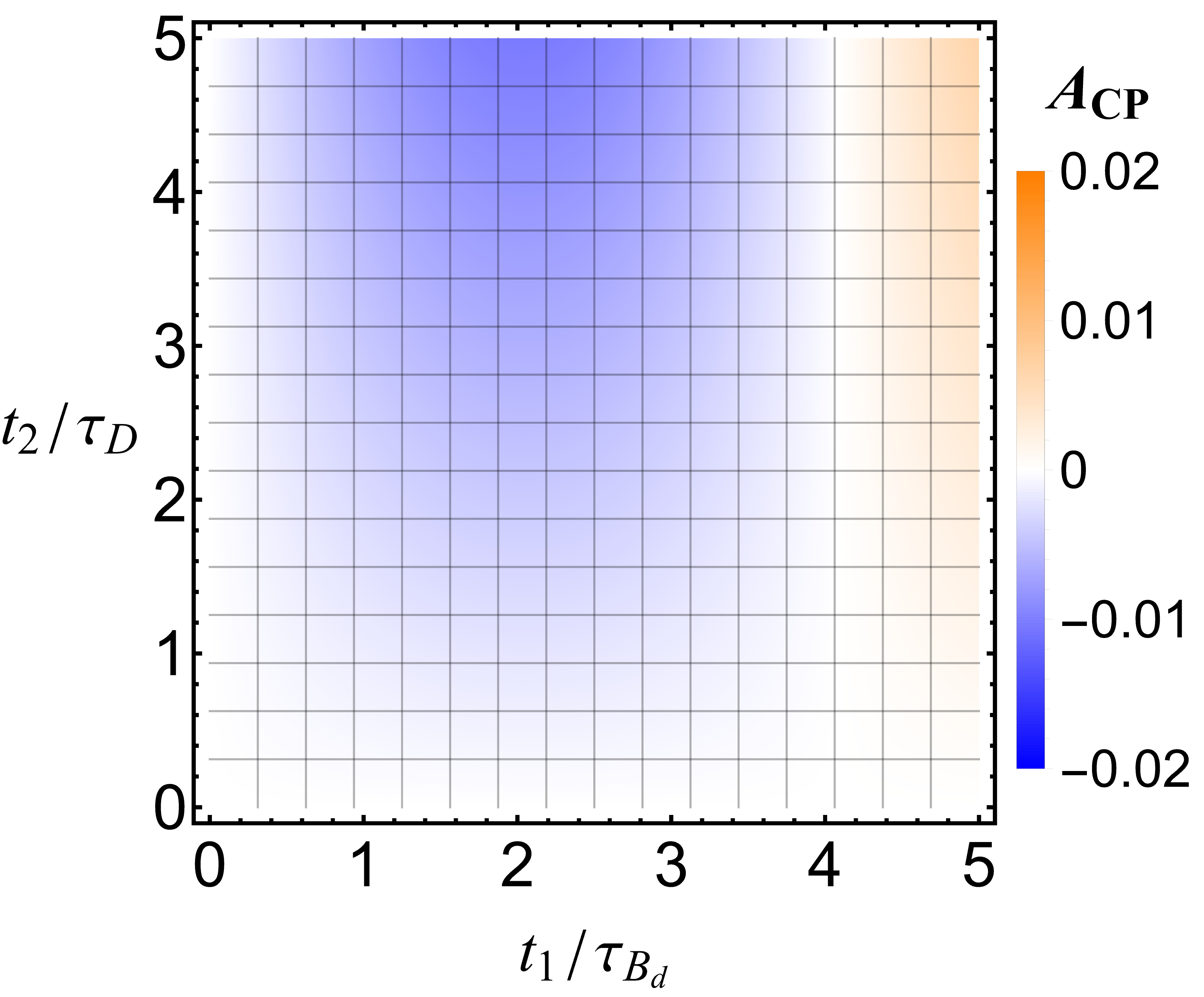}
    \hspace{0.1cm}
    \includegraphics[keepaspectratio,width=4.2cm]{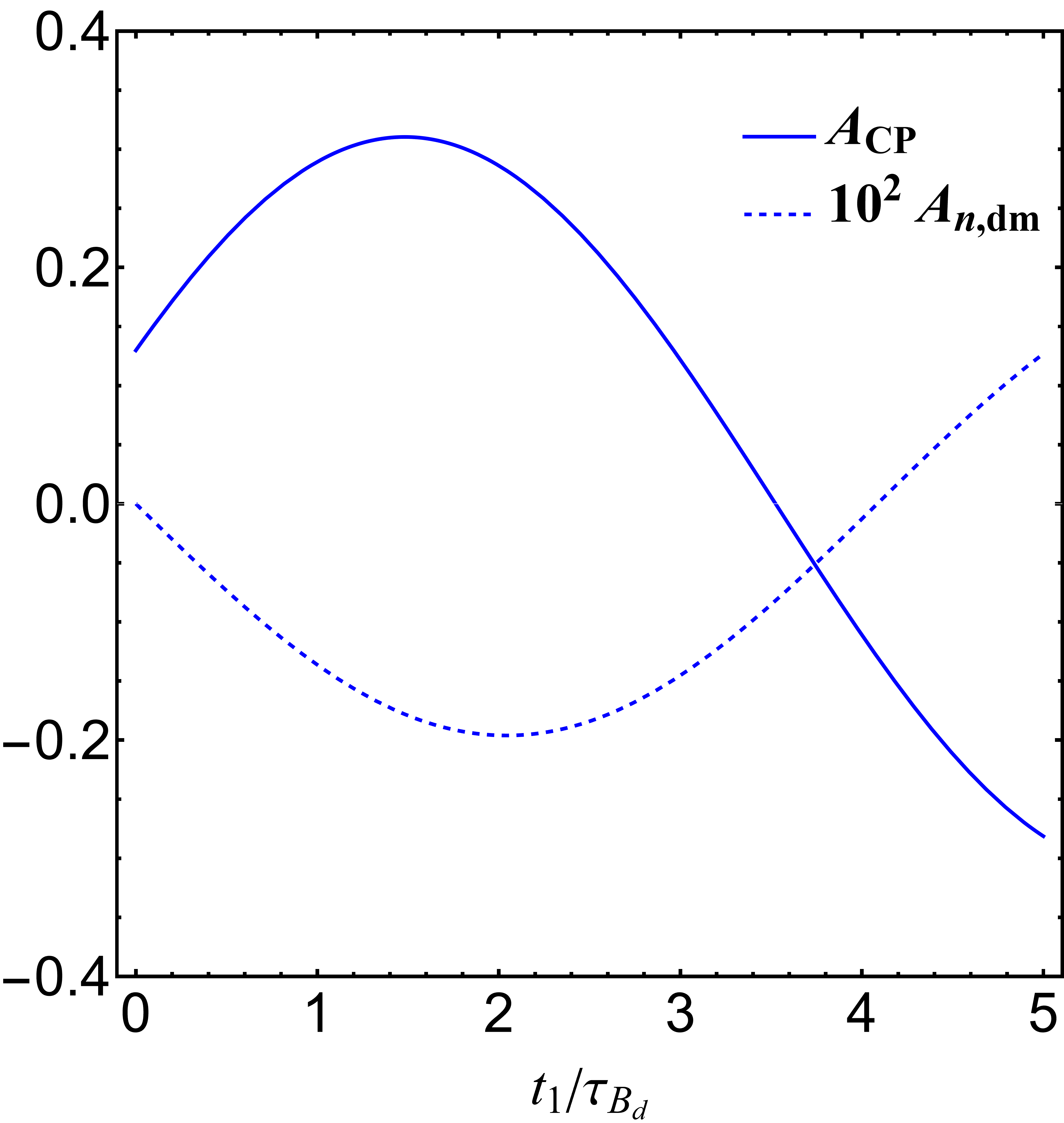}
     \hspace{0.1cm}
    \includegraphics[keepaspectratio,width=4.2cm]{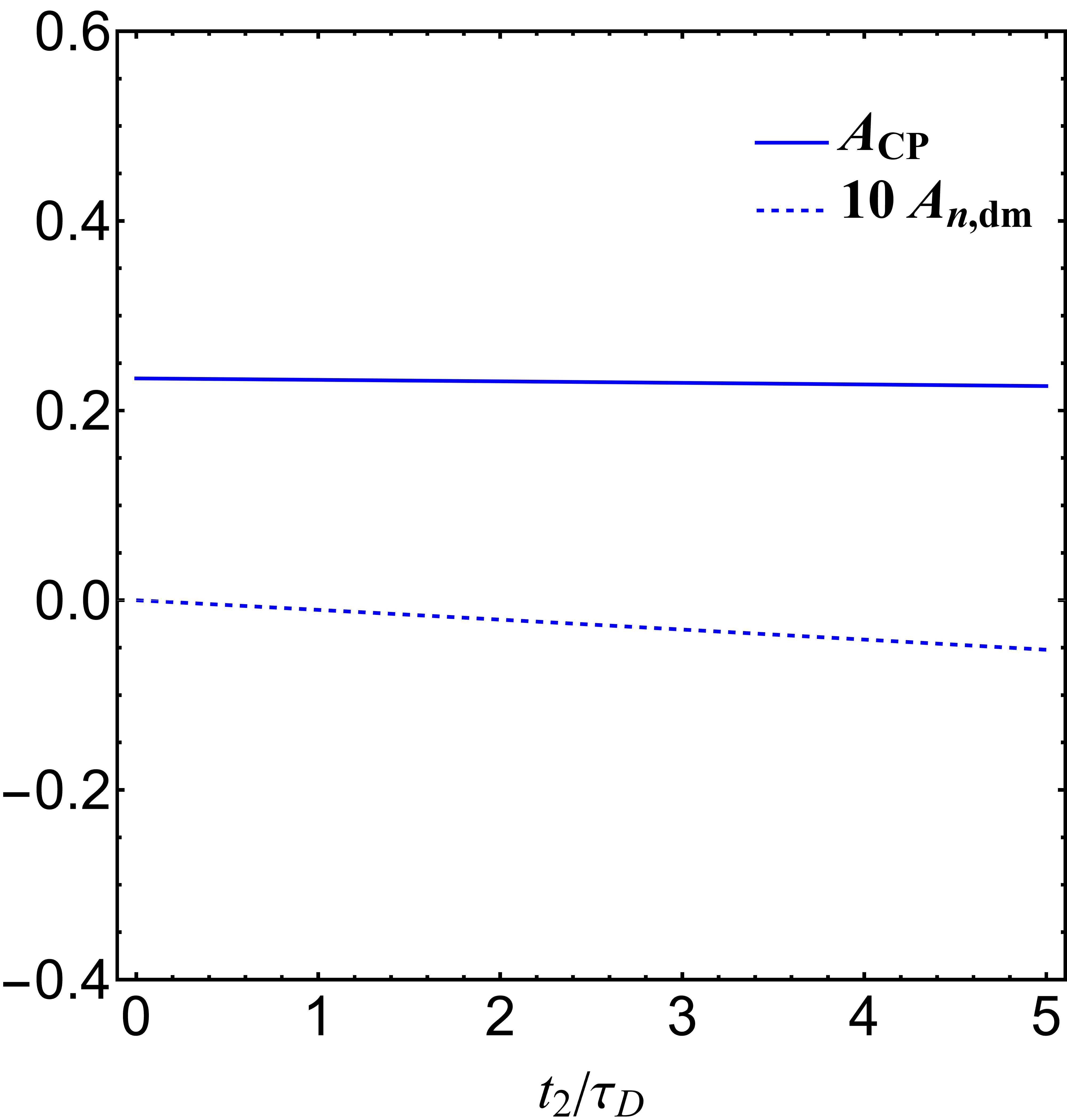}
    \caption{Time dependence of the CP asymmetry $A_{\rm CP}$ in $B^0_d(t_1)\to D(t_2) K^0_S \to (\pi^+\pi^-)K^0_S$. The left panel displays the two-dimensional time dependence. The middle panel and the right panel display the dependence on $t_1$ (with $t_2$ integrated from 0 to $5\tau_{D}$) and $t_2$ (with $t_1$ integrated from $0$ to $3\tau_{B_d}$), respectively.}\label{fig:d4}
\end{figure*}

We perform the numerical analysis assuming the parameter values $r_B = 0.366$ and $\delta_B = 164^\circ$ as in Sect.~\ref{sec:3.1.3}. The results are depicted in Fig.~\ref{fig:d4}. On the left panel, the two-dimensional time-dependent CP violation exhibits a distribution akin to the decay channel $B^0_d \to J/\psi K \to J/\psi (\pi^+\pi^-)$, but with peak values significantly lower. Integrating $t_2$ from 0 to $5\tau_D$, we present the $t_1$-dependent results in the middle panel, revealing tiny double-mixing CP violation. This is primarily attributed to the small magnitudes of $x_D$ and $y_D$, both approximately $10^{-3}$. These small values reduces both $\sinh{\frac{1}{2}\Delta \Gamma_D t_2}$ and $\sin{\Delta m_D t_2}$. The right panel of Fig.~\ref{fig:d4} shows the $t_2$ dependence of $A_{\rm CP}$ when integrating $t_1$ from 0 to $3\tau_{B_d}$. The near-linear dependency of $A_{dm}$ on $t_2$ is primarily due to its denominator being dominated by the $\cosh{y_D t_2}$ term and its numerator by the $\sinh{x_D t_2}$ term. It is concluded that the CP asymmetry in this channel is dominated by the part induced by the interference between the $b\to c$ and $b\to u$ amplitudes, and the part induced by the interference between decays with and without $B$ mixing, while the double-mixing CP violation is negligible owing to tiny $D$-mixing effect.

\subsubsection{\texorpdfstring{$B^0_d\to  D K^0_S\to (K^-\pi^+) K^0_S$}{}}

The last $B^0_d$ decay channel selected is $B^0_d\to  D K^0_S\to (K^-\pi^+) K^0_S$. The observation of a similar channel $\bar{B}^0_d \to D^0\bar{K}^{\star0}$ was reported by~\cite{Belle:2002jgk, BaBar:2005osj}, but the branching ratio and relevant parameters were measured using self-tagging methods and thus the mixing effect was not considered. In this channel, the intermediate $D$ meson can be either $D^0$ or $\bar D^0$, although the latter has a suppressed contribution. Therefore, this channel is categorized as the third type.

To present the result of the CP asymmetry, we introduce the amplitude ratio of the primary decay as in~\eqref{eq19} and the ratio of the secondary decay as 
\begin{gather}\label{eq23}
    \frac{A\pqty{\bar{D}^0 \to K^- \pi^+}}{A\pqty{D^0 \to K^- \pi^+}}=-r_De^{-i\delta_D},
\end{gather}
where the numerical values for magnitude ratio $r_D$ and the strong phase $\delta_D$ are listed in Table~\ref{input}. The direct CP violation in the $D$ meson decays is neglected due to the minimal weak phases between the first two generations of quarks. The double-mixing term in~\eqref{eq4} is expressed as
\begin{align}\label{eq315-1}
        A_{n,dm}(t_1,t_2)=& -\frac{e^{-\pqty{\Gamma_{B_d} t_1+\Gamma_D t_2}}}{2D(t_1,t_2)}\Big[\Big(\left|\frac{q_D}{p_D}\right|+\left|\frac{p_D}{q_D}\right|\Big)\nonumber\\
        &\times\sin{(\omega_1-\phi_{B_d}-\phi_D)}\sinh{\frac{1}{2}\Delta \Gamma_D t_2}\nonumber \\
        &+ \Big(\left|\frac{q_D}{p_D}\right|-\left|\frac{p_D}{q_D}\right|\Big)\cos{(\omega_1-\phi_{B_d}-\phi_D)}\nonumber\\
        &\times\sin{\Delta m_D t_2}\Big]\sin{\Delta m_{B_d} t_1} \;,
\end{align}
where the contributions below $\mathcal{O}(10^{-3})$ have been neglected. It turns out that the corrections from the doubly Cabibbo-suppressed $D^0 \to K^+\pi^-$ decay are negligible, so practically only the interference term between $B^0_d \to \bar{D}^0 \to D^0$ and $B^0_d \to \bar{B}^0_d \to D^0$ is retained.
With also the contributions below $\mathcal{O}(10^{-3})$ neglected, the non-double-mixing CP asymmetry reads 
\begin{align}\label{eq:nondmb5}
A_{non-dm,1}(t_1,t_2)=&\ r_B\Big\{\Big(\abs{\frac{p_D}{q_D}}\cos{\pqty{\delta_B+\omega_2-\phi_D}}-\abs{\frac{q_D}{p_D}}\nonumber\\
&\times\cos{\pqty{\delta_B-\omega_2+\phi_D}}\Big)\sinh{\frac{1}{2}\Delta \Gamma_D t_2}\nonumber\\
&+\Big(\abs{\frac{p_D}{q_D}}\sin{\pqty{\delta_B+\omega_2-\phi_D}}-\abs{\frac{q_D}{p_D}}\nonumber\\
&\times\sin{\pqty{\delta_B-\omega_2+\phi_D}}\Big)\sin{\Delta m_D t_2}\nonumber\\
&+2r_D \sin{\pqty{\delta_B-\delta_D}}\sin{\omega_2}\Big(\cosh{\frac{1}{2}\Delta \Gamma_{D}t_2}\nonumber\\
&+\cos{\Delta m_{D} t_2}\Big)\Big\}f(t_1,t_2), \nonumber \\
A_{non-dm,2}(t_1,t_2)=& -2\frac{e^{-\Gamma_{B_d}t_1}}{D(t_1,t_2)}\abs{g_{+,D}(t_2)}^2\sin{\Delta m_{B_d} t_1}\times\nonumber\\
&\Big\{r_B\cos{\delta_B}\sin{(\omega_1-\omega_2-\phi_{B_d})}\nonumber\\
&+r_D\cos{\delta_D}\sin{\pqty{\omega_1-\phi_{B_d}}}\Big\},
\end{align}
where $f(t_1,t_2) \equiv e^{-\Gamma_D t_2}\abs{g_{+,B_d}(t_1)}^2/4D(t_1,t_2)$. The denominator $D(t_1,t_2)$ can be derived by combining~\eqref{eq4} with~\eqref{A1-7} - \eqref{A1-12}, while maintaining consistency with all the approximations used in the numerator. The parameters $r_B$ and $\delta_B$ have been defined in~\eqref{eq19}. The non-double-mixing CP asymmetry contains $r_D$ in the expression~\eqref{eq:nondmb5}, indicating that the doubly Cabibbo-suppressed $D^0 \to K^+\pi^-$ contribution is considerable here.

\begin{table}
\caption{Two-dimensional time integration of $A_{\rm CP}(t_1, t_2)$ in $B^0_d(t_1) \to D(t_2) K^0_S \to (K^-\pi^+) K^0_S$.}\label{ACP5}
\begin{center}
{
\begin{tabular}{c c c}
\hline
$t_1/\tau_{B_d}$ & $t_2/ \tau_D$ & $A_{\rm CP}$\\
\hline
0$\sim$3 & 0$\sim$ 5 & -60.71\%\\
0$\sim$2 & 0$\sim$ 5 & -69.20\%\\
0$\sim$3 & 0$\sim$ 4 & -60.71\%\\
0$\sim$2 & 0$\sim$ 4 &-69.20\%\\
\hline
\end{tabular}
}
\end{center}
\end{table}

\begin{figure*}[htbp]
    \centering
    \includegraphics[keepaspectratio,width=5.3cm]{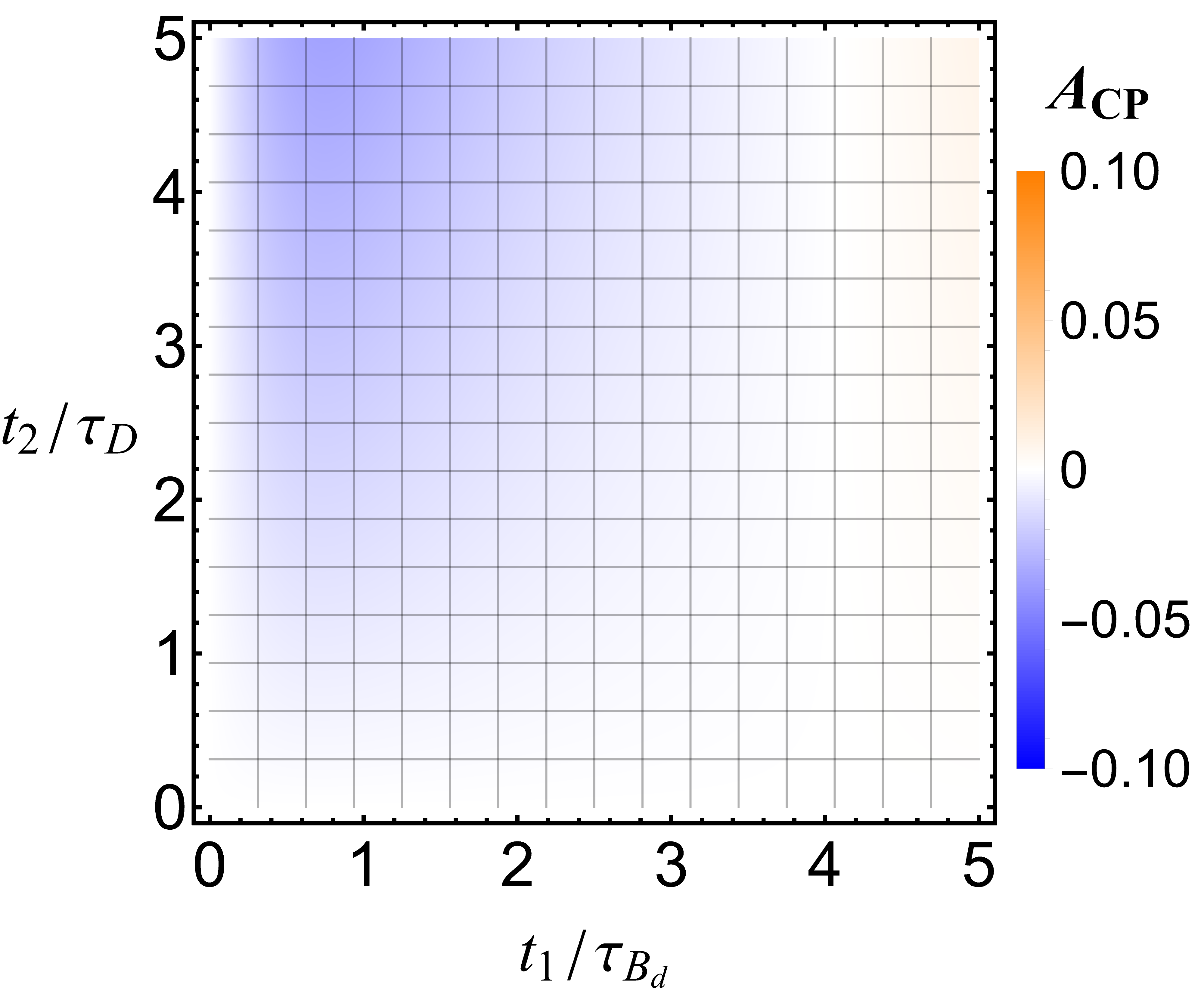}
    \hspace{0.1cm}
    \includegraphics[keepaspectratio,width=4.2cm]{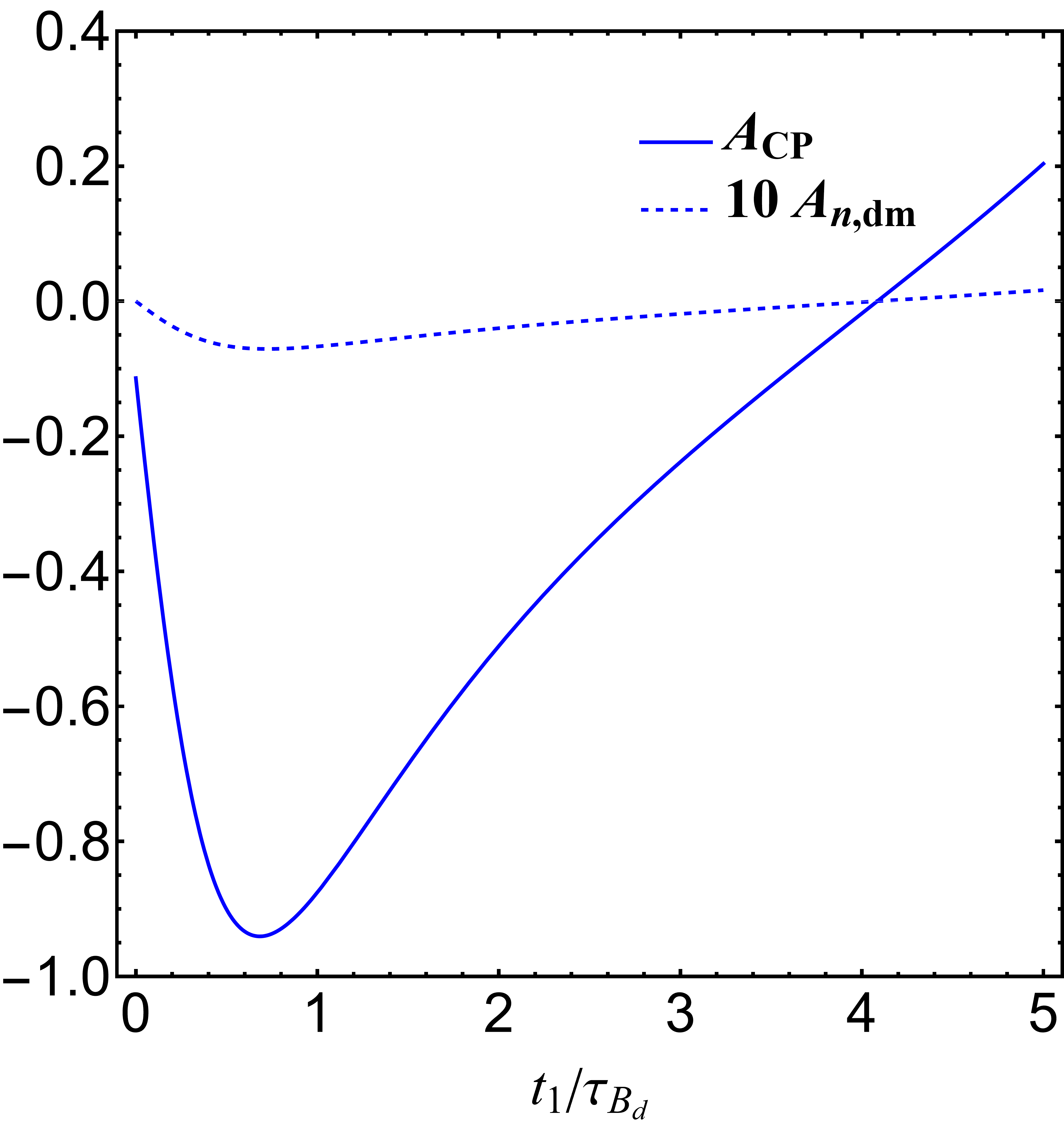}
     \hspace{0.1cm}
    \includegraphics[keepaspectratio,width=4.2cm]{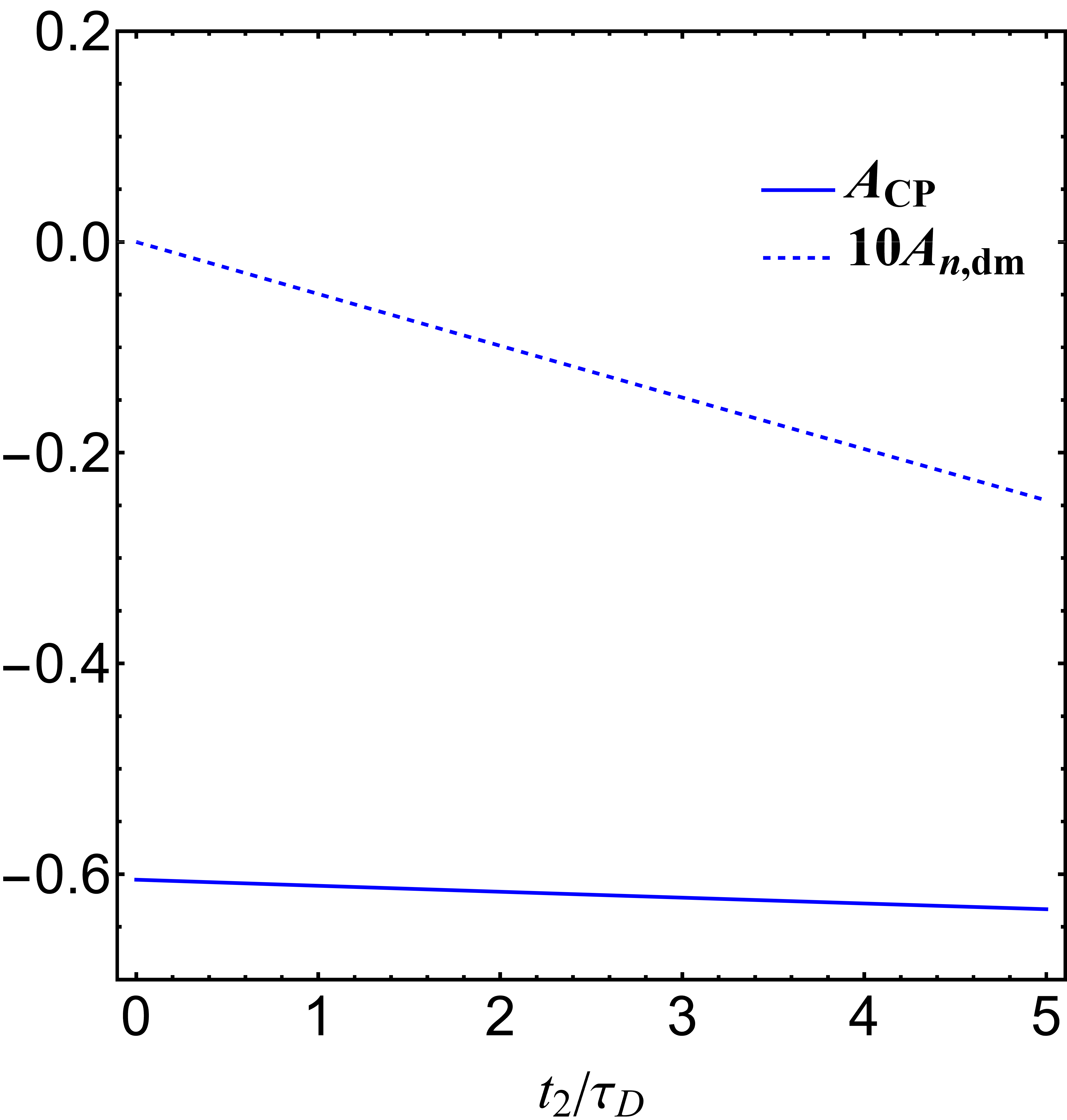}
    \caption{Time dependence of the CP asymmetry $A_{\rm CP}$ in $B^0_d(t_1) \to D(t_2) K^0_S \to (K^-\pi^+) K^0_S$. The left panel displays the two-dimensional time dependence. The middle panel and the right panel display the dependence on $t_1$ (with $t_2$ integrated from 0 to $5\tau_{D}$) and $t_2$ (with $t_1$ integrated from $0$ to $3\tau_{B_d}$), respectively.}\label{fig:d5}
\end{figure*}

We perform numerical analysis on this channel. As depicted in the left panel of Fig.~\ref{fig:d5}, the peak of the two-dimensional time-dependent double-mixing results is on the order of $10^{-2}$, occurring at later $t_2$ and earlier $t_1$ times. When integrating $t_2$ from 0 to $5\tau_D$, the peak value of the double-mixing effect contributes only 0.8\% to the total. The dependencies of $A_{\rm CP}$ and $A_{dm}$ on $t_1$ are similar, differing only in amplitude magnitude. Upon integrating $t_1$ from 0 to $3\tau_{B_d}$, the $t_2$ dependence of $A_{\rm CP}$ is shown in the right panel. The dependence of $A_{dm}$ on $t_2$ is nearly linear, attributed to its nearly hyperbolic sine dependence on $t_2$. However, due to the small value of $y_D$, the overall variation of $A_{dm}$ with $t_2$ remains nearly linear. The double-mixing CP violation in the decay channel of this meson is relatively more significant than in the $\pi^+\pi^-$ final state, due to the presence of the $r_D$ factor which suppresses the dominant non-double-mixing CP violation terms. The dominant contribution to the non-double-mixing CP violation arises from the interference between the $B^0_d \to \bar{D}^0$ and $B^0_d \to D^0$ path, {\it i.e.}, the third line of $A_{non-dm,1}(t_1,t_2)$. A slightly smaller contribution comes from the interference between the $B^0_d \to D^0$ and $B^0_d \to \bar{B}^0_d \to D^0$ paths, {\it i.e.}, the first term of $A_{non-dm,2}(t_1,t_2)$ which involves the factor $r_B$. The two-dimensional time integration of $A_{\rm CP}(t_1,t_2)$ for this channel, presented in Table \ref{ACP5}, suggests that the CP asymmetry $A_{\rm CP}$ is relatively insensitive to the decay of the $D$ meson.

\subsection{\texorpdfstring{$B_s^0$}{} decays}

In this subsection, we extend our investigation to the $B^0_s$ meson, by considering decays with the same final states as the $B_d^0$ decays. For simplicity, we retain the approximation:
\begin{equation}\label{eq25}
    q_{B_s}/p_{B_s}=e^{-i\phi_{B_s}}.
\end{equation}
Unlike the $B^0_d$ meson, the $B_s^0$ meson exhibits a non-negligible difference in decay widths, as indicated in Table~\ref{input}. Consequently, the double-mixing CP violation for the $B_s^0$ meson comprises both a component proportional to $\sin{\Delta m_{B_s} t_1}$, 
represented by $A_{n, dm}$, and a component proportional to $\sinh{\frac{1}{2} \Delta \Gamma_{B_s} t_1}$, represented by $A_{h, dm}$. Although the underlying calculation process is similar to that for the $B^0_d$ meson, the results for the $B^0_s$ meson are inherently more complicated, implying more fruitful phenomenology.

\subsubsection{\texorpdfstring{$B^0_s \to J/\psi K \to J/\psi f_+ $}{}}

We first consider the process $B^0_s \to J/\psi K \to J/\psi f_+ $, which originates from the $\bar{b} \to \bar{c}c\bar{d}$ quark-level transition. The first observation of such a mode $B_s^0\to J/\psi K_S^0$ was reported by CDF along with the measured branching ratio~\cite{CDF:2011fhd}. In this channel, both the transitions $B^0_s \to J/\psi \bar{K}^0$ and $B^0_s \to \bar{B}^0_s \to J/\psi K^0$ can occur, with final state $f_+$ selected as $\pi^+\pi^-$ being common to both $K^0$ and $\bar{K}^0$, resulting in four possible paths. This makes these channels fall into the second category of our previously defined classifications. 

In the calculation of the CP violation of this channel, direct CP violation in the primary and secondary decays are neglected. The involved phase difference between the decay amplitude $\braket{J/\psi K^0}{\bar{B}^0_s}$ and its charge conjugation $\braket{J/\psi \bar{K}^0}{B^0_s}$ is given by $-e^{i\omega_{B_s}} = -\pqty{V_{cb}V^\star_{cd}}/\pqty{V^\star_{cb}V_{cd}}$, with its numerical value detailed in Table \ref{input}. The negative sign arises from the convention $CP\ket{J/\psi K^0} = -\ket{J/\psi \bar{K}^0}$. The double-mixing CP asymmetry is given by
\begin{align}\label{eq27}
    A_{n,dm}(t_1, t_2)=&\ 2\frac{e^{-\pqty{\Gamma_{B_s} t_1 +\Gamma_K t_2}}}{D(t_1,t_2)}\Big\{\sinh{\frac{1}{2}\Delta\Gamma_K t_2}\sin{\Phi_4 }\Big(\abs{\frac{q_K}{p_K}}\nonumber\\
    &+\abs{\frac{p_K}{q_K}}\Big)+\sin{\Delta m_K t_2}\cos{\Phi_4 }\Big(\abs{\frac{q_K}{p_K}}-\nonumber\\
    &\abs{\frac{p_K}{q_K}}\Big)\Big\}
    \sin{\Delta m_{B_s} t_1},
\end{align}
where $\Phi_4 \equiv \omega_{B_s}-\phi_{B_s}-\phi_K$. It is induced by the interference between $B^0_s \to \bar{K}^0 \to \pi^+\pi^-$ and $B^0_s \to \bar{B}^0_s \to K^0 \to \bar{K}^0 \to \pi^+\pi^-$, as well as between the $B^0_s \to \bar{K}^0 \to K^0 \to \pi^+\pi^-$ and $B^0_s \to \bar{B}^0_s \to K^0 \to \pi^+\pi^-$ pathways. The denominator $D(t_1,t_2)$ is obtained from~\eqref{eq14-7} -~\eqref{eq14-10}, where the substitutions $\abs{p_K/q_K} \leftrightarrow \abs{q_K/p_K}$ and $\phi_K \to -\phi_K$ need to be performed.

\begin{figure*}[htbp]
    \centering
    \includegraphics[keepaspectratio,width=5.3cm]{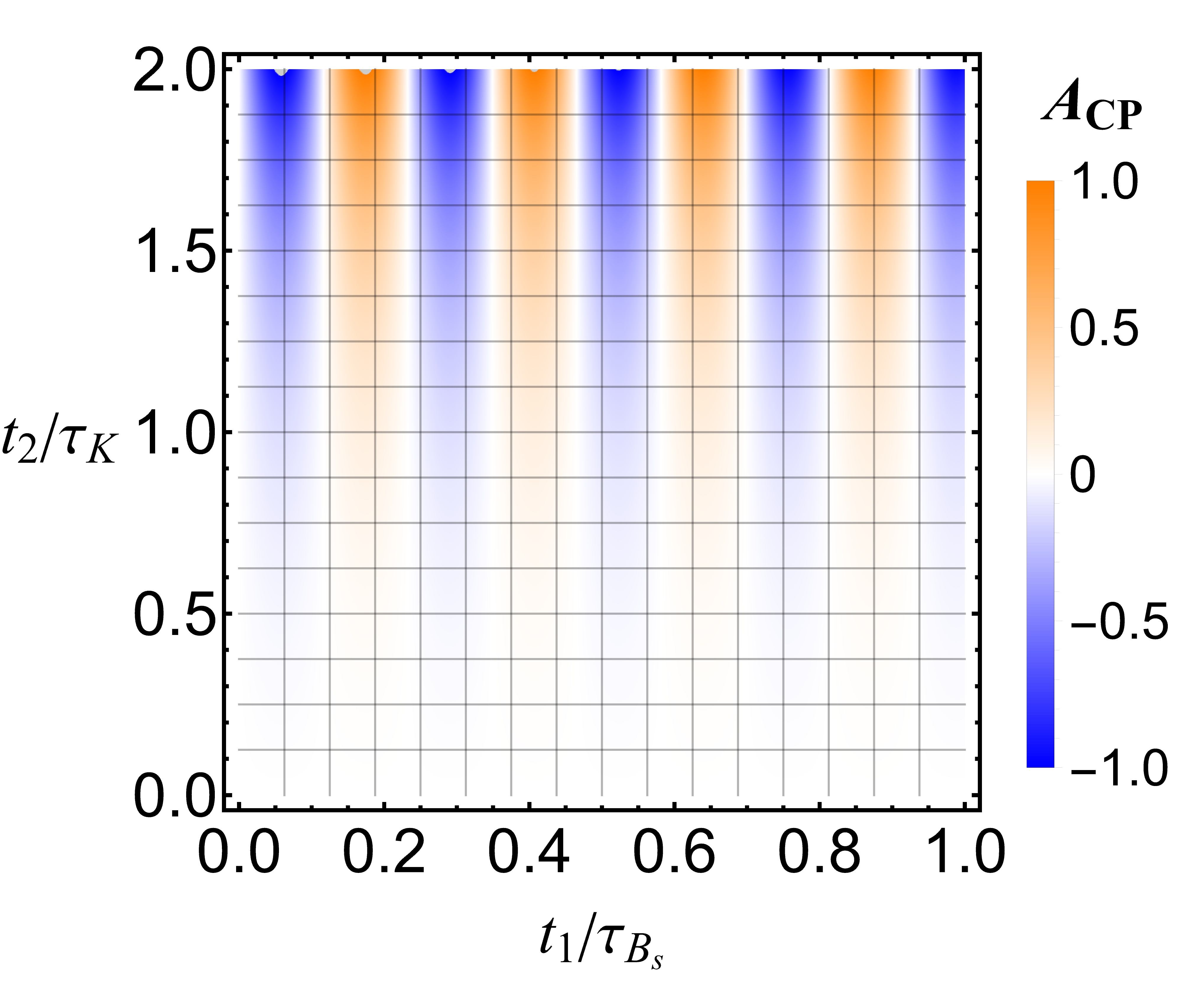}
    \hspace{0.1cm}
    \includegraphics[keepaspectratio,width=4.2cm]{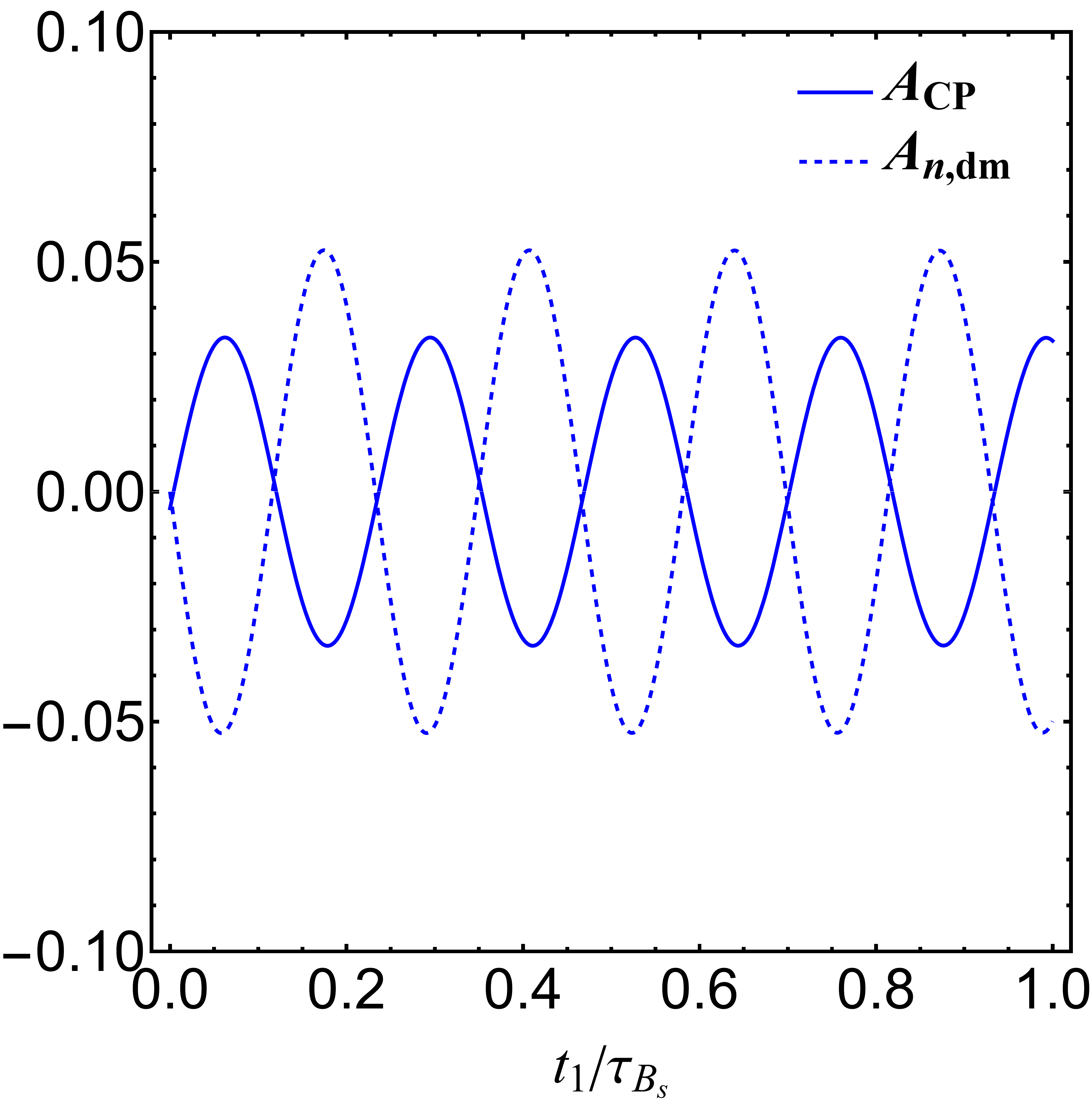}
     \hspace{0.1cm}
    \includegraphics[keepaspectratio,width=4.2cm]{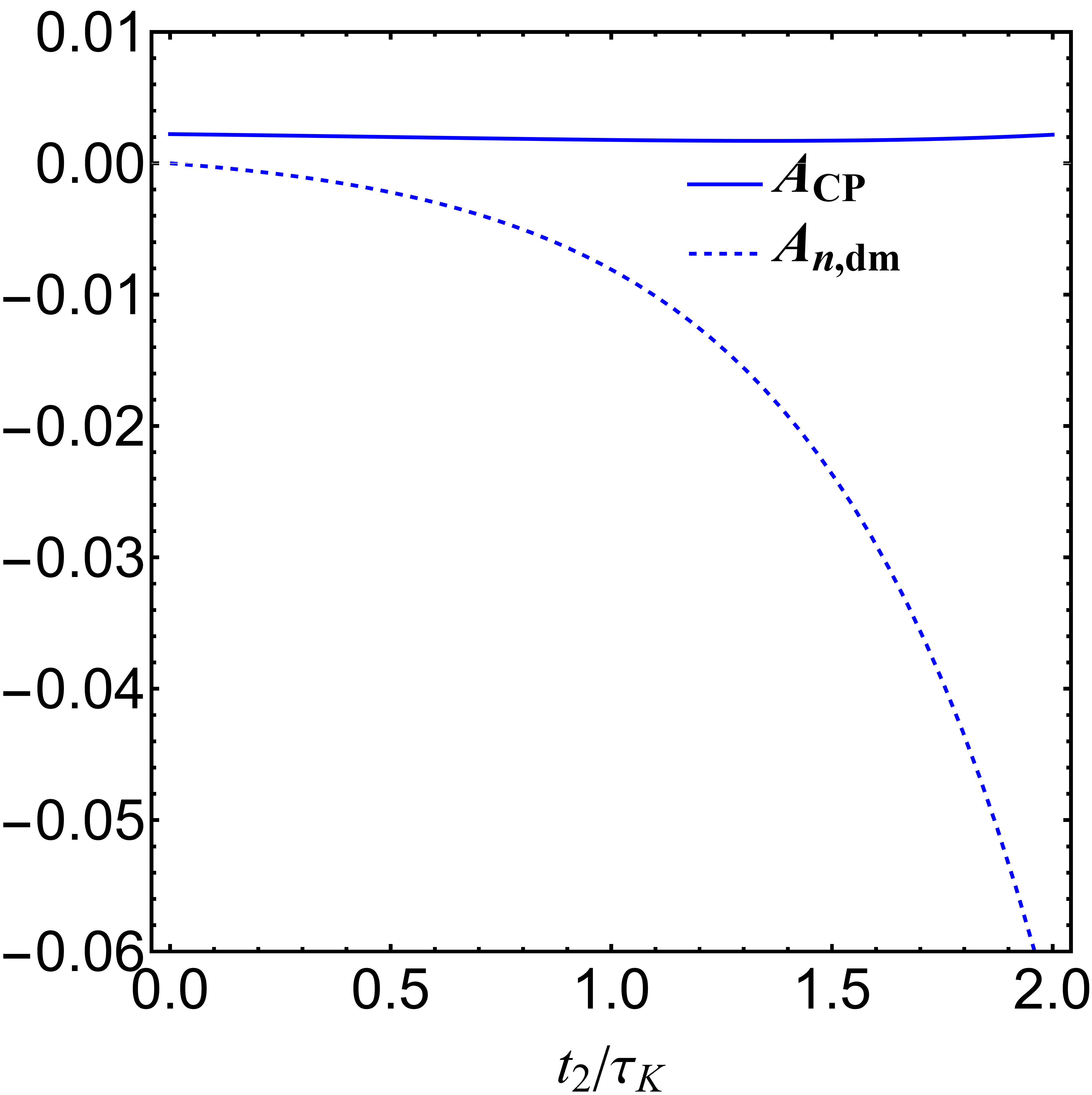}
    \caption{Time dependence of the CP asymmetry $A_{\rm CP}$ in $B^0_s(t_1) \to J/\psi K(t_2) \to J/\psi f_+$. The left panel displays the two-dimensional time dependence. The middle panel and the right panel display the dependence on $t_1$ (with $t_2$ integrated from 0 to $2\tau_{K}$) and $t_2$ (with $t_1$ integrated from $0$ to $\tau_{B_s}$), respectively.}\label{fig:s1}
\end{figure*}

We have conducted numerical analyses of this decay channel, obtaining the CP asymmetry and its time dependence, as illustrated in Fig.~\ref{fig:s1}. The left panel shows that the maximum value of the two-dimensional time-dependent CP asymmetry can exceed 50\%, a substantial value. Integrating the time $t_2$ from 0 to $2\tau_K$, the middle panel shows the $t_1$ dependence of $A_{\rm CP}$. The rapid oscillation observed is due to the large $x_s$ value of the $B_s^0$ meson. The total CP asymmetry is dominated by the $S_n$ term, which can be further separated into the double-mixing part $A_{n, dm}$ and the non-double-mixing part, both of which have the sine dependence on $t_1$, as defined in~\eqref{eq4}. The double-mixing part has a magnitude approximately twice the total CP asymmetry but with an opposite sign. Integrating $t_1$ from 0 to $\tau_{B_s}$ reveals the $t_2$ dependence of $A_{\rm CP}$, as shown in the right panel. In this case, $A_{\rm CP}$ is almost zero because the cancellation between the double-mixing and non-double-mixing terms in $S_n$, while $A_{n,dm}$ varies approximately as $-\exp(y_K t_2)\sinh{y_K t_2}$ with respect to $t_2$.

\subsubsection{\texorpdfstring{$B^0_s \to J/\psi K \to  J/\psi (\pi^\pm\ell^\mp\nu_\ell)$}{}}

The decay process $B^0_s \to J/\psi K \to  J/\psi (\pi^-\ell^+\nu_\ell)$ can occur via the two paths $B^0_s \to J/\psi \bar{K}^0 \to J/\psi K^0 \to J/\psi(\pi^-\ell^+\nu_\ell)$ and $B^0_s\to \bar B^0_s \to J/\psi K^0 \to J/\psi(\pi^-\ell^+\nu_\ell)$; while; the process $B^0_s \to J/\psi K \to  J/\psi (\pi^+\ell^-\bar{\nu}_\ell)$ can occur via $B^0_s \to J/\psi \bar{K}^0 \to J/\psi(\pi^+\ell^-\bar \nu_\ell)$ and $B^0_s\to \bar B^0_s \to J/\psi K^0 \to J/\psi \bar{K}^0 \to J/\psi$
$(\pi^+\ell^-\bar \nu_\ell)$. Therefore, they both align with the first category in our prior discussion.

In the calculation of the CP asymmetry of these channels, we still neglect the direct CP violation in both the primary and secondary decays. Then, we have $\braket{f}{K^0} = \braket{\bar{f}}{\bar K^0}$ and the weak phase difference between the involved primary decay amplitudes is $\omega_{B_s}$ as introduced in previously. For the decay channel with the final state $\pi^-\ell^+\nu_\ell$, the corresponding double-mixing terms are given by
\begin{align}
    A_{h,dm}(t_1, t_2)=& -\frac{e^{-\pqty{\Gamma_{B_s} t_1 + \Gamma_K t_2}}}{2D(t_1,t_2)}\Big\{\sinh{\frac{1}{2}\Delta \Gamma_K t_2}\cos{\Phi_4 }\times\nonumber\\
    &\pqty{\abs{\frac{p_K}{q_K}}-\abs{\frac{q_K}{p_K}}}+\sin{\Delta m_K t_2}\sin{\Phi_4}\Big(\abs{\frac{p_K}{q_K}}\nonumber\\
    &+\abs{\frac{q_K}{p_K}}\Big)\Big\}\sinh{\frac{1}{2}\Delta \Gamma_{B_s} t_1}, \label{eq29-1}\\
    A_{n,dm}(t_1, t_2) =& -\frac{e^{-\pqty{\Gamma_{B_s} t_1 + \Gamma_K t_2}}}{2D(t_1,t_2)}\Big\{-\sinh{\frac{1}{2}\Delta \Gamma_K t_2}\Big(\abs{\frac{p_K}{q_K}}\nonumber\\
    &+\abs{\frac{q_K}{p_K}}\Big)\sin{\Phi_4}+\sin{\Delta m_K t_2}\cos{\Phi_4 }\Big(\abs{\frac{p_K}{q_K}}\nonumber\\
    &-\abs{\frac{q_K}{p_K}}\Big)\Big\}\sin{\Delta m_{B_s} t_1}.\label{eq29-2}
\end{align}
Combining~\eqref{eq4} with~\eqref{eq12} - \eqref{eq13-2} and making the substitutions $\abs{p_K/q_K} \leftrightarrow \abs{q_K/p_K}$ and $\phi_K \to -\phi_K$ yields the denominator $D(t_1,t_2)$.

\begin{figure*}[htbp]
    \centering
    \includegraphics[keepaspectratio,width=5.3cm]{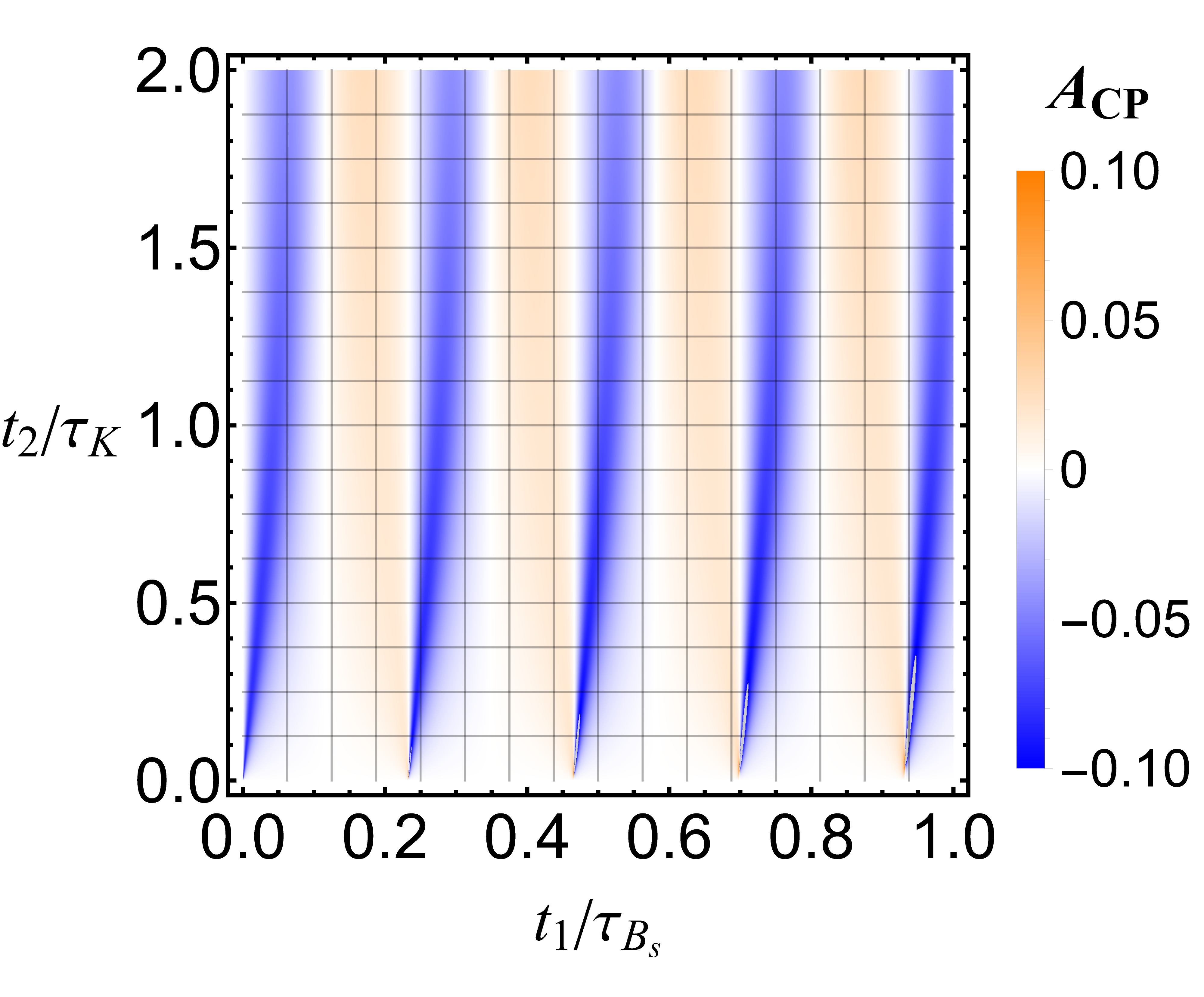}
    \hspace{0.1cm}
    \includegraphics[keepaspectratio,width=4.2cm]{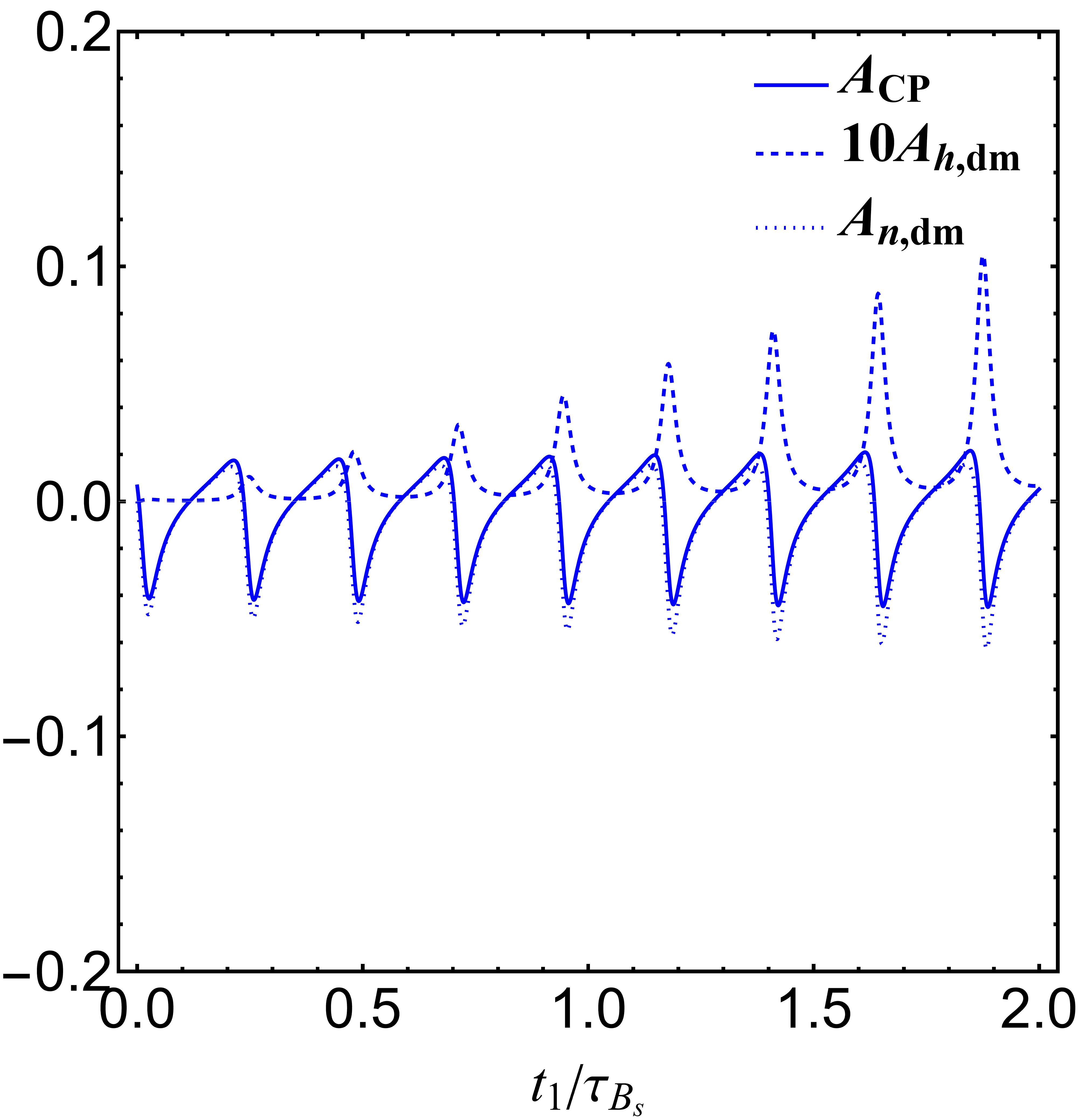}
     \hspace{0.1cm}
    \includegraphics[keepaspectratio,width=4.2cm]{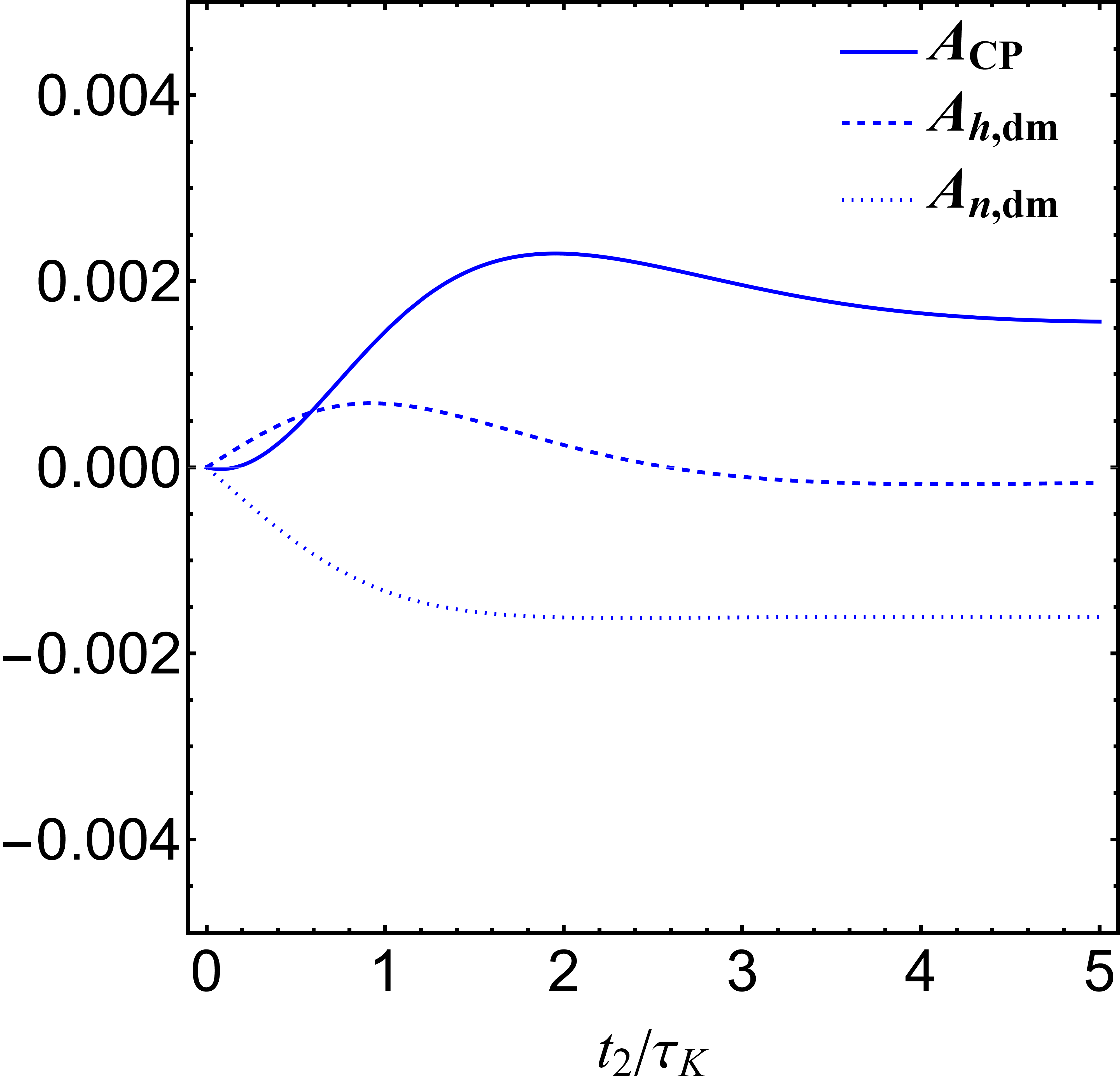}
    \caption{Time dependence of the CP asymmetry $A_{\rm CP}$ in $B^0_s(t_1) \to J/\psi  K(t_2) \to J/\psi (\pi^-\ell^+\nu_\ell)$. The left panel displays the two-dimensional time dependence. The middle panel and the right panel display the dependence on $t_1$ (with $t_2$ integrated from 0 to $\tau_{K}$) and $t_2$ (with $t_1$ integrated from $0$ to $2\tau_{B_s}$), respectively.}\label{fig:s2}
\end{figure*}

The numerical results for the $\pi^-\ell^+\nu_\ell$ channel are illustrated in Fig.~\ref{fig:s2}. The left panel shows the two-dimensional time dependence of the CP asymmetry, whose maximal values are of $\mathcal{O}(10\%)$. Integrating time $t_2$ from 0 to $\tau_K$, it is observed from the middle panel that the total asymmetry is dominated by the $A_{n, dm}$ term. This is attributable to the sine dependence on $t_1$ and hyperbolic sine dependence on $t_2$ of $A_{n, dm}$, particularly for the large $x_s$. Conversely, the numerator of $A_{h, dm}$ exhibits a hyperbolic sine dependence on $t_1$ and a sine dependence on $t_2$. In the range of $t_1$ from 0 to $2\tau_{B_s}$, the numerator of $A_{h, dm}$ increases almost linearly while the denominator behaves like a sine function, so $A_{h, dm}$ experiences modulation over time, as reflected in its increasing peak value. Integrating over $t_1$ from 0 to $2\tau_{B_s}$, we observe the $t_2$ dependence of $A_{\rm CP}$ in the right panel of Fig.~\ref{fig:s2}. Now, the contributions of the $C$ term and $A_{n, dm}$ term are comparable, slightly surpassing the $A_{h, dm}$ term.

When the final state is considered as $\pi^+\ell^-\bar{\nu}_\ell$, the results for the double-mixing terms are similar, but with an extra negative sign in the term $A_{h, dm}$. The corresponding expressions are
\begin{align}
A_{h,dm}(t_1, t_2)=& -\frac{e^{-\pqty{\Gamma_{B_s} t_1 + \Gamma_K t_2}}}{2D(t_1,t_2)}\Big\{\sinh{\frac{1}{2}\Delta \Gamma_K t_2} \Big[\cos{\Phi_4 }\times\nonumber\\
&\Big(\abs{\frac{q_K}{p_K}}-\abs{\frac{p_K}{q_K}}\Big)\Big]-\sin{\Delta m_K t_2}\Big[\sin{\Phi_4 }\Big(\abs{\frac{q_K}{p_K}}\nonumber\\
&+\abs{\frac{p_K}{q_K}}\Big)\Big]\Big\}\sinh{\frac{1}{2}\Delta \Gamma_{B_s} t_1}, \label{eq29-3}
\end{align}
\begin{align}
A_{n,dm}(t_1, t_2) =& -\frac{e^{-\pqty{\Gamma_{B_s} t_1 + \Gamma_K t_2}}}{2D(t_1,t_2)}\Big\{-\sinh{\frac{1}{2}\Delta \Gamma_K t_2} \Big[\sin{\Phi_4 }\nonumber\\
&\times\Big(\abs{\frac{q_K}{p_K}}+\abs{\frac{p_K}{q_K}}\Big)\Big]-\sin{\Delta m_K t_2}\Big[\cos{\Phi_4 }\times\nonumber\\
&\Big(\abs{\frac{q_K}{p_K}}-\abs{\frac{p_K}{q_K}}\Big)\Big]\Big\}\sin{\Delta m_{B_s} t_1}.\label{eq29-4}
\end{align}
The denominator $D(t_1,t_2)$ can be obtained by combining~\eqref{eq4} with~\eqref{eq12} - \eqref{eq13-2}, while also performing the substitutions $C^\prime_+(t_2)$
$  \leftrightarrow C^\prime_-(t_2)$, $\abs{p_K/q_K} \leftrightarrow \abs{q_K/p_K}$ and $\phi_K \to -\phi_K$, and adding an extra negative sign in front of $S^\prime_n(t_2)$. 

\begin{figure*}[htbp]
    \centering
    \includegraphics[keepaspectratio,width=5.3cm]{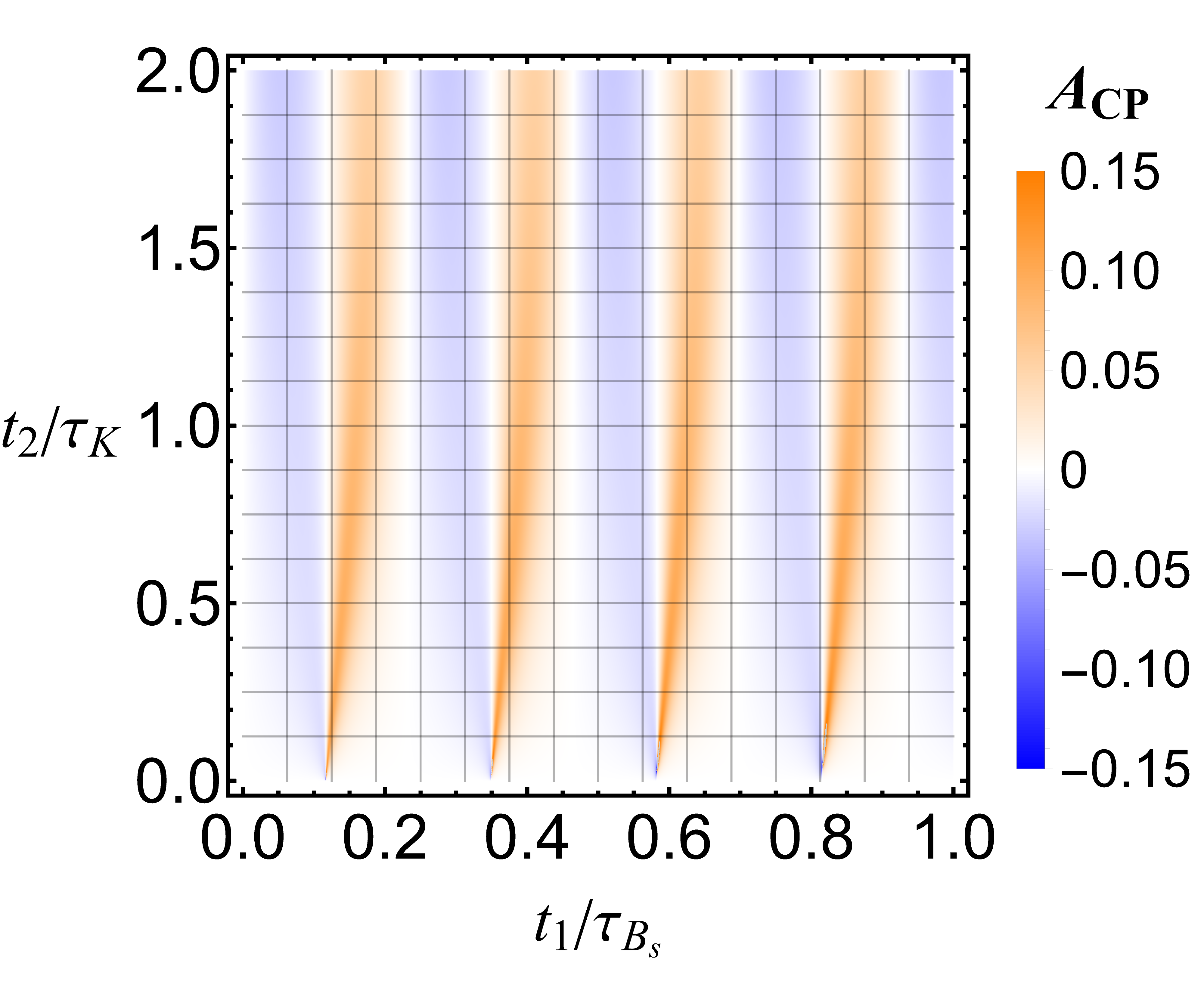}
    \hspace{0.1cm}
    \includegraphics[keepaspectratio,width=4.2cm]{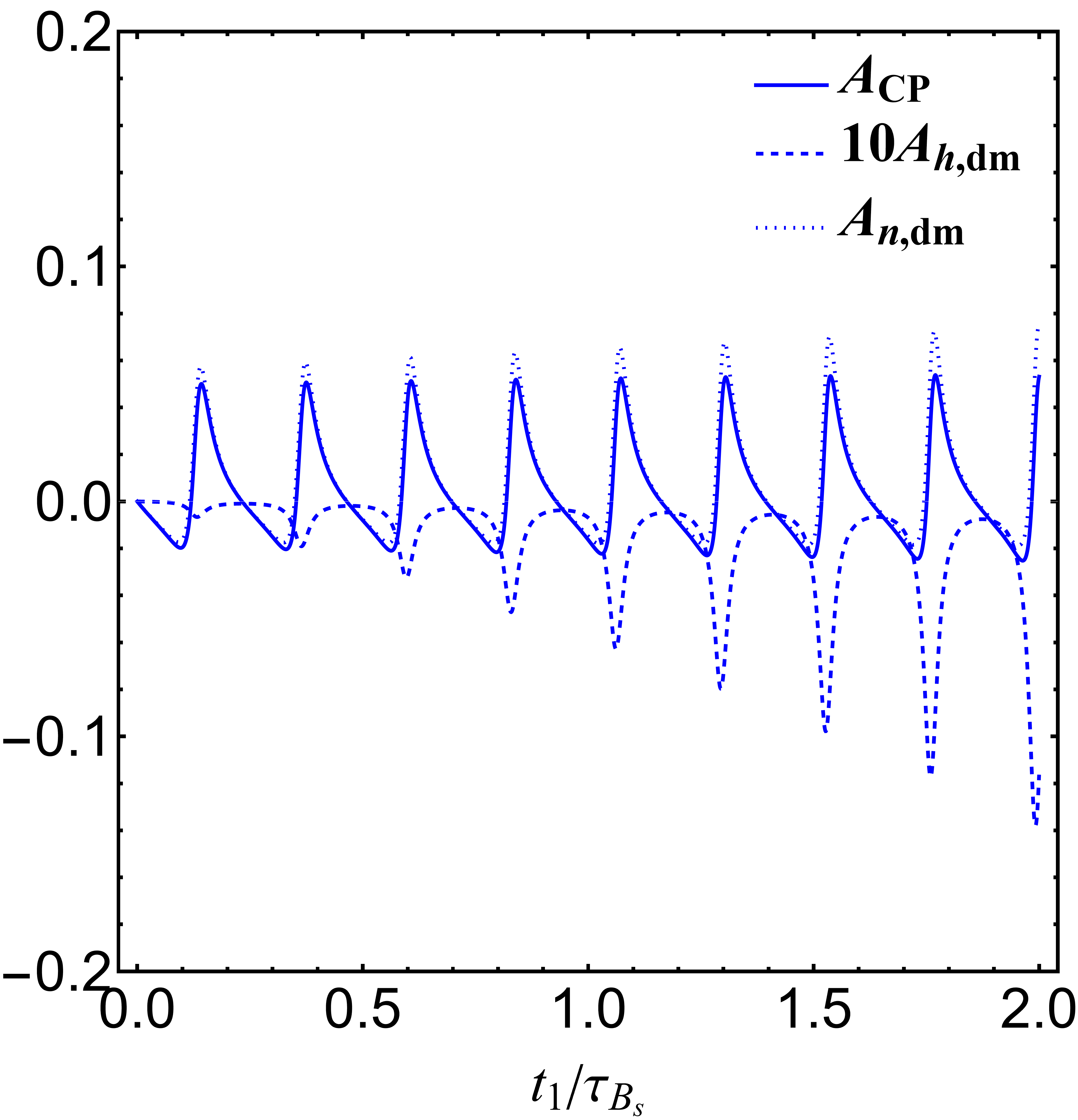}
     \hspace{0.1cm}
    \includegraphics[keepaspectratio,width=4.2cm]{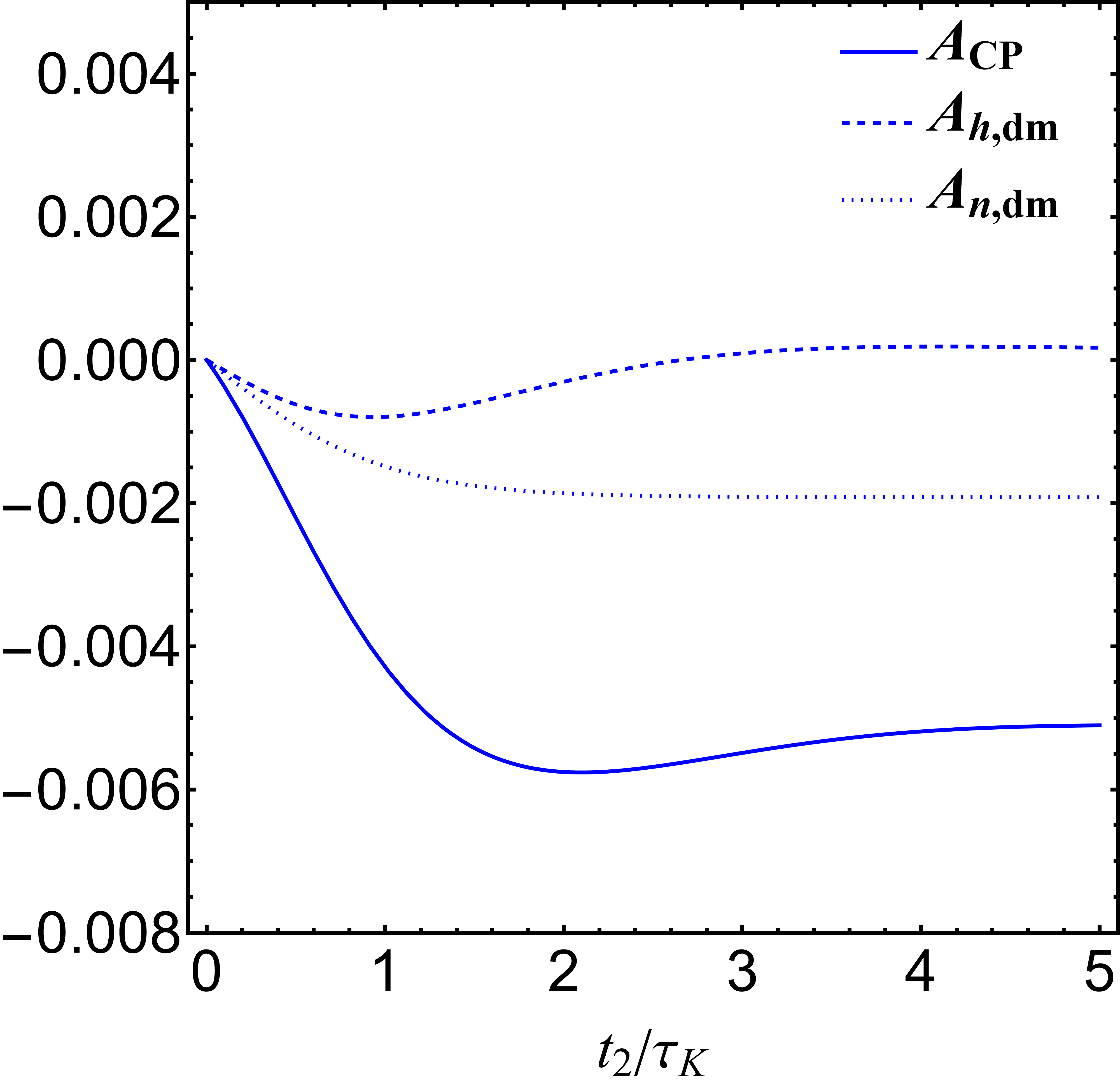}
    \caption{Time dependence of the CP asymmetry $A_{\rm CP}$ in $B^0_s(t_1) \to J/\psi K(t_2) \to J/\psi (\pi^+\ell^-\bar{\nu}_\ell)$. The left panel displays the two-dimensional time dependence. The middle panel and the right panel display the dependence on $t_1$ (with $t_2$ integrated from 0 to $\tau_{K}$) and $t_2$ (with $t_1$ integrated from $0$ to $2\tau_{B_s}$), respectively.}\label{fig:s2n}
\end{figure*}

Our numerical analysis for this channel is depicted in Fig.~\ref{fig:s2n}. The left panel presents the two-dimensional time dependence. The difference between the results for this channel and the $\pi^-\ell^+\nu_\ell$ channel lies in the modulation of the denominator. When $t_2$ is fixed, the dominant term in the denominator has a hyperbolic cosine dependence on $t_1$, which is the same in the results for the two different channels, while the sine term is a secondary contribution, and it contributes to the two channels with a sign difference. The middle panel shows the CP asymmetry with $t_2$ integrated from 0 to $\tau_K$. While the $A_{n, dm}$ term still dominates, the growth magnitude of the $A_{h, dm}$ term is larger compared to the $\pi^-\ell^+\nu_\ell$ channel. Integrating over $t_1$ and analyzing $t_2$ dependence in the right panel, the $A_{h, dm}$ term emerges as dominant. Additionally, both the $A_{\rm CP}$ and $A_{n, dm}$ terms in the $\pi^+\ell^-\bar\nu_\ell$ channel display a sign difference from the $\pi^-\ell^+\nu_\ell$ channel.

\subsubsection{\texorpdfstring{$B^0_s \to \rho^0 K \to \rho^0 (\pi^\pm\ell^\mp\nu)$}{}}

The decay processes $B^0_s \to \rho^0 K \to \rho^0 (\pi^\pm\ell^\mp\nu)$ resemble to $B^0_s \to J/\psi K^0_{L,S}(\to \pi^-\ell^+\nu_\ell)$, differing primarily in the phase difference $\omega_5$. This difference arises from the phase between the decay amplitude $\braket{\rho^0\bar{K}^0}{B^0_s}$ and its charge conjugate $\braket{\rho^0 K^0}{\bar{B}^0_s}$, contrasting with the phase difference $\omega_{B_s}$ in the former. Considering $CP\ket{\rho^0 K^0} = -\ket{\rho^0 \bar{K}^0}$, we define
\begin{align}\label{eq30}
     \frac{\braket{\rho^0K^0}{\bar{B}^0_s}}{\braket{\rho^0\bar{K}^0}{B^0_s}}=- e^{i\omega_5},\,\omega_5=\arg{\frac{V_{ub}V^\star_{ud}}{V^\star_{ub}V_{ud}}},
\end{align}
where again we have neglected the direct CP asymmetry in these processes.

For the final state $\pi^-\ell^+\nu_\ell$, two possible paths exist: $B^0_s \to \rho^0 \bar{K}^0 \to \rho^0 K^0 \to \rho^0 (\pi^-\ell^+\nu_\ell)$ and $B^0_s \to \bar{B}^0_s \to \rho^0 K^0 \to \rho^0(\pi^-\ell^+\nu_\ell)$. Therefore, it is categorized as the first type in our classification. The results for the double-mixing terms are 
\begin{align}
       A_{h,dm}(t_1, t_2)=& -\frac{e^{-\pqty{\Gamma_{B_s} t_1 + \Gamma_K t_2}}}{2D(t_1,t_2)}\Big\{\pqty{\abs{\frac{p_K}{q_K}}-\abs{\frac{q_K}{p_K}}}\cos{\Phi_5 }\nonumber\\
       &\times\sinh{\frac{1}{2}\Delta \Gamma_K t_2}-\pqty{\abs{\frac{p_K}{q_K}}+\abs{\frac{q_K}{p_K}}}\sin{\Phi_5 }\nonumber\\
       &\times\sin{\Delta m_K t_2}\Big\}\sinh{\frac{1}{2}\Delta \Gamma_{B_s} t_1},
\end{align}
\begin{align}
  A_{n,dm}(t_1, t_2) =& -\frac{e^{-\pqty{\Gamma_{B_s} t_1 + \Gamma_K t_2}}}{2D(t_1,t_2)}\Big\{\pqty{\abs{\frac{p_K}{q_K}}+\abs{\frac{q_K}{p_K}}}\sin{\Phi_5 }\nonumber\\
  &\times\sinh{\frac{1}{2}\Delta \Gamma_K t_2}+\pqty{\abs{\frac{p_K}{q_K}}-\abs{\frac{q_K}{p_K}}}\cos{\Phi_5 }\nonumber\\
  &\times\sin{\Delta m_K t_2}\Big\}\sin{\Delta m_{B_s} t_1},\label{eq32}
\end{align}
where $\Phi_5 \equiv \phi_{B_s}+\phi_K-\omega_5$. $D(t_1,t_2)$ can be obtained by replacing $\omega_{B_s}$ in the denominator of~\eqref{eq29-1} with $\omega_5$.

\begin{figure*}[htbp]
    \centering
    \includegraphics[keepaspectratio,width=5.3cm]{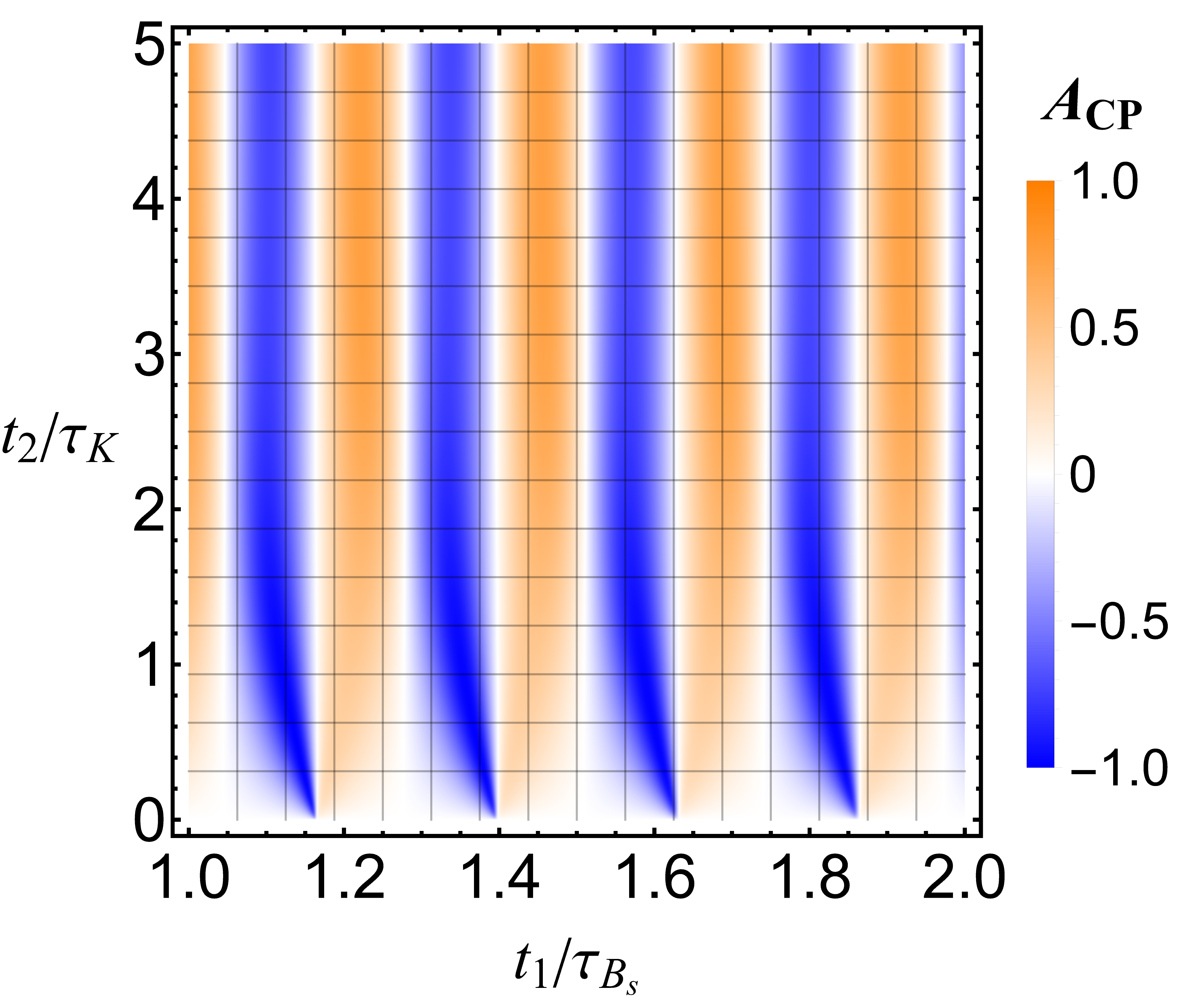}
    \hspace{0.1cm}
    \includegraphics[keepaspectratio,width=4.2cm]{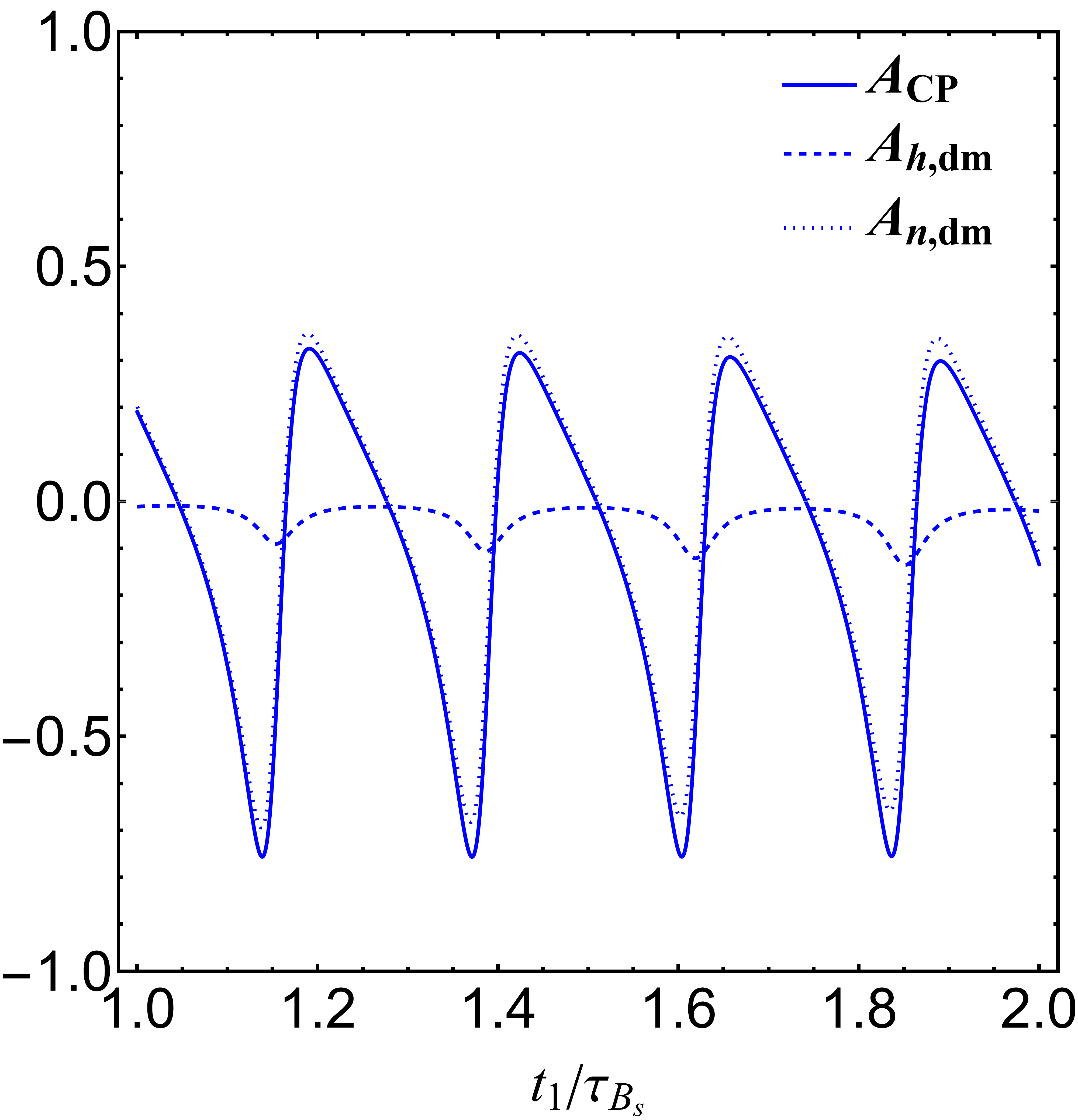}
     \hspace{0.1cm}
    \includegraphics[keepaspectratio,width=4.2cm]{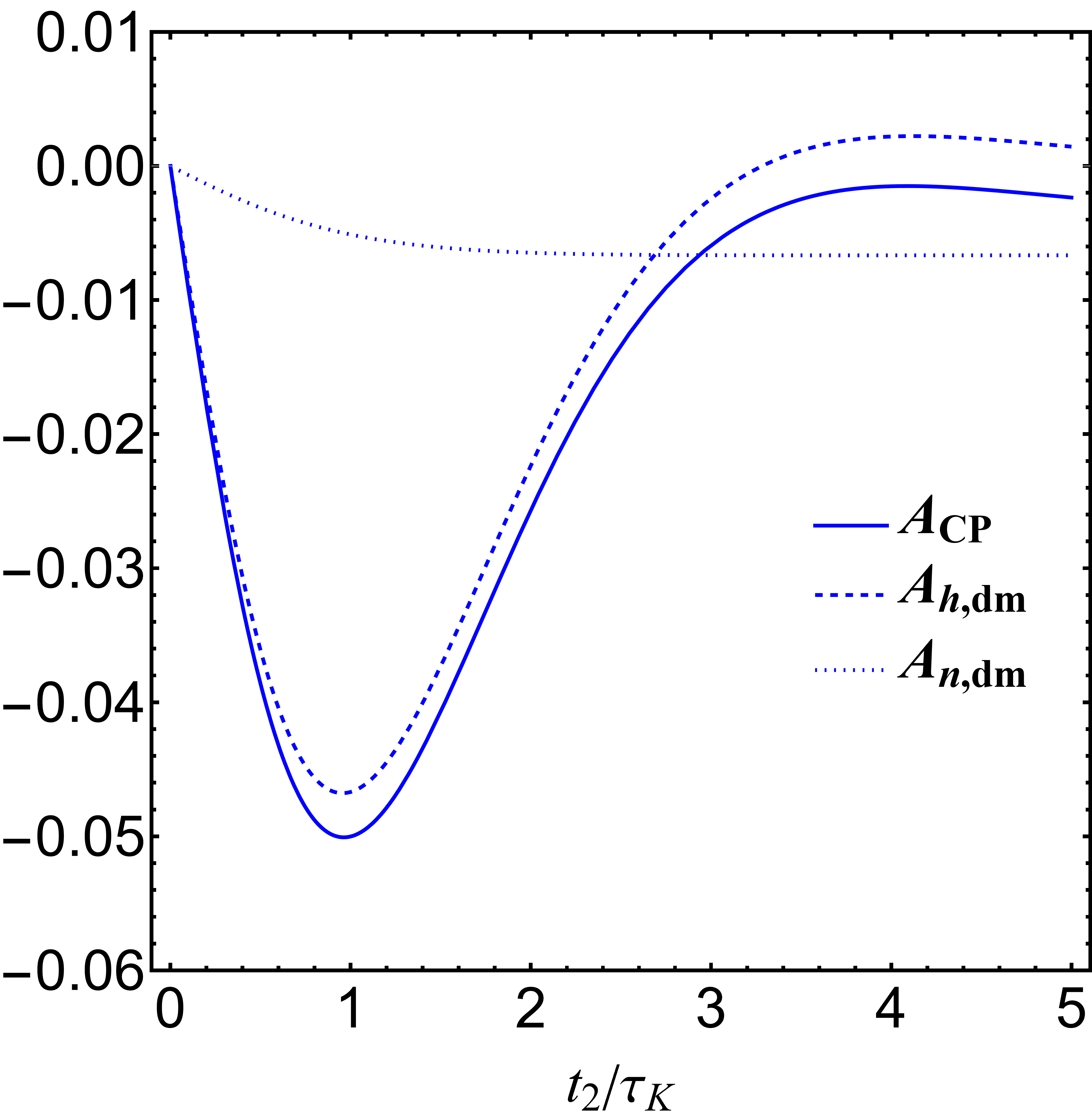}
    \caption{Time dependence of the CP asymmetry $A_{\rm CP}$ in $B^0_s(t_1) \to \rho^0  K(t_2) \to \rho^0 (\pi^-\ell^+\nu_\ell)$. The left panel displays the two-dimensional time dependence. The middle panel and the right panel display the dependence on $t_1$ (with $t_2$ integrated from 0 to $\tau_{K}$) and $t_2$ (with $t_1$ integrated from $\tau_{B_s}$ to $5\tau_{B_s}$), respectively.}\label{fig:s3}
\end{figure*}

We have conducted the numerical analysis of both the total CP asymmetry $A_{\rm CP}$ and the double-mixing CP asymmetry $A_{dm}$, which are depicted in Fig.~\ref{fig:s3}. The two-dimensional time-dependent CP violation is depicted in the left panel, showing a peak value exceeding $50\%$. By integrating $t_2$ from 0 to $\tau_K$, we obtain the $t_1$-dependent results displayed in the middle panel, where the double-mixing CP violation is notably significant. Here, $A_{\rm CP}$ predominantly arises from the double-mixing CP violation, particularly from the $A_{n, dm}$ term, which sums the contributions of the decays of $K_S^0$ and $K_L^0$. Integrating $t_1$ from $\tau_{B_s}$ to $5\tau_{B_s}$ yields the $t_2$-dependent results, showcased in the right panel. In this scenario, $A_{\rm CP}$ still mainly originates from the double-mixing CP violation, with the term $A_{h, dm}$ being dominant, particularly governed by the $K_S^0-K_L^0$ interference. The study of the $K^0_S - K^0_L$ interference effect is of significant importance. In collider experiments such as LHCb, the detection
of the $K_L$ component is notoriously challenging due to its considerably longer lifetime.
Conversely, the interference effect, with an effective lifetime only twice that of the $K_S$, offers a significantly improved detection rate and an opportunity to probe $K_L$ physics~\cite{AlvesJunior:2018ldo}. It is worth noting that the significance of $K^0_S - K^0_L$ interference has been discussed in previous studies~\cite{Grossman:2011zk}, which explores the CP violation in $D^\pm \to K_S \pi^\pm$. Within our theoretical framework, the results of this study can be recovered by calculating CP violation in the $D^0 \to K\pi$ process with turning off the mixing effects of $D^0$.

It is important to note that the penguin contribution to $A_{CP}$ in $B^0_s \to \rho^0 K \to \rho^0(\pi\ell\nu)$ with the integration of $t_1$, does yield a considerable effect. A comprehensive
analysis is required to investigate the contribution from direct CP violation, which will be addressed in a future study.

The results for the double-mixing terms of the $\pi^+\ell^-\bar{\nu}_\ell$ channel are similar to those of the $\pi^-\ell^+{\nu}_\ell$ channel with an extra overall negative sign in the term $A_{h, dm}$. They are calculated as
\begin{align}
A_{h,dm}(t_1, t_2)=& -\frac{e^{-\pqty{\Gamma_{B_s} t_1 + \Gamma_K t_2}}}{2D(t_1,t_2)}\Big\{\sinh{\frac{1}{2}\Delta \Gamma_K t_2} \Big[\cos{\Phi_5 }\nonumber\\
&\times\Big(\abs{\frac{q_K}{p_K}}-\abs{\frac{p_K}{q_K}}\Big)\Big]+\sin{\Delta m_K t_2}\Big[\sin{\Phi_5 }\nonumber\\
&\times\Big(\abs{\frac{q_K}{p_K}}+\abs{\frac{p_K}{q_K}}\Big)\Big]\Big\}\sinh{\frac{1}{2}\Delta \Gamma_{B_s} t_1}, 
\end{align}

\begin{align}
A_{n,dm}(t_1, t_2) =& -\frac{e^{-\pqty{\Gamma_{B_s} t_1 + \Gamma_K t_2}}}{2D(t_1,t_2)}\Big\{\sinh{\frac{1}{2}\Delta \Gamma_K t_2} \Big[\sin{\Phi_5 }\nonumber\\
&\times\Big(\abs{\frac{q_K}{p_K}}+\abs{\frac{p_K}{q_K}}\Big)\Big] -\sin{\Delta m_K t_2}\Big[\cos{\Phi_5 }\nonumber\\
&\times\Big(\abs{\frac{q_K}{p_K}}-\abs{\frac{p_K}{q_K}}\Big)\Big]\Big\}\sin{\Delta m_{B_s} t_1},
\end{align}
where the denominator $D(t_1,t_2)$ can be obtained from the denominator of~\eqref{eq29-3} by replacing $\omega_{B_s}$ with $\omega_5$.

\begin{figure*}[htbp]
    \centering
    \includegraphics[keepaspectratio,width=5.3cm]{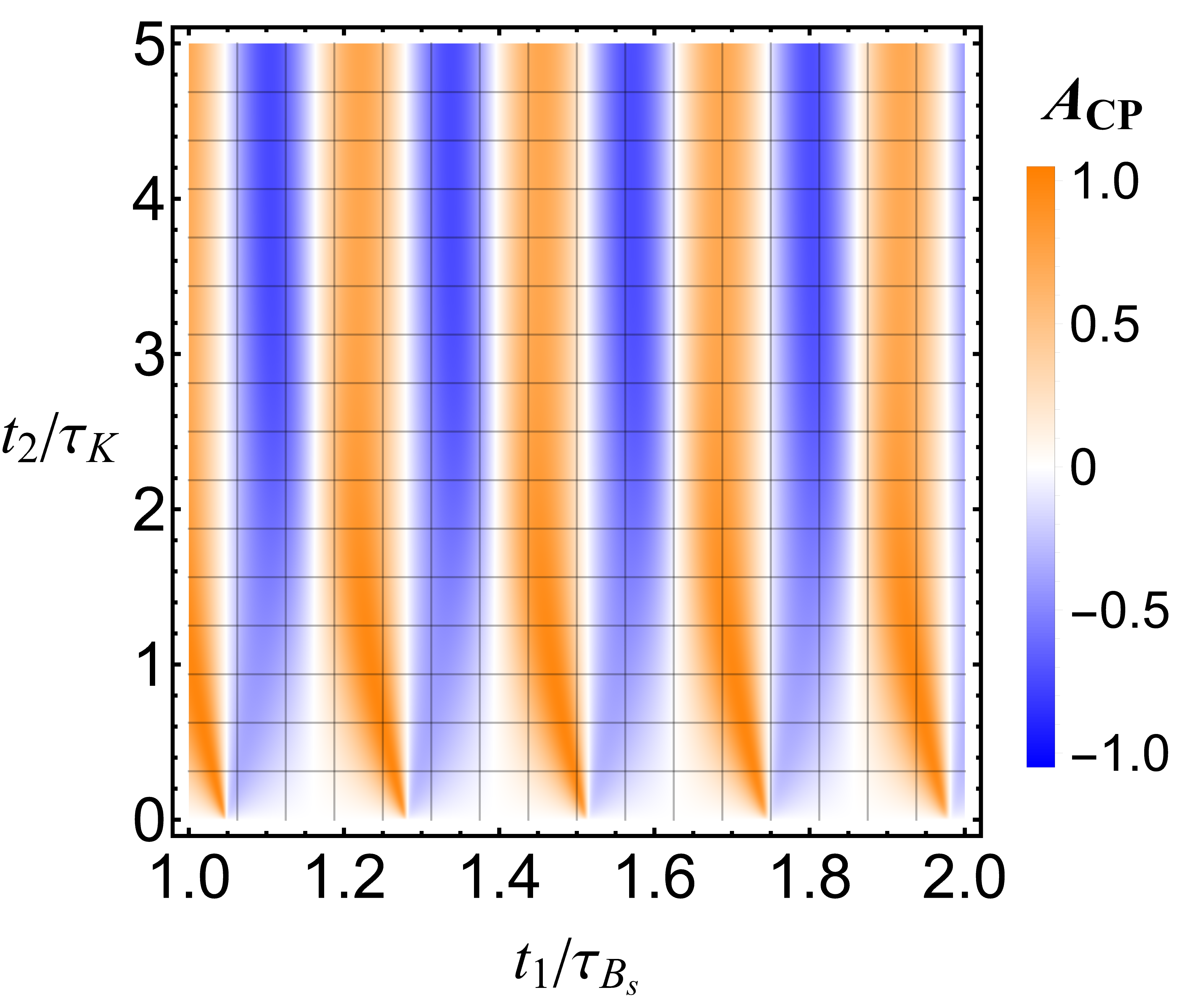}
    \hspace{0.1cm}
    \includegraphics[keepaspectratio,width=4.2cm]{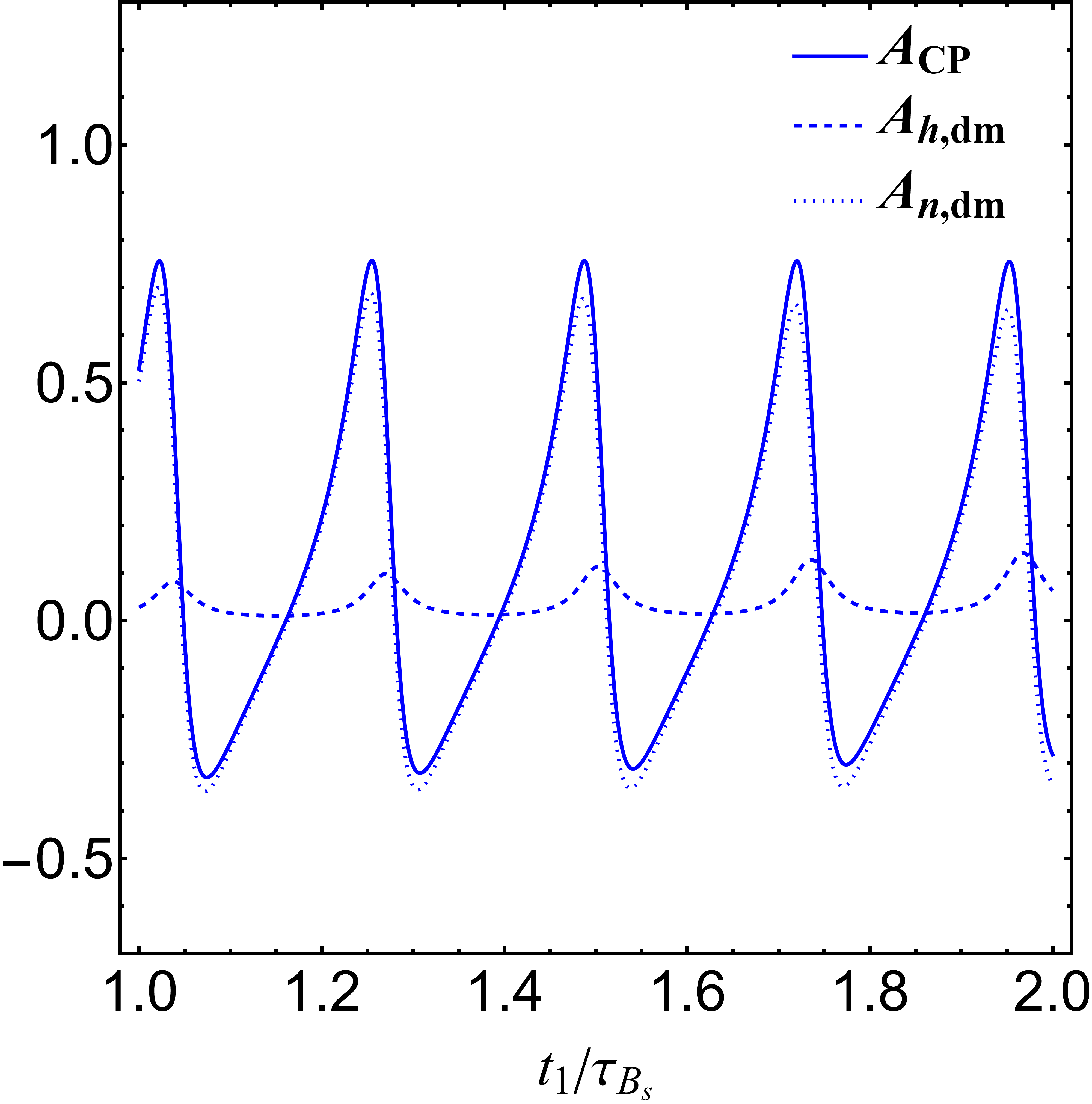}
     \hspace{0.1cm}
    \includegraphics[keepaspectratio,width=4.2cm]{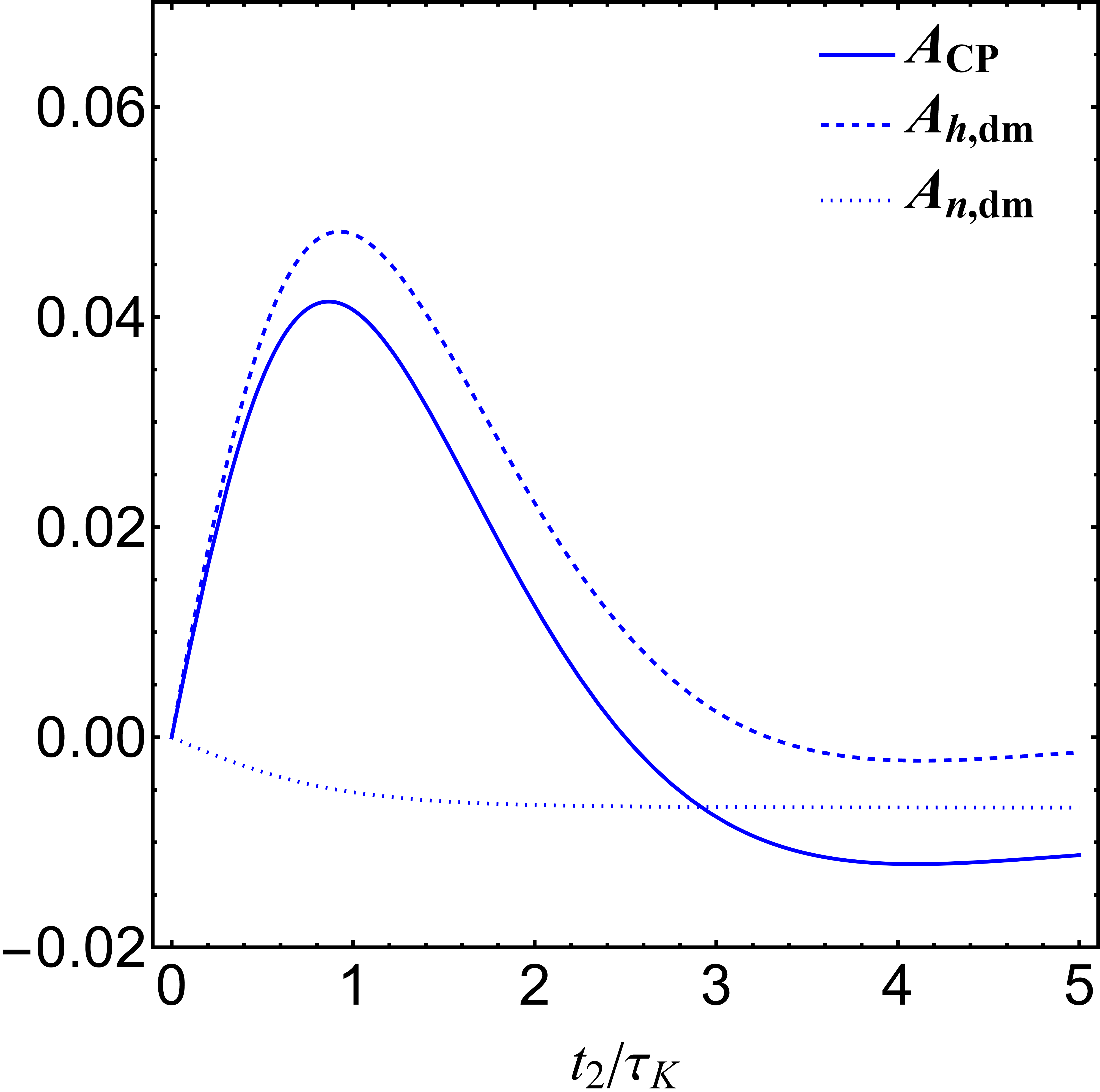}
    \caption{Time dependence of the CP asymmetry $A_{\rm CP}$ in $B^0_s(t_1) \to \rho^0  K(t_2) \to \rho^0 (\pi^+\ell^-\bar{\nu}_\ell)$. The left panel displays the two-dimensional time dependence. The middle panel and the right panel display the dependence on $t_1$ (with $t_2$ integrated from 0 to $\tau_{K}$) and $t_2$ (with $t_1$ integrated from $\tau_{B_s}$ to $5\tau_{B_s}$), respectively.}\label{fig:s3n}
\end{figure*}

Our analysis for this channel reveals a smaller CP violation compared to the $\pi^-\ell^+\nu_\ell$ final state, as displayed in Fig.~\ref{fig:s3n}. The left panel displays the two-dimensional time dependence of the CP asymmetry. Integrating $t_2$ from 0 to $\tau_K$, the $t_1$-dependent results, depicted in the middle panel, highlight the dominant role of separate contributions from $K_S^0$ and $K_L^0$ in the double-mixing CP violation. When integrating $t_1$ from $\tau_{B_s}$ to $5\tau_{B_s}$, the $t_2$-dependent results, as shown in the right panel, are predominantly influenced by the $K_S^0-K_L^0$ interference. Notably, the value of the $t_1$-integrated CP violation is almost entirely positive, contrasting with the result for the $\pi^-\ell^+\nu_\ell$ final state. This distinction enables the experimental differentiation of these two processes.

We emphasize that the best approach for the first experimental attempt to measure the double-mixing CP violation would be to combine the two decay channels mentioned above. Considering the CP violation defined by the differences of the summed branching ratios 

\begin{align}
   { \frac{\mathcal{B}_1 - \mathcal{B}_2}{\mathcal{B}_1 + \mathcal{B}_2}} \; , \label{eq39-3}
\end{align}
where
\begin{align}
   \mathcal{B}_1 & = \mathcal{B}[B^0_s/ \bar B^0_s \to \rho^0 K \to \rho^0 (\pi^-\ell^+\nu_\ell)],\nonumber\\
   \mathcal{B}_2 & = \mathcal{B}[B^0_s/ \bar B^0_s \to \rho^0 K \to \rho^0 (\pi^+\ell^-\bar\nu_\ell)]\; , 
\end{align}
initial tagging for $B$ mesons in experiments is not necessary and the corresponding efficiency loss is prevented. It contains the double-mixing and non-double-mixing terms, which are calculated to be 
\begin{align}
A_{h,dm}(t_1,t_2) =& -\frac{e^{-\pqty{\Gamma_{B_s}t_1+\Gamma_K t_2}}}{D(t_1,t_2)}\Big\{\sinh{\frac{1}{2}\Delta \Gamma_K t_2}\cos{\Phi_5 }\nonumber\\
&\times\pqty{\abs{\frac{p_K}{q_K}}-\abs{\frac{q_K}{p_K}}}-\sin{\Delta m_K t_2}\sin{\Phi_5 }\nonumber\\
&\times\pqty{\abs{\frac{p_K}{q_K}}+\abs{\frac{q_K}{p_K}}}\Big\}\sinh{\frac{1}{2}\Delta \Gamma_{B_s} t_1},\label{eqc1}
\end{align}
\begin{align}
A_{non-dm}(t_1,t_2) = &\ \frac{e^{-\Gamma_{B_s}t_1}}{D(t_1,t_2)}\Big(\abs{\frac{p_K}{q_K}}^2-\abs{\frac{q_K}{p_K}}^2\Big)\abs{g_{-,K}(t_2)}^2\nonumber\\
&\times \cosh{\frac{1}{2}\Delta \Gamma_{B_s} t_1},\label{eqc2}
\end{align}
\begin{align}
D(t_1,t_2)=& \ e^{-\Gamma_{B_s}t_1}\Big\{\Big(\abs{\frac{p_K}{q_K}}^2+\abs{\frac{q_K}{p_K}}^2\Big)\abs{g_{-,K}(t_2)}^2\nonumber\\
&+2\abs{g_{+,K}(t_2)}^2 \Big\}\cosh{\frac{1}{2}\Delta \Gamma_{B_s} t_1}-e^{-\pqty{\Gamma_{B_s}t_1+\Gamma_K t_2}}\nonumber\\
&\times\Big\{\sinh{\frac{1}{2}\Delta \Gamma_K t_2}\cos{\Phi_5 }\pqty{\abs{\frac{p_K}{q_K}}+\abs{\frac{q_K}{p_K}}}-\nonumber\\
&\sin{\Delta m_K t_2}\sin{\Phi_5 }\pqty{\abs{\frac{p_K}{q_K}}-\abs{\frac{q_K}{p_K}}}\Big\}\sinh{\frac{1}{2}\Delta \Gamma_{B_s} t_1} . \label{eqc3}
\end{align}
Apparently, the non-double-mixing term is negligible and the total CP violation is dominated by the double-mixing contribution. 

\begin{figure*}[htbp]
    \centering
    \includegraphics[keepaspectratio,width=5.3cm]{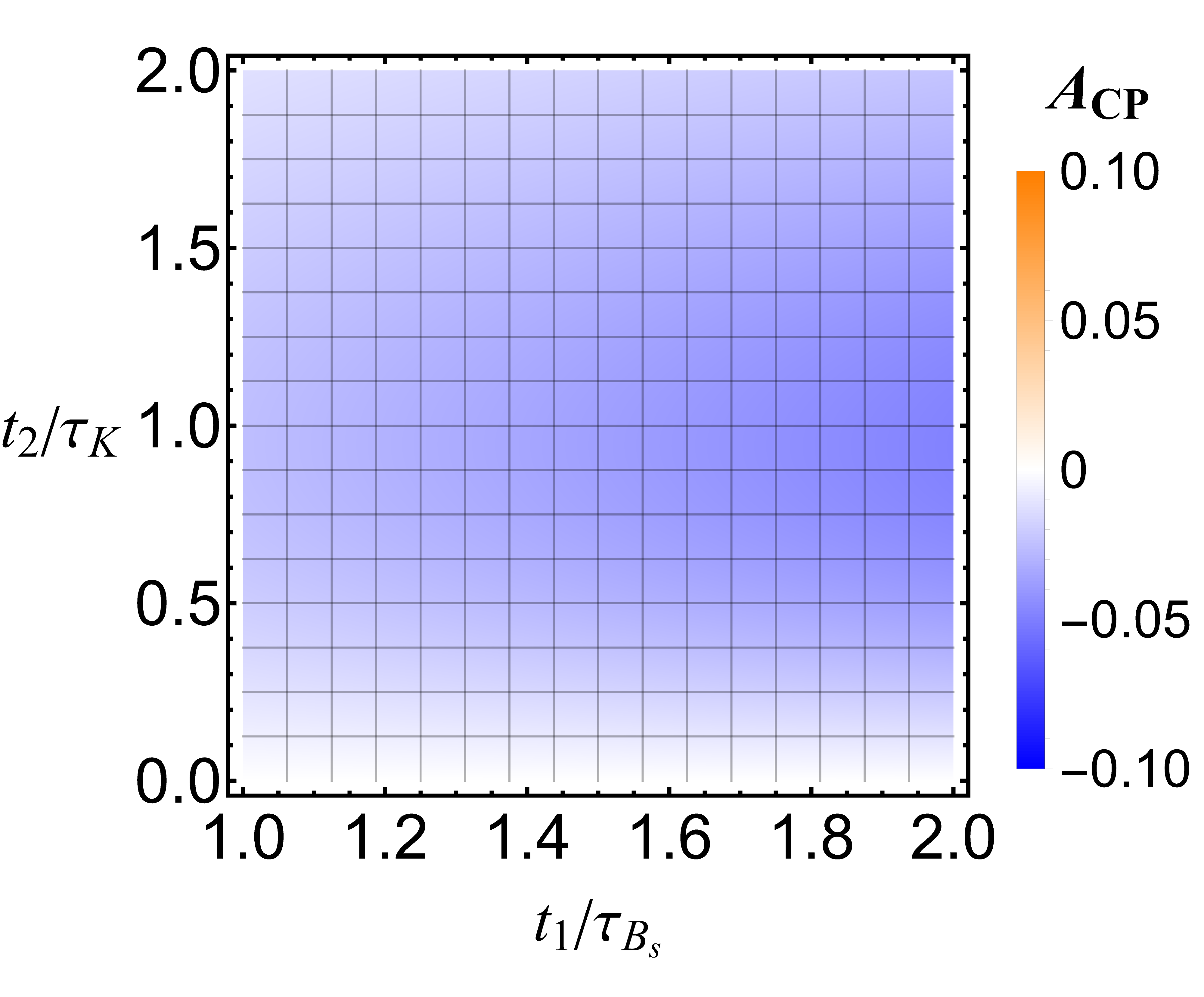}
    \hspace{0.1cm}
    \includegraphics[keepaspectratio,width=4.2cm]{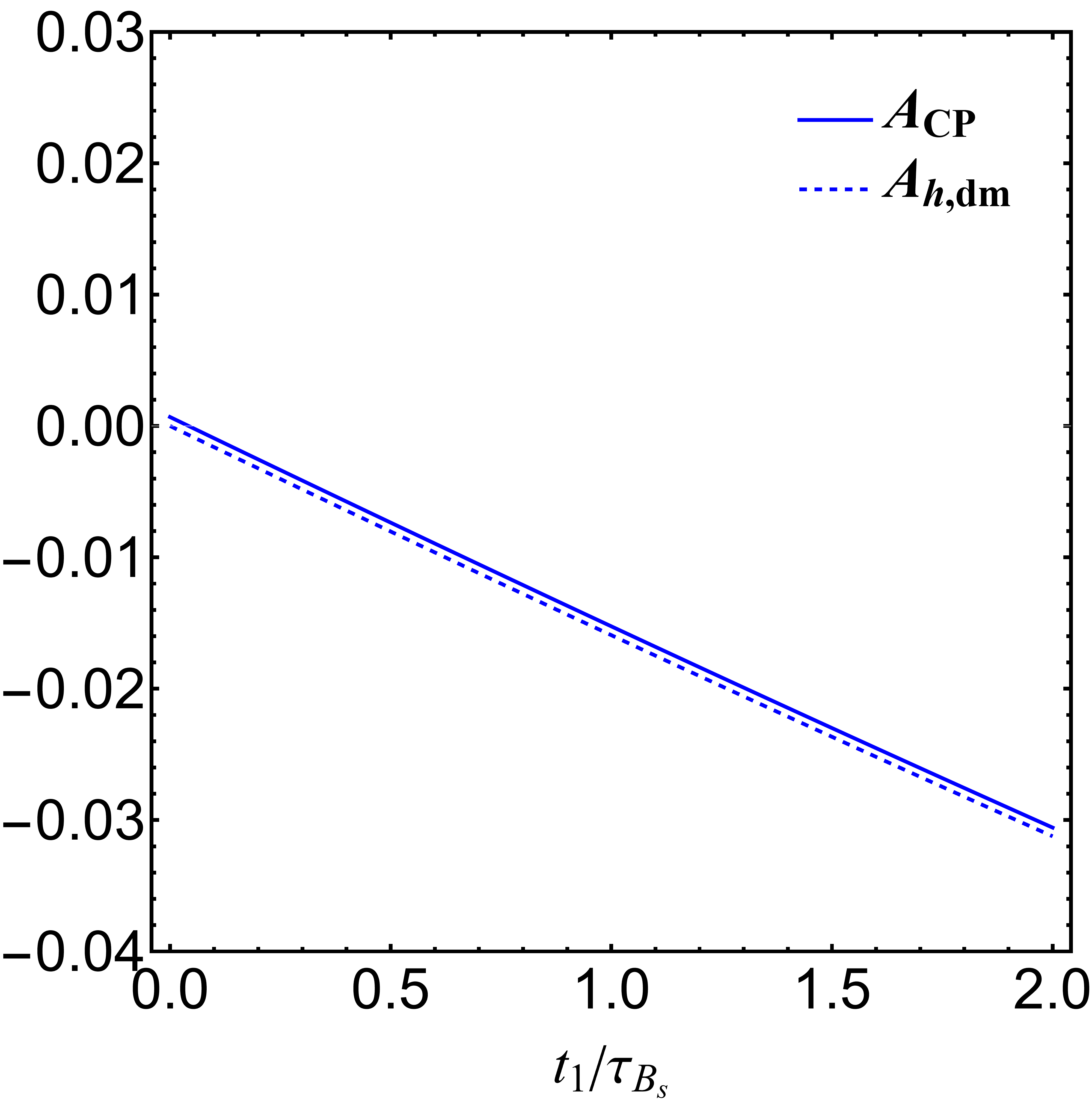}
     \hspace{0.1cm}
    \includegraphics[keepaspectratio,width=4.2cm]{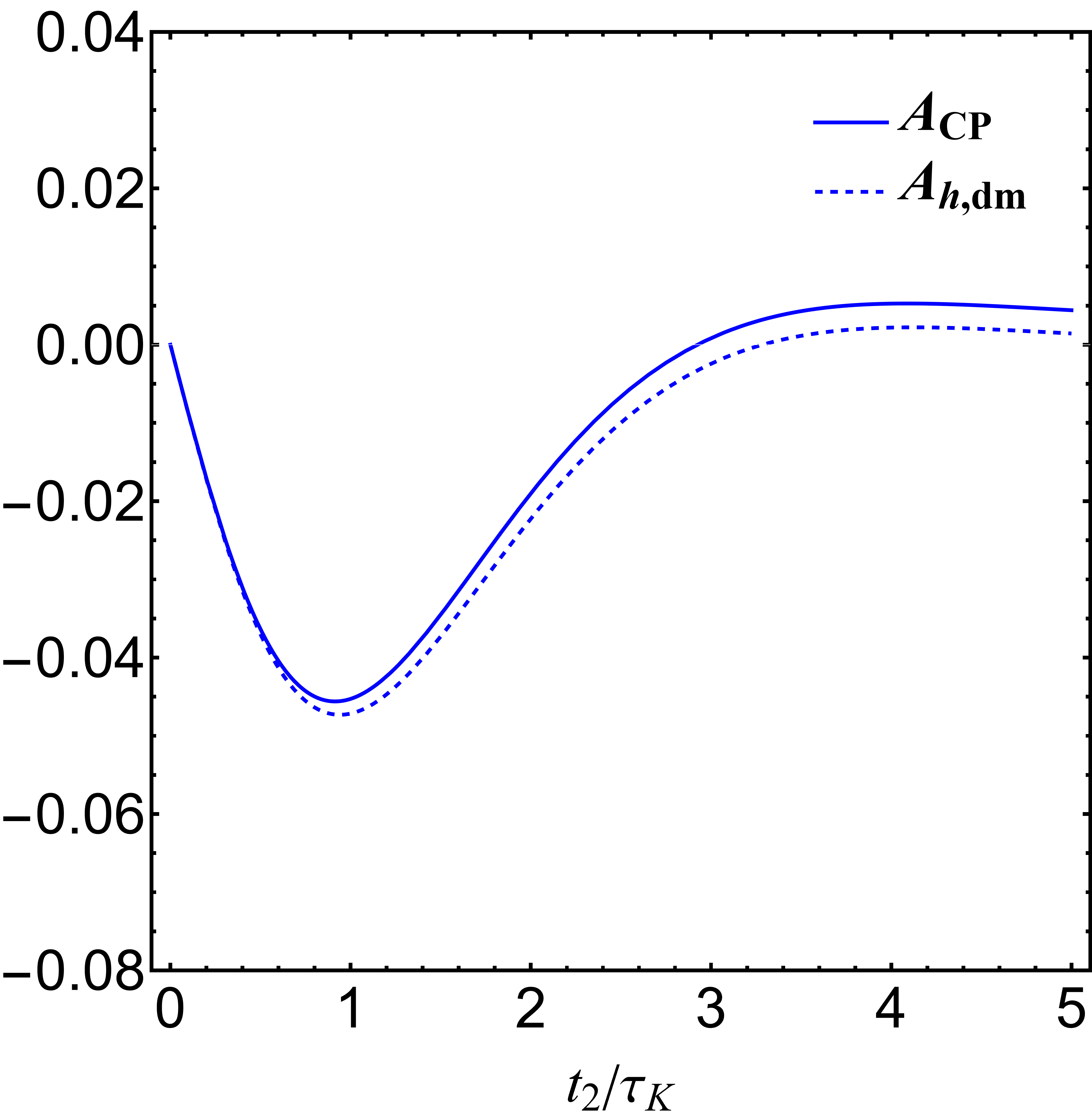}
    \caption{Time dependence of the CP asymmetry $A_{\rm CP}$ in $B^0_s(t_1) \to \rho^0  K(t_2) \to \rho^0 (\pi\ell\nu)$ defined by ~\eqref{eq39-3}. The left panel displays the two-dimensional time dependence. The middle panel and the right panel display the dependence on $t_1$ (with $t_2$ integrated from 0 to $\tau_{K}$) and $t_2$ (with $t_1$ integrated from $\tau_{B_s}$ to $5\tau_{B_s}$), respectively.}\label{fig:s3c}
\end{figure*}

\begin{table}[htbp]
\caption{The time integrated $A_{\rm CP}$ in $B^0_s \to \rho^0 K \to \rho^0 (\pi\ell\nu_\ell)$.}\label{ACP6}
\begin{center}
{
\begin{tabular}{ccc}
\hline
$t_1/\tau_{B_s}$ & $t_2/ \tau_K$ & $A_{\rm CP}$\\
\hline
0$\sim$3 & 0$\sim$ 2 & -1.33\%\\
0$\sim$2 & 0$\sim$ 2 & -1.06\%\\

0$\sim$3 & 0$\sim$ 1 & -1.28\%\\
0$\sim$2 & 0$\sim$ 1 & -1.03\%\\
0$\sim$3 & 0$\sim$ 0.5 & -0.83\% \\
0$\sim$2 & 0$\sim$ 0.5 & -0.67\%\\
\hline
\end{tabular}
}
\end{center}
\end{table}

We conducted a numerical analysis of this process, and the results are depicted in Fig.~\ref{fig:s3c}. The left panel exhibits that the total CP asymmetry has a hyperbolic tangent dependence on $t_1$, consequently behaving linearly when $t_1$ is not too large. When integrating $t_2$ from 0 to $\tau_K$, the total CP violation is predominantly contributed from the double-mixing CP violation, as displayed in the middle panel. It exhibits a linear dependence on $t_1$, which is consistent with the two-dimensional time dependence. The right panel shows the variation of CP violation with respect to $t_2$ after integrating $t_1$ from $\tau_{B_s}$ to $5\tau_{B_s}$. As $t_2$ increases, it first reaches a maximum and then rapidly decreases to nearly zero. This is because the primary contribution of the double-mixing CP violation in the numerator comes from the interference term between $K_S^0$ and $K_L^0$, {\it i.e.}, the second term in ~\eqref{eqc3}, which has a sine dependence on $t_2$. Meanwhile, the denominator is mainly contributed by the term that depends on the hyperbolic cosine function of $t_2$, which continues to increase with time. The time-integrated results for the CP asymmetry~\eqref{eq39-3} for this case are presented in Table~\ref{ACP6}, for experimental convenience. With different integrated time ranges, the values are basically around -1\%.

\subsubsection{\texorpdfstring{$B^0_s \to D K \to (K^-\pi^+) (\pi^-\ell^+\nu_\ell)$}{}}

We now focus on the decay channel $B^0_s \to D K \to (K^-\pi^+)$
$ (\pi^-\ell^+\nu_\ell)$. This process involves four paths, even when we neglect the small mixing of the neutral $D$ meson: $B^0_s \to D^0 \bar{K}^0$
$ \to D^0 K^0 \to f$, $B^0_s \to \bar{D}^0\bar{K}^0 \to \bar{D}^0 K^0 \to f$, $B^0_s \to \bar{B}^0_s \to D^0 K^0 $
$\to f$ and $B^0_s \to \bar{B}^0_s \to \bar{D}^0 K^0 \to f$, where $f \equiv (K^-\pi^+)(\pi\ell\nu)$. This situation is very unique and does not belong to any of the categories we previously defined, but the expression for the CP violation can be derived analogously. We define the ratios between the decay amplitudes as
\begin{align}\label{eq:rdelta}
\frac{A\pqty{\bar{B}^0_s \to D^0 K^0}}{A\pqty{B^0_s \to \bar{D}^0 \bar{K}^0}}&=e^{i\omega_3},\quad \omega_3=\arg{\frac{V_{cb}^{}V^\star_{ud}}{V^\star_{cb}V_{ud}^{}}}, \nonumber \\ 
\frac{A\pqty{B^0_s \to D^0 \bar{K}^0}}{A\pqty{B^0_s \to \bar{D}^0 \bar{K}^0}}&=r e^{i\pqty{\delta_4+\omega_4}},\quad \omega_4=\arg{\frac{V_{ub}^\star V^{}_{cd}}{V^\star_{cb}V_{ud}^{}}},
\end{align}
where $\omega_{3, 4}$ are the weak phases and $\delta_4$ is the strong phase. Then, the double-mixing and non-double-mixing CP violation terms defined in \eqref{eq4} are calculated as
\begin{widetext}
\begin{align}
A_{h,dm}(t_1,t_2)=& \frac{N_1}{D(t_1,t_2)}\nonumber\\
=&\ \frac{e^{-\pqty{\Gamma_{B_s} t_1 +\Gamma_K t_2}}}{2D(t_1,t_2)}\abs{\frac{p_K}{q_K}}\Big\{\sinh{\frac{\Delta\Gamma_K t_2}{2}}\left[r\cos{\pqty{\delta_4+\Phi_6}}-r_D \cos{\pqty{\Phi_7 -\delta_D}}\right]+\sin{\Delta m_K t_2}\Big[-r\sin{\pqty{\delta_4+\Phi_6}}\nonumber\\
&+r_D \sin{\pqty{\Phi_7 -\delta_D}}\Big]\Big\}\sinh{\frac{\Delta \Gamma_{B_s}t_1}{2}}-\frac{e^{-\pqty{\Gamma_{B_s} t_1 +\Gamma_K t_2}}}{2D(t_1,t_2)}\abs{\frac{q_K}{p_K}}\Big\{\sinh{\frac{\Delta\Gamma_K t_2}{2}}\left[r\cos{\pqty{\delta_4-\Phi_6}}-r_D \cos{\pqty{\Phi_7 +\delta_D}}\right]\nonumber\\
&+\sin{\Delta m_K t_2}\left[-r\sin{\pqty{\delta_4-\Phi_6}}-r_D \sin{\pqty{\Phi_7 + \delta_D}}\right]\Big\}\sinh{\frac{\Delta \Gamma_{B_s}t_1}{2}}-\frac{e^{-\pqty{\Gamma_{B_s} t_1 +\Gamma_K t_2}}}{2D(t_1,t_2)}\abs{\frac{p_K}{q_K}}rr_D\Big\{\sinh{\frac{\Delta \Gamma_K t_2}{2}}\nonumber\\
&\times\pqty{r\cos{\pqty{\Phi_8 -\delta_D}}-r_D \cos{\pqty{\delta_4-\Phi_6}}}\sin{\Delta m_K t_2}\pqty{r\sin{\pqty{\Phi_8 -\delta_D}}-r_D \sin{\pqty{\delta_4-\Phi_6 }}}\Big\}\sinh{\frac{\Delta \Gamma_{B_s}t_1}{2}}\nonumber\\
&+\frac{e^{-\pqty{\Gamma_{B_s} t_1 +\Gamma_K t_2}}}{2D(t_1,t_2)}rr_D\abs{\frac{q_K}{p_K}}\sinh{\frac{\Delta \Gamma_{B_s}t_1}{2}}\Big\{r\pqty{\sinh{\frac{\Delta \Gamma_K t_2}{2}}\cos{\pqty{\Phi_8 +\delta_D}}-\sin{\Delta m_K t_2}\sin{\pqty{\Phi_8 +\delta_D}} }\nonumber\\
&-r_D\pqty{\sinh{\frac{\Delta \Gamma_K t_2}{2}}\cos{\pqty{\delta_4+\Phi_6}}+\sin{\Delta m_K t_2}\sin{\pqty{\delta_4+\Phi_6}}}\Big\},\label{BsD1}
\end{align}
\begin{align}
A_{n,dm}(t_1,t_2)=&\frac{N_2}{D(t_1,t_2)}\nonumber\\
=&\ \frac{e^{-\pqty{\Gamma_{B_s} t_1 +\Gamma_K t_2}}}{2D(t_1,t_2)}\abs{\frac{p_K}{q_K}}\Big\{\sinh{\frac{\Delta\Gamma_K t_2}{2}}\left[r\sin{\pqty{\delta_4+\Phi_6}}-r_D \sin{\pqty{\Phi_7 -\delta_D}}\right]\sin{\Delta m_K t_2}\Big[r\cos{\pqty{\delta_4+\Phi_6}}\nonumber\\
&-r_D \cos{\pqty{\Phi_7-\delta_D}}\Big]\Big\}\sin{\Delta m_{B_s}t_1}-\frac{e^{-\pqty{\Gamma_{B_s} t_1 +\Gamma_K t_2}}}{2D(t_1,t_2)}\abs{\frac{q_K}{p_K}}\Big\{\sinh{\frac{\Delta\Gamma_K t_2}{2}}\left[r\sin{\pqty{\delta_4-\Phi_6}}+r_D \sin{\pqty{\Phi_7+\delta_D}}\right]\nonumber\\
&+\sin{\Delta m_K t_2}\left[r\cos{\pqty{\delta_4-\Phi_6}}-r_D \cos{\pqty{\Phi_7 +\delta_D}}\right]\Big\}\sin{\Delta m_{B_s}t_1}-\frac{e^{-\pqty{\Gamma_{B_s} t_1 +\Gamma_K t_2}}}{2D(t_1,t_2)}\abs{\frac{p_K}{q_K}}rr_D\Big\{\sinh{\frac{\Delta\Gamma_K t_2}{2}}\nonumber\\
&\times\left[-r \sin{\pqty{\Phi_8 -\delta_D}}+r_D\sin{\pqty{\delta_4-\Phi_6}} \right]+\sin{\Delta m_K t_2}\left[r\cos{\pqty{\Phi_8 -\delta_D}}-r_D \cos{\pqty{\delta_4-\Phi_6}}\right]\Big\}\sin{\Delta m_{B_s}t_1}\nonumber\\
&+\frac{e^{-\pqty{\Gamma_{B_s} t_1 +\Gamma_K t_2}}}{2D(t_1,t_2)}rr_D\abs{\frac{q_K}{p_K}}\sin{\Delta m_{B_s}t_1}\Big\{r\pqty{\sinh{\frac{\Delta \Gamma_K t_2}{2}}\sin{\pqty{\Phi_8 +\delta_D}}+\sin{\Delta m_K t_2}\cos{\pqty{\Phi_8 +\delta_D}}}\nonumber\\
&+r_D\pqty{\sinh{\frac{\Delta \Gamma_K t_2}{2}}\sin{\pqty{\delta_4+\Phi_6}}-\sin{\Delta m_K t_2}\cos{\pqty{\delta_4+\Phi_6}}}\Big\},\label{BsD2}
\end{align}
\begin{align}
A_{non-dm}(t_1,t_2)=&\frac{N_3}{D(t_1,t_2)}\nonumber\\
=& \Big\{\abs{g_{+,{B_s}}(t_1)}^2\abs{g_{-,K}(t_2)}^2\Big\{\abs{\frac{p_K}{q_K}}^2\left[r^2+r^2_D-2r_Dr\cos{\pqty{\delta_4+\delta_D+\omega_4}} \right]-\abs{\frac{q_K}{p_K}}^2\Big[r^2+r^2_D-2r_Dr\nonumber\\
&\times\cos{\pqty{\delta_4+\delta_D-\omega_4}} \Big]\Big\}-2rr_D\abs{g_{+,K}(t_2)}^2\left[\cos{\pqty{\delta_4-\omega_4-\delta_D}}- \cos{\pqty{\delta_4+\omega_4-\delta_D}}\right]\abs{g_{-,{B_s}}(t_1)}^2\Big\}\frac{1}{D(t_1,t_2)},\label{BsD3}
\end{align}
\end{widetext}
where $\Phi_6 \equiv \phi_{B_s}+\phi_K-\omega_3+\omega_4$, $\Phi_7 \equiv \phi_{B_s}+\phi_K-\omega_3$ and $\Phi_8 \equiv -\phi_{B_s}-\phi_K+\omega_3-2\omega_4$. The denominator $D(t_1,t_2)$ can be divided into three parts $N_1^\prime + N_2^\prime + N_3^\prime$, which can be obtained by make some replacements to $N_1$, $N_2$, and $N_3$ in \eqref{BsD1} - \eqref{BsD3}. $N_1^\prime$ and $N_2^\prime$ can be obtained by changing the sign in front of the exponential in the fourth line and the third-to-last line of ~\eqref{BsD1} and ~\eqref{BsD2} to an opposite sign. As for $N_3^\prime$, in addition to changing the minus sign in front of $\abs{q_K/p_K}^2$ in the third line and the minus sign between the two cosines in the fourth line of ~\eqref{BsD3} to a positive sign in $N_3$, we still need to incorporate an additional term $2\abs{g_{-,{B_s}}(t_1)}^2\abs{g_{+, K}(t_2)}^2\pqty{r_D^2r^2+1}$ to the modified $N_3$.  

\begin{figure*}[htbp]
    \centering
    \includegraphics[keepaspectratio,width=5.3cm]{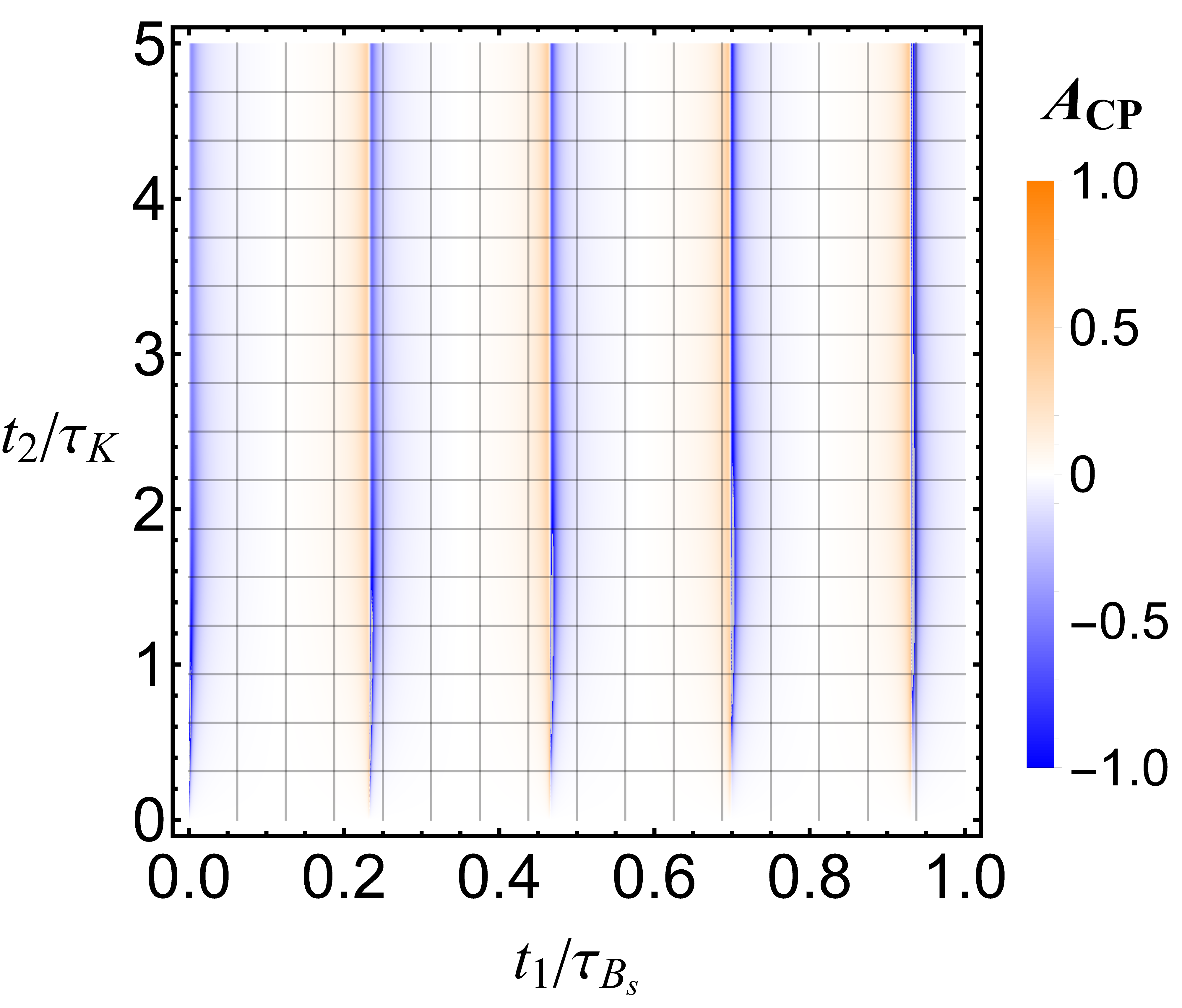}
    \hspace{0.1cm}
    \includegraphics[keepaspectratio,width=4.2cm]{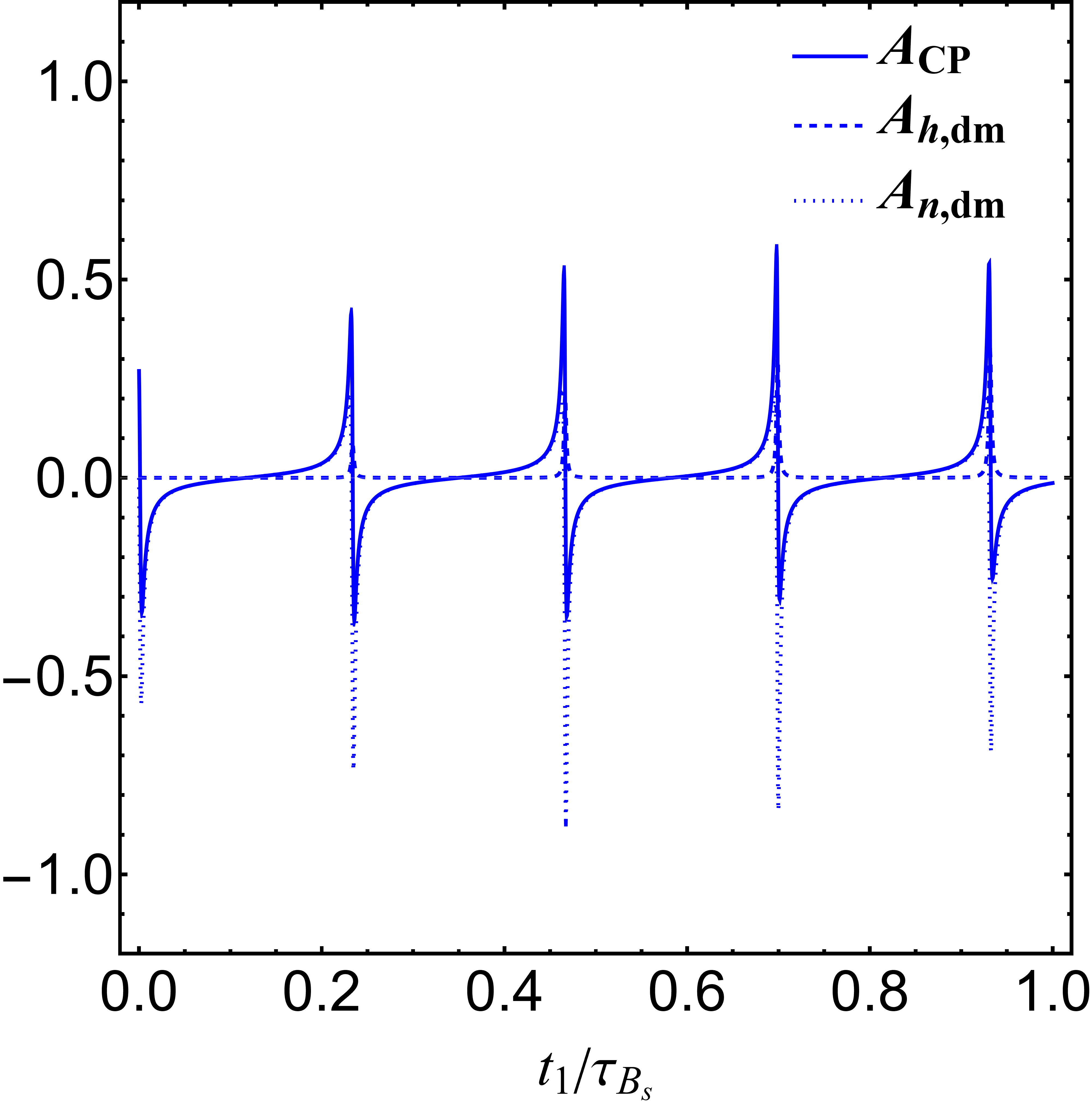}
     \hspace{0.1cm}
    \includegraphics[keepaspectratio,width=4.2cm]{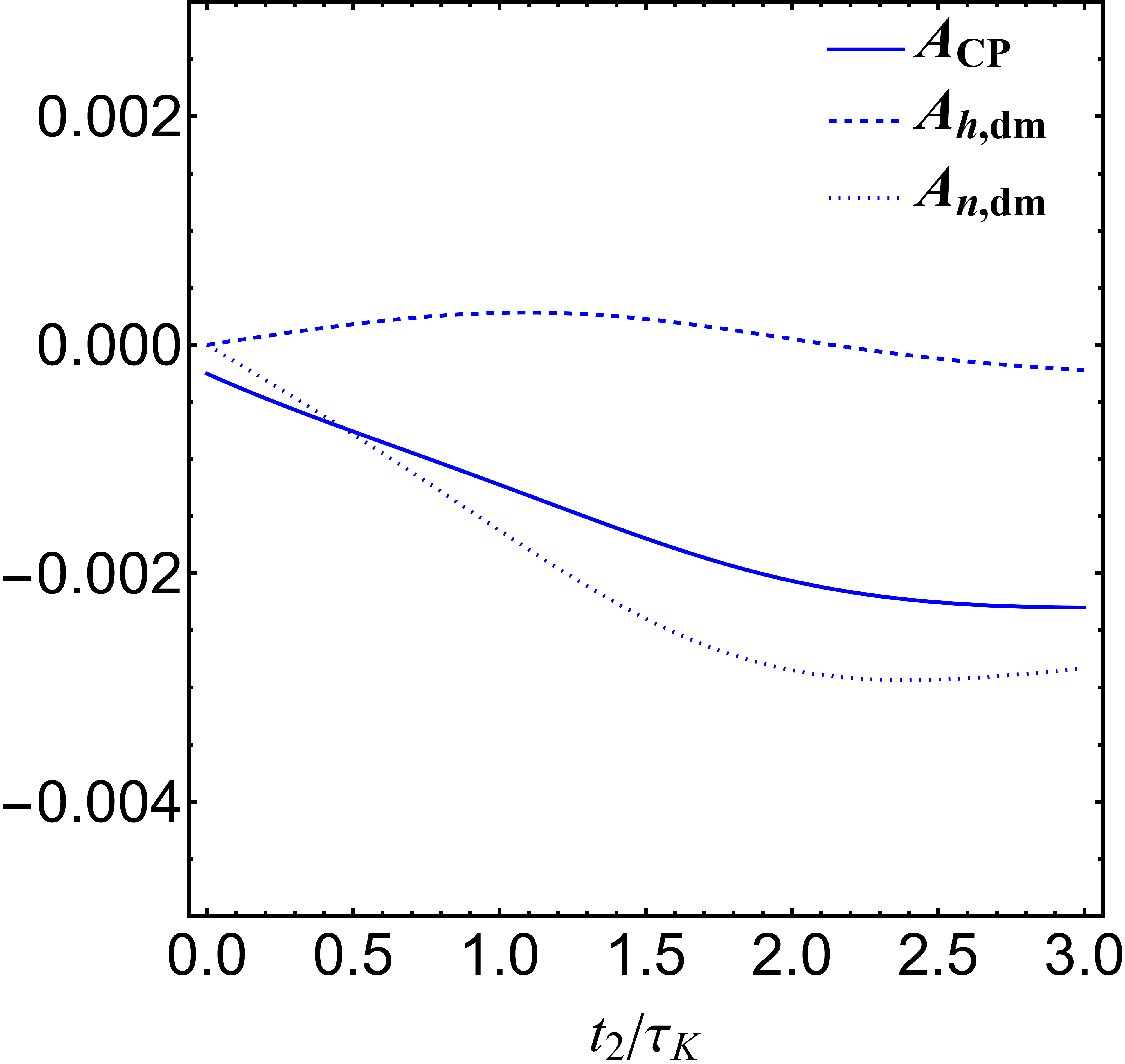}
    \caption{Time dependence of the CP asymmetry $A_{\rm CP}$ in $B^0_s(t_1) \to D K(t_2) \to (K^-\pi^+) (\pi^-\ell^+\nu_\ell)$. The left panel displays the two-dimensional time dependence. The middle panel and the right panel display the dependence on $t_1$ (with $t_2$ integrated from 0 to $2\tau_{K}$) and $t_2$ (with $t_1$ integrated from $0$ to $\tau_{B_s}$), respectively.}\label{fig:s4t}
\end{figure*}

The numerical result of the CP asymmetry in this process is shown in Fig.~\ref{fig:s4t}. The two-dimensional time-dependent CP violation is shown in the left panel. While its peaks surpass 50\%, it is only present in the proximity of specific moments. The result of integrating $t_2$ from 0 to $2\tau_K$ is shown in the middle panel. The contribution of $A_{n,dm}$ dominates except at the positions of positive peaks, as the values of $A_{h,dm}$ and $A_{non-dm}$ are both positive and form upward pulses at each peak position, while $A_{n,dm}$ rapidly rises at each peak position and then directly becomes a negative pulse. Additionally, as the sum of $A_{h,dm}$ and $A_{non-dm}$ is positive when $A_{n,dm}$ is negative, and their contribution is smaller than $A_{n,dm}$, the overall CP violation $A_{CP}$ has larger positive peaks and smaller negative peaks compared to $A_{n,dm}$. The result of integrating $t_1$ from 0 to $\tau_{B_s}$ is shown in the right panel. $A_{h,dm}$ is one order of magnitude smaller than $A_{n,dm}$. The interference between $B^0_s \to D^0 \bar{K}^0 \to D^0 K^0$ and $B^0_s \to \bar{B}^0_s \to D^0 K^0$ is suppressed by the factor $r$. Considering the contribution from $\bar{D}^0$, the interference between $B^0_s \to \bar{D}^0 \bar{K}^0 \to \bar{D}^0 K^0$ and $B^0_s \to \bar{B}^0_s \to D^0 K^0$ is suppressed by $r_D$, while the remaining interference contributions are all suppressed by factors of $rr_D$ or even more suppressed.

\subsubsection{\texorpdfstring{$B^0_s\to D K^0_S\to (\pi^+\pi^-) K^0_S$ and $B^0_s\to D K^0_S\to(K^-\pi^+)$
$ K^0_S$}{}}

We now analyze the decay modes $B^0_s \to D K^0_S\to f_D K^0_S$, with $f_D = \pi^+\pi^-, K^-\pi^+$. Measurements of similar channels have been performed by experiments. The branching ratio of $B^0_s \to \bar{D}^0 \bar{K}^{\star 0}$ has been reported by LHCb~\cite{LHCb:2011yev, LHCb:2014ioa}, but the final state is self-tagged by the flavor of the $K^{\star 0}$, which excludes time-dependent analysis. The CP violation for $A_{\rm CP}(B^0_s \to f_D K^{\star 0})$ has been measured~\cite{LHCb:2014pae}. Considering the possible decays $B^0_s \to D^0 K^0_S$, $B^0_s \to \bar{D}^0 K^0_S$, $D^0 \to f_D$, and $\bar{D}^0 \to f_D$, there are eight pathways for these two process, fitting them into what is defined as the third type.

\begin{figure*}[htbp]
    \centering
    \includegraphics[keepaspectratio,width=5.3cm]{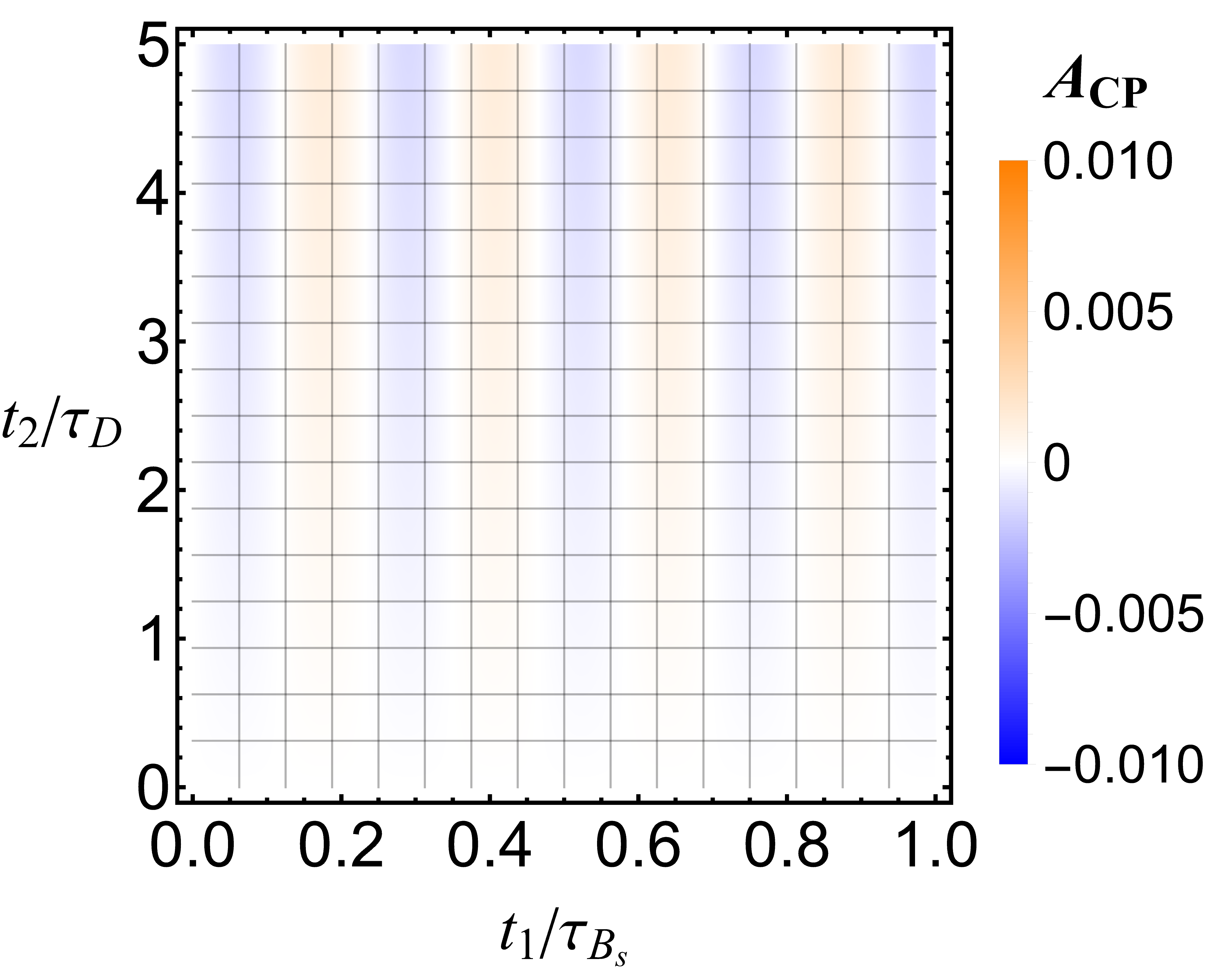}
    \hspace{0.1cm}
    \includegraphics[keepaspectratio,width=4.2cm]{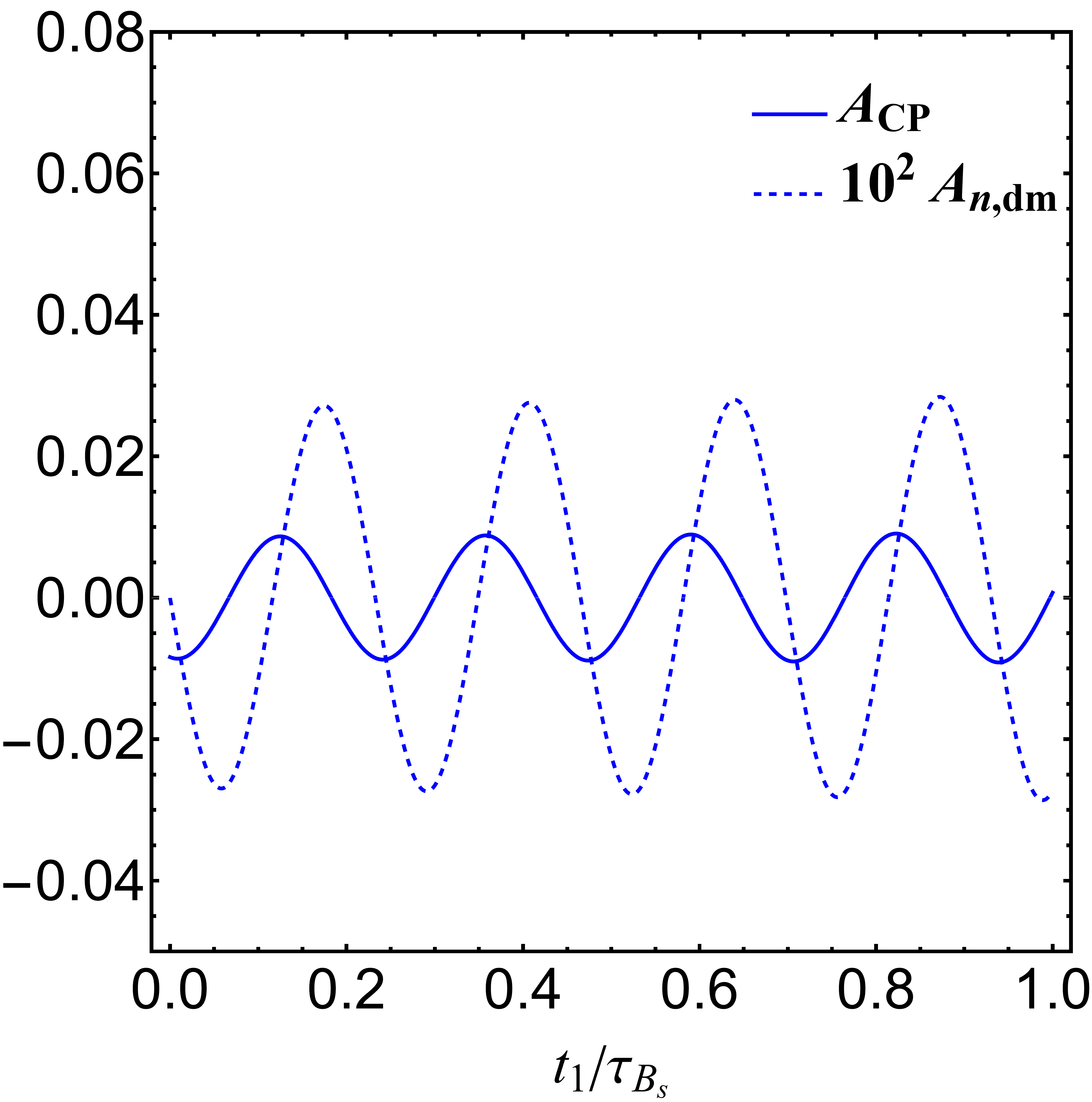}
     \hspace{0.1cm}
    \includegraphics[keepaspectratio,width=4.2cm]{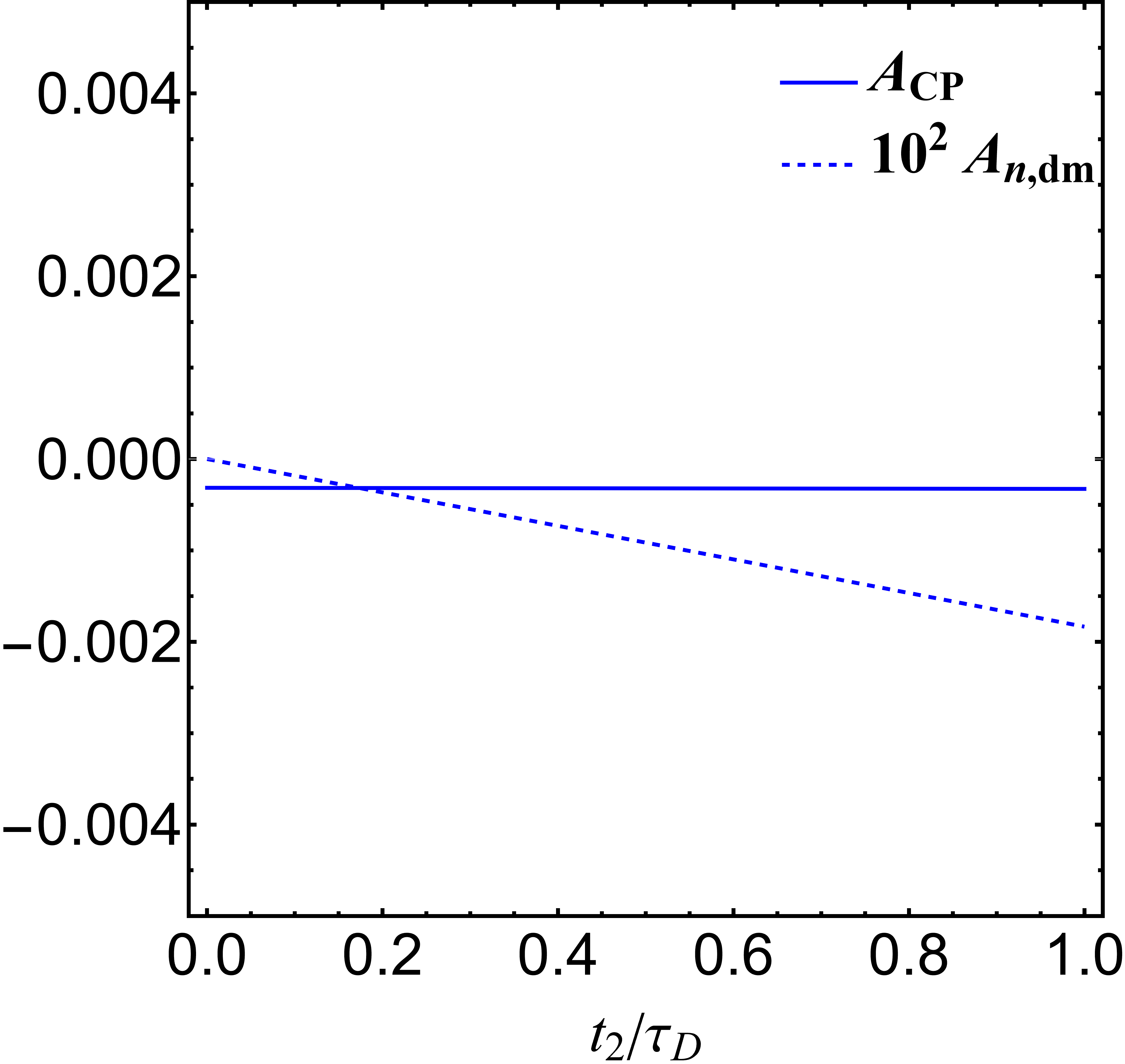}
    \caption{Time dependence of the CP asymmetry $A_{\rm CP}$ in $B^0_s(t_1) \to D(t_2) K^0_S \to (\pi^+\pi^-) K^0_S$. The left panel displays the two-dimensional time dependence. The middle panel and the right panel display the dependence on $t_1$ (with $t_2$ integrated from 0 to $5\tau_{D}$) and $t_2$ (with $t_1$ integrated from $0$ to $\tau_{B_s}$), respectively.}\label{fig:s5}
\end{figure*}

\begin{figure*}[htbp]
    \centering
    \includegraphics[keepaspectratio,width=5.3cm]{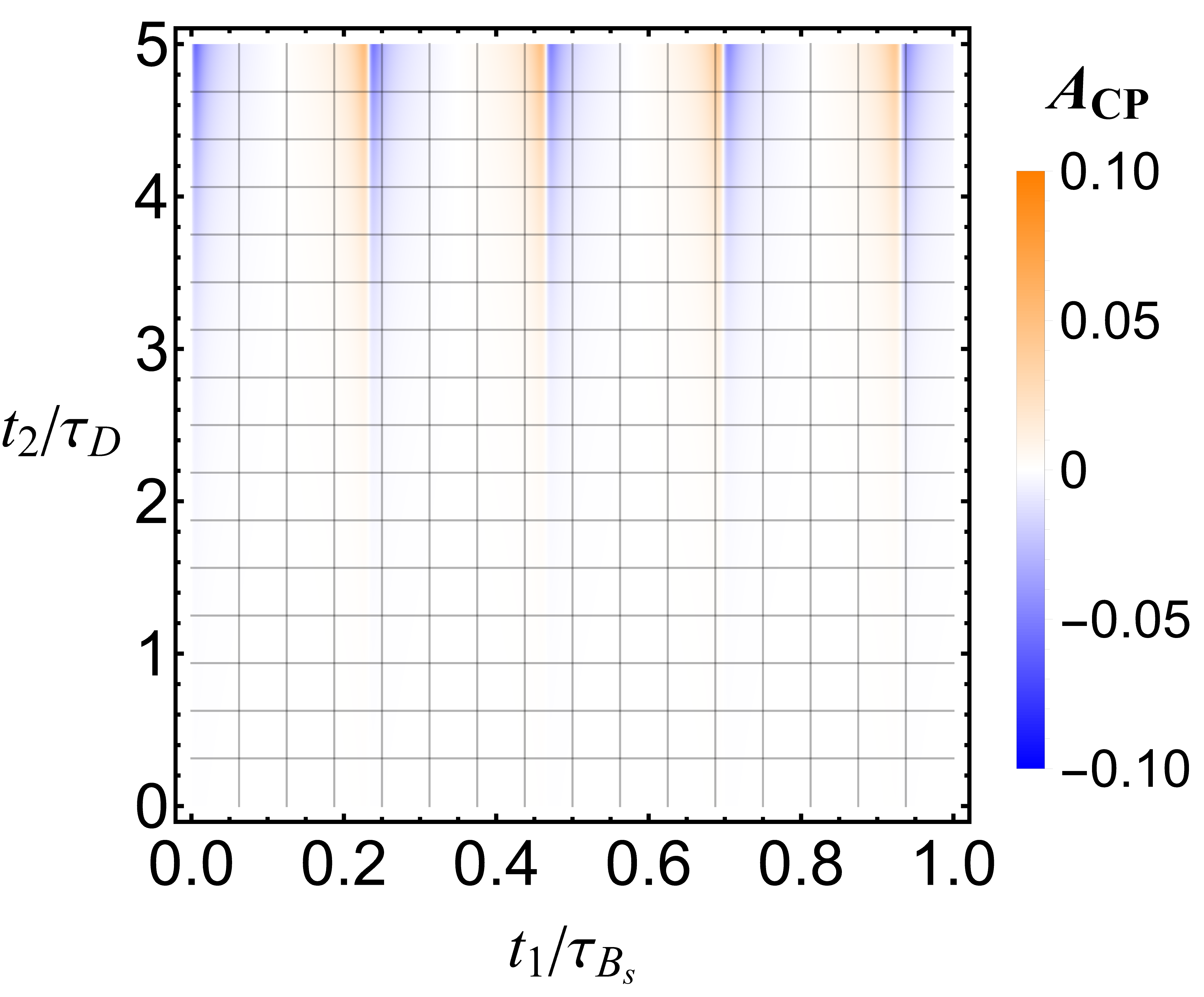}
    \hspace{0.1cm}
    \includegraphics[keepaspectratio,width=4.2cm]{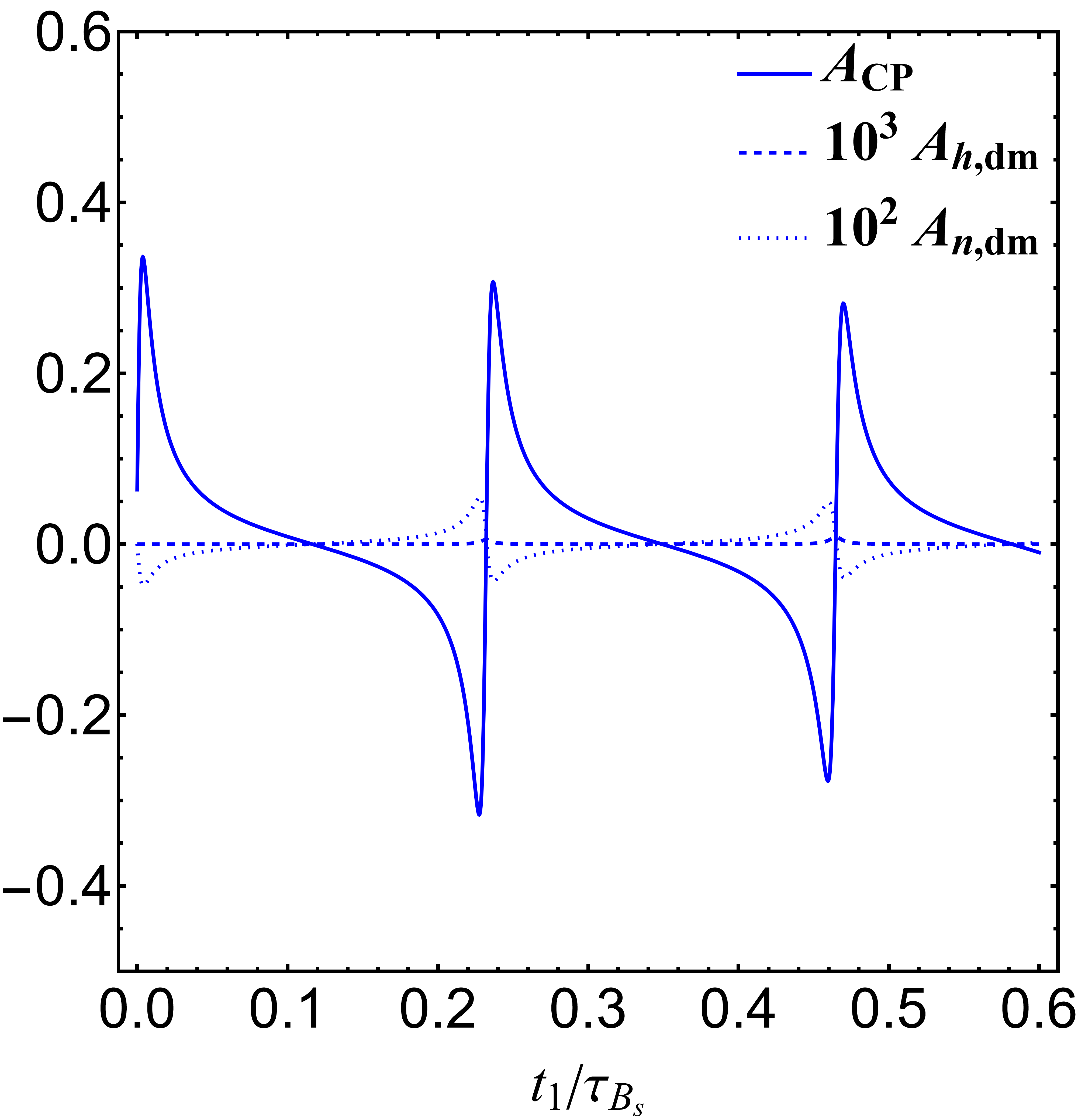}
     \hspace{0.1cm}
    \includegraphics[keepaspectratio,width=4.2cm]{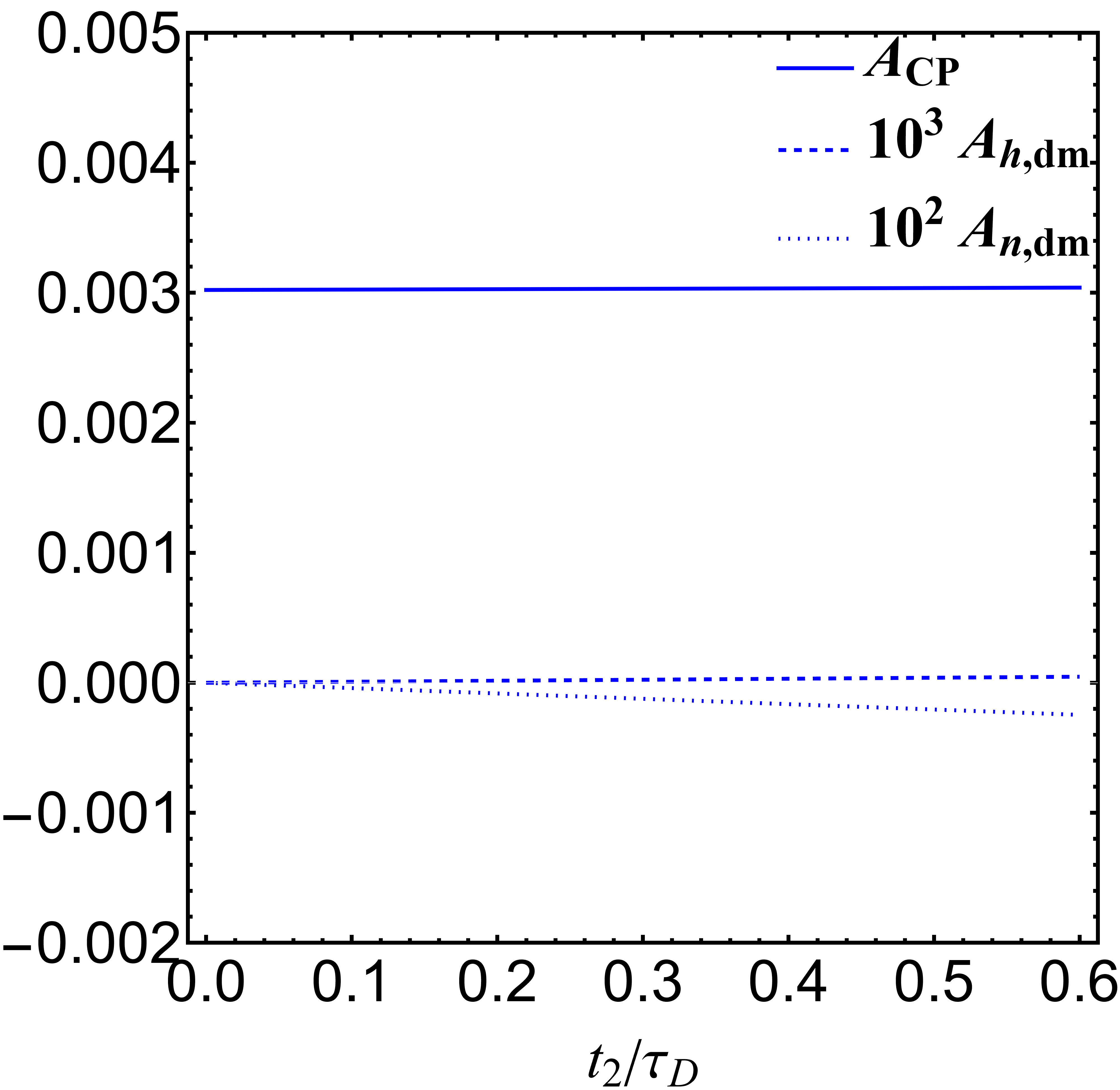}
    \caption{Time dependence of the CP asymmetry $A_{\rm CP}$ in $B^0_s(t_1) \to D(t_2) K^0_S \to (K^-\pi^+) K^0_S$. The left panel displays the two-dimensional time dependence. The middle panel and the right panel display the dependence on $t_1$ (with $t_2$ integrated from 0 to $5\tau_{D}$) and $t_2$ (with $t_1$ integrated from $0$ to $\tau_{B_s}$), respectively.}\label{fig:s6}
\end{figure*}

For the channel with $f_D = \pi^+\pi^-$, we derive the results for all the components of CP violation by setting $r_1 = r$, $\delta_1 = \delta_4$, $\theta_1 = \omega_4$, $\theta_2 = 0$, $r_2 = 1$, $\delta_2 = 0$ and $\theta_3 = \omega_3$ in~\eqref{eq15}, and then we obtain all the terms, which are shown in ~\eqref{A1-1} -~\eqref{A1-12}. For $r$ and $\delta_4$ defined in~\eqref{eq:rdelta}, we numerically use $r=0.02$ and $\delta_4 = 194^\circ$ as a benchmark. The values for $\omega_3$ and $\omega_4$ are given in Table~\ref{input}. In this case, the dominant term in the double-mixing CP violation arises from the interference between the $B^0_s \to \bar{D}^0 K^0_S \to (\pi^+\pi^-) K^0_S$ and $B^0_s \to \bar{B}^0_s \to D^0 K^0_S \to \bar{D}^0 K^0_S \to (\pi^+\pi^-)K^0_S$ amplitudes as well as that between the $B^0_s \to \bar{D}^0 K^0_S \to D^0 K^0_S \to (\pi^+\pi^-)K^0_S$ and $B^0_s \to \bar{B}^0_s \to D^0 K^0_S \to (\pi^+\pi^-)K^0_S$ amplitudes. Our numerical results for this channel are presented in Fig.~\ref{fig:s5}, with the two-dimensional time dependence in the left panel. Its peak value is $\mathcal{O}(10^{-3})$. Integrating $t_2$ from 0 to $5\tau_D$ yields the $t_1$-dependent result, shown in the middle panel, revealing the double-mixing CP violation effect is of the order $\mathcal{O}(10^{-4})$. This is due to the small values $x_D$ and $y_D$ of the $D$ meson as listed in Table~\ref{input}. The $t_2$-integrated CP asymmetry is dominated by terms arising from the interference between the $B^0_s \to \bar{D}^0 K^0_S \to (\pi^+\pi^-)K^0_S$ and $B^0_s \to \bar{B}^0_s \to D^0 K^0_S \to (\pi^+\pi^-)K^0_S$ amplitudes, which have a cosine or hyperbolic cosine dependence on $t_2$ and have a sine dependence on $t_1$, namely $\sin{\Delta m_{B_s}t_1}$. Integrating $t_1$ from 0 to $\tau_{B_s}$, the magnitude of the double-mixing term $A_{n, dm}$ climbs slowly as $t_2$ increases attributed to small mixing parameters of $D$ meson, and the total asymmetry $A_{\rm CP}$ does not show visible changes before $t_2$ becomes too large, as depicted in the right panel.

For the channel with $f_D = K^-\pi^+$, the double-mixing and non-double-mixing contributions to the CP asymmetry can be obtained by taking~\eqref{eq315-1} and~\eqref{eq:nondmb5} with the substitutions $\phi_{B_d}\to \phi_{B_s}$, $r_B \to r$ and $\delta_B \to \delta_4$, and the replacement of the denominator $D(t_1,t_2)$ accordingly. For $r$ and $\delta_4$, we still numerically use $r=0.02$ and $\delta_4 = 194^\circ$. The presence of the suppression factor $r_D$ and the strong phase $\delta_D$ results in a pattern distinct from that of the $\pi^+\pi^-$ channel. The numerical analysis for the double-mixing CP asymmetry $A_{dm}$ is depicted in Fig.~\ref{fig:s6}. The left panel shows the two-dimensional time dependence, indicating the insignificance of the CP violation with $t_2 < 3\tau_D$, below the threshold of current experimental detection capacities. This double-mixing effect is also negligible in the integrated results over $t_2$ or $t_1$, presented on the middle and the right panel, respectively.

\section{Conclusions}\label{Conclusion}

It was proposed recently by~\cite{Shen:2023nuw} that a long-overlooked CP violation effect exists in a decay chain with two neutral mesons involved, namely the double-mixing CP violation. This violation is induced by the interference among different oscillation paths of two neutral flavored mesons. In this work, we have analyzed the double-mixing CP violation in a series of cascade decay processes $M^0_1(t_1) \to M_2(t_2) \to f$, where the primary neutral meson $M^0_1$ is either $B^0_{d,s}$ mesons and the secondary neutral meson is $D^0$ or $K^0$ mesons. The involvement of two time parameters, $t_1$ and $t_2$, makes the two-dimensional time analysis of double-mixing CP violation allowable. We performed numerical analyses for each decay channel, providing the two-dimensional time-dependent double-mixing CP violation and the CP violation integrated over $t_1$ or $t_2$.

Our results show that the double-mixing CP violation can be very significant, with instances exceeding 50\% -- notably in the decay channel $B^0_s \to \rho^0 K \to \rho^0 (\pi\ell\nu)$. In similar cases with kaons reconstructed by semi-leptonic states, the total CP violation almost entirely comes from the double-mixing CP violation. Moreover, our findings indicate that the double-mixing CP violation's dependency varies distinctly with $t_1$ and $t_2$: when integrating over $t_2$, the direct decays of $K^0_S$ and $K^0_L$ predominantly contribute to the CP violation, whereas integrating over $t_1$ highlights the dominance of $K^0_S-K^0_L$ interference. We emphasize that one of the first experimental attempts to measure the double-mixing CP violation would be to combine the two decay channels $B^0_s \to \rho^0 K \to \rho^0 (\pi^-\ell^+\nu_\ell)$ and $B^0_s \to \rho^0 K \to \rho^0 (\pi^+\ell^-\bar \nu_\ell)$. This method does not require initial tagging of neutral $B$ mesons, thus avoiding efficiency losses. 

In comparison, the significance of double-mixing CP violation is less in decay channels involving $D^0$ as $M^0_2$, mainly due to the tiny mixing observed in neutral $D$ mesons, such as in the decay channel $B^0_s \to D K^0_S \to (\pi^+\pi^-) K^0_S$. Nonetheless, double-mixing CP violation offers a major advantage by facilitating the extraction of CKM phase angles without theoretical inputs. For example, the decay channel $B^0_d \to D K \to (K^-\pi^+)(\pi\ell\nu)$ involves both strong and weak phases, with the overall weak phase being approximated by $2\beta+\gamma$. This allows for the simultaneous extraction of strong and weak phases from experimental data on double-mixing CP violation, thereby directly testing the CKM mechanism.

It is noteworthy that the predominant neutral meson oscillation involved in the double-mixing CP violation cannot be directly ascertained, as the extent of such CP violation hinges on the specific decay channel. The presence of the double-mixing CP violation necessitates the concurrent mixing of two distinct neutral mesons, with various factors potentially influencing its magnitude, including the mixing phase angles of the mesons and the selection of spectator mesons.

For the decay channel $ B^0_d(t_1) \to M_2(t_2) $, the double-mixing CP violation is more significant when $ M_2 $ is $K^0 $ compared to when $ M_2 $ is $ D^0 $. A primary reason for this is that the double-mixing CP violation is linked to the mixing parameter $ x $ of the neutral mesons through relationships such as $ \sin{x_K \Gamma_K t_2} $ and $ \sin{x_D \Gamma_D t_2} $, where the mixing parameter of $ K^0 $ ($ x_K \approx 0.946 $) is significantly larger than that of $ D^0$ ($ x_D \sim 10^{-3} $). Moreover, the decay channel $ B^0_d(t_1) \to D(t_2) $ involves more sources of interference compared to $ B^0_d(t_1) \to K(t_2) $, which could lead to an increase in the denominator of the double-mixing CP violation, resulting in a smaller effect. While in the aforementioned scenario, the double-mixing CP violation in $B^0_d(t_1) \to K(t_2)$ is more significant than in $B^0_d(t_1) \to D(t_2)$, it does not imply that the oscillation of a specific neutral meson consistently dominates the contribution to the double-mixing CP violation in all cases. For instance, the comparison of contributions of $B^0_d$ and $B^0_s$ to the double-mixing CP violation in the decay channels $B^0_d(t_1) \to K(t_2)$ and $B^0_s(t_1) \to K(t_2)$ is not straightforward. The double-mixing CP violation in $B^0_d(t_1) \to DK(t_2) \to (K^-\pi^+)(\pi^+\ell^-\bar{\nu}_\ell)$ is smaller than in $B^0_s(t_1) \to \rho^0 K(t_2) \to \rho^0(\pi^+\ell^-\bar{\nu}_\ell)$, while a different pattern emerges in the comparison between $B^0_d(t_1) \to J/\psi K(t_2) \to J/\psi(\pi^+\pi^-)$ and $B^0_s(t_1) \to J/\psi K(t_2) \to J/\psi$ $(\pi^+\pi^-)$, with the former exhibiting evidently larger double-mixing CP violation than the latter.

\begin{acknowledgements}
The authors are grateful to Cai-Dian L\"u, Wen-Bin Qian, Liang Sun, and Yue-Hong Xie for useful discussions. This work is supported by the Natural Science Foundation of China under grant No.~12375086.
\end{acknowledgements}

\appendix
\begin{widetext}
\section{Formulae for Category 3}
Here we present relevant analytical expressions for the terms of the double-mixing CP violation of {\bf Category 3} in Sect.~\ref{sec:formulae}.
Based on the definitions of the parameters provided in~\eqref{eq15}, we can derive the individual terms in~\eqref{eq4} as 
\begin{align}
C_+(t_2) = &\ e^{-\Gamma_2 t_2}\Big\{r_2\sinh{\frac{\Delta \Gamma_2 t_2}{2}}\left[r_1^2\pqty{\abs{\frac{q_2}{p_2}}\cos{\Omega_3}-\abs{\frac{p_2}{q_2}}\cos{\Omega_4}}+\abs{\frac{p_2}{q_2}}\cos{\Omega_3}-\abs{\frac{q_2}{p_2}}\cos{\Omega_4} \right]-r_2\sin{\Delta m_2 t_2}\Big[r_1^2\Big(\abs{\frac{q_2}{p_2}}\sin{\Omega_3}\nonumber\\
&-\abs{\frac{p_2}{q_2}}\sin{\Omega_4}\Big)-\abs{\frac{p_2}{q_2}}\sin{\Omega_3}+\abs{\frac{q_2}{p_2}}\sin{\Omega_4} \Big]+r_1\sinh{\frac{\Delta \Gamma_2 t_2}{2}}\Big[r_2^2\pqty{\abs{\frac{q_2}{p_2}}\cos{\Omega_1}-\abs{\frac{p_2}{q_2}}\cos{\Omega_2}}+\abs{\frac{p_2}{q_2}}\cos{\Omega_1}\nonumber\\
&-\abs{\frac{q_2}{p_2}}\cos{\Omega_2} \Big]-r_1\sin{\Delta m_2 t_2}\left[r_2^2\pqty{\abs{\frac{q_2}{p_2}}\sin{\Omega_1}-\abs{\frac{p_2}{q_2}}\sin{\Omega_2}}-\abs{\frac{p_2}{q_2}}\sin{\Omega_1}+\abs{\frac{q_2}{p_2}}\sin{\Omega_2} \right]    \Big\} \nonumber\\
&+2r_1 r_2 \abs{g_{+,2}(t_2)}^2\Big[\cos{\pqty{\Omega_1-\Omega_3}}-\cos{\pqty{\Omega_2-\Omega_4}}\Big]+ 2r_1 r_2 \abs{g_{-,2}(t_2)}^2\Big[\cos{\pqty{\Omega_1+\Omega_3}}-\cos{\pqty{\Omega_2+\Omega_4}}\Big]\nonumber\\
& +\pqty{r_1^2r_2^2-1}\abs{g_{-,2}(t_2)}^2\pqty{\abs{\frac{q_2}{p_2}}^2-\abs{\frac{p_2}{q_2}}^2},\label{A1-1}
\end{align}
where $\Omega_1 \equiv \delta_1+\theta_1-\phi_2$, $\Omega_2 \equiv \delta_1-\theta_1+\phi_2 $, $\Omega_3 \equiv \delta_2  -\phi_2$ and $\Omega_4 \equiv \delta_2 +\phi_2$, 
\begin{align}
C_-(t_2) =&\ \abs{g_{-,2}(t_2)}^2\left\{r_1^2\pqty{\abs{\frac{q_1}{p_1}}^2\abs{\frac{p_2}{q_2}}^2-\abs{\frac{p_1}{q_1}}^2\abs{\frac{q_2}{p_2}}^2}+r_2^2\pqty{\abs{\frac{q_1}{p_1}}^2\abs{\frac{q_2}{p_2}}^2-\abs{\frac{p_1}{q_1}}^2\abs{\frac{p_2}{q_2}}^2 }\right\}+2r_1r_2 \abs{\frac{q_1}{p_1}}^2\Big\{ \abs{g_{+,2}(t_2)}^2\nonumber\\
&\times\cos{\pqty{\Omega_2+\Omega_3}}+\abs{g_{-,2}(t_2)}^2\cos{\pqty{\Omega_2-\Omega_3}}\Big\}-2r_1 r_2 \abs{\frac{p_1}{q_1}}^2\left\{\abs{g_{+,2}(t_2)}^2\cos{\pqty{\Omega_1+\Omega_4}}+\abs{g_{-,2}(t_2)}^2\cos{\pqty{\Omega_1-\Omega_4}}\right\}\nonumber\\
&+e^{-\Gamma_2 t_2}\left\{r_2\abs{\frac{q_1}{p_1}}^2\abs{\frac{q_2}{p_2}}\pqty{\sinh{\frac{\Delta \Gamma_2 t_2}{2}}\cos{\Omega_3}-\sin{\Delta m_2 t_2}\sin{\Omega_3}}+r_1^2r_2\abs{\frac{q_1}{p_1}}^2\abs{\frac{p_2}{q_2}}\Big(\sinh{\frac{\Delta \Gamma_2 t_2}{2}}\cos{\Omega_3}+\sin{\Delta m_2 t_2}\right.\nonumber\\
&\left.\sin{\Omega_3}\Big)-r_2\abs{\frac{p_1}{q_1}}^2\abs{\frac{p_2}{q_2}}\pqty{\sinh{\frac{\Delta \Gamma_2 t_2}{2}}\cos{\Omega_4}-\sin{\Delta m_2 t_2}\sin{\Omega_4}}-r_1^2r_2\abs{\frac{p_1}{q_1}}^2\abs{\frac{q_2}{p_2}}\Big(\sinh{\frac{\Delta \Gamma_2 t_2}{2}}\cos{\Omega_4}+\sin{\Delta m_2 t_2}\right.\nonumber\\
&\left.\sin{\Omega_4}\Big)+r_1\sinh{\frac{\Delta \Gamma_2 t_2}{2}}\left[\abs{\frac{q_1}{p_1}}^2\cos{\Omega_2}\pqty{r_2^2\abs{\frac{q_2}{p_2}}+\abs{\frac{p_2}{q_2}}}-\abs{\frac{p_1}{q_1}}^2\cos{\Omega_1}\pqty{r_2^2\abs{\frac{p_2}{q_2}}+\abs{\frac{q_2}{p_2}}}\right] + r_1\sin{\Delta m_2 t_2} \right.\nonumber\\
&\left.\times\Big[\abs{\frac{q_1}{p_1}}^2\sin{\Omega_2}\pqty{r_2^2\abs{\frac{q_2}{p_2}}-\abs{\frac{p_2}{q_2}}}-\abs{\frac{p_1}{q_1}}^2\sin{\Omega_1}\pqty{r_2^2\abs{\frac{p_2}{q_2}}-\abs{\frac{q_2}{p_2}}}\Big] \right\}+\pqty{r_1^2r_2^2+1}\pqty{\abs{\frac{q_1}{p_1}}^2-\abs{\frac{p_1}{q_1}}^2}\abs{g_{+,2}(t_2)}^2,\label{A1-2}
\end{align}
\begin{align}
S_{n,1}(t_2)=&\ \frac{e^{-\Gamma_2 t_2}}{2}\Big\{g(-,-,-,-)\times\sinh{\frac{\Delta \Gamma_2 t_2}{2}}\Big[r_1^2\sin{\pqty{\Omega_5-2\Omega_6}}+\sin{\Omega_5}\Big]+g(-,+,-,+)\times\sin{\Delta m_2 t_2}\Big[r_1^2\cos{\pqty{\Omega_5-2\Omega_6}}\nonumber\\
&-\cos{\Omega_5}\Big]\Big\}+r_1r_2\abs{\frac{q_2}{p_2}} \sin{\pqty{\delta_1-\Omega_5+\Omega_6}}e^{-\Gamma_2 t_2}\left\{ \abs{\frac{q_1}{p_1}}\left[\sinh{\frac{\Delta \Gamma_2 t_2}{2}}\cos{\Omega_3} - \sin{\Delta m_2 t_2}\sin{\Omega_3} \right] \right.\nonumber\\
&\left.+\abs{\frac{p_1}{q_1}}\left[\sinh{\frac{\Delta \Gamma_2 t_2}{2}}\cos{\Omega_4} + \sin{\Delta m_2 t_2}\sin{\Omega_4} \right] \right\}-r_1r_2\abs{\frac{p_2}{q_2}} \sin{\pqty{\delta_1+\Omega_5-\Omega_6}}e^{-\Gamma_2 t_2}\nonumber\\
&\times\left\{ \abs{\frac{q_1}{p_1}}\left[\sinh{\frac{\Delta \Gamma_2 t_2}{2}}\cos{\Omega_3} + \sin{\Delta m_2 t_2}\sin{\Omega_3} \right]+\abs{\frac{p_1}{q_1}}\left[\sinh{\frac{\Delta \Gamma_2 t_2}{2}}\cos{\Omega_4} - \sin{\Delta m_2 t_2}\sin{\Omega_4} \right] \right\},\label{A1-3}
\end{align}
where $\Omega_5 \equiv \theta_3-\phi_1-\phi_2$, $\Omega_6 \equiv \theta_1-\phi_2$, and the $g(-,+,-,+)$ is defined as
\begin{align}
g(-,+,-,+) = -\abs{\frac{q_1}{p_1}}\abs{\frac{p_2}{q_2}}+r_2^2 \abs{\frac{q_1}{p_1}}\abs{\frac{q_2}{p_2}}-r_2^2\abs{\frac{p_1}{q_1}}\abs{\frac{p_2}{q_2}}+\abs{\frac{p_1}{q_1}}\abs{\frac{q_2}{p_2}}, \nonumber
\end{align}
and others $g$ functions with different sign arguments can be obtained by flip signs in the above expression accordingly, 
\begin{align}
S_{n,2}(t_2)=& -r_1^2r_2\abs{\frac{q_1}{p_1}}\left\{  \abs{g_{+,2}(t_2)}^2\sin{\pqty{\Omega_3+\Omega_5-2\Omega_6}}-\abs{g_{-,2}(t_2)}^2\sin{\pqty{\Omega_3-\Omega_5+2\Omega_6}}\right\} +r_2\abs{\frac{q_1}{p_1}}\Big\{\abs{g_{+,2}(t_2)}^2\times\nonumber\\
&\sin{\pqty{\Omega_3-\Omega_5}}-\abs{g_{-,2}(t_2)}^2\sin{\pqty{\Omega_3+\Omega_5}} \Big\}+r_1^2 r_2\abs{\frac{p_1}{q_1}}\Big\{\abs{g_{+,2}(t_2)}^2\sin{\pqty{\Omega_4-\Omega_5+2\Omega_6}}-\abs{g_{-,2}(t_2)}^2\nonumber\\
&\times\sin{\pqty{\Omega_4+\Omega_5-2\Omega_6}} \Big\}-r_2\abs{\frac{p_1}{q_1}}\left\{\abs{g_{+,2}(t_2)}^2\sin{\pqty{\Omega_4+\Omega_5}}-\abs{g_{-,2}(t_2)}^2\sin{\pqty{\Omega_4-\Omega_5}}\right\}\nonumber\\
&+r_1\sin{\pqty{\delta_1-\Omega_5+\Omega_6}}f_1(q,p)-r_1\sin{\pqty{\delta_1+\Omega_5-\Omega_6}}f_1(p,q),\label{A1-4}
\end{align}
where 
\begin{align}
f_1(q,p) = \abs{g_{+,2}(t_2)}^2\pqty{\abs{\frac{q_1}{p_1}}+r_2^2\abs{\frac{p_1}{q_1}}}+\abs{\frac{q_2}{p_2}}^2\abs{g_{-,2}(t_2)}^2\pqty{r_2^2\abs{\frac{q_1}{p_1}}+\abs{\frac{p_1}{q_1}}}, \nonumber
\end{align}
and the function $f_1(p,q)$ is obtained by exchanging $p_{1,2} \leftrightarrow q_{1,2}$ in $f_1(q,p)$, 
\begin{align}
S_{h,1}(t_2) =&\ \frac{e^{-\Gamma_2 t_2}}{2}\Big\{ g(+,+,-,-)\sinh{\frac{\Delta \Gamma_2 t_2}{2}}\Big[r_1^2\cos{\pqty{\Omega_5-2\Omega_6}}+\cos{\Omega_5}\Big]-g(+,-,-,+)\sin{\Delta m_2 t_2}\Big[r_1^2\sin{\pqty{\Omega_5-2\Omega_6}}\nonumber\\
&-\sin{\Omega_5} \Big]\Big\}+r_1r_2\abs{\frac{q_2}{p_2}} \cos{\pqty{\delta_1-\Omega_5+\Omega_6}}e^{-\Gamma_2 t_2}\left\{ \abs{\frac{q_1}{p_1}}\left[\sinh{\frac{\Delta \Gamma_2 t_2}{2}}\cos{\Omega_3} - \sin{\Delta m_2 t_2}\sin{\Omega_3} \right] \right.\nonumber\\
&\left.-\abs{\frac{p_1}{q_1}}\left[\sinh{\frac{\Delta \Gamma_2 t_2}{2}}\cos{\Omega_4} + \sin{\Delta m_2 t_2}\sin{\Omega_4} \right] \right\}+r_1r_2\abs{\frac{p_2}{q_2}} \cos{\pqty{\delta_1+\Omega_5-\Omega_6}}e^{-\Gamma_2 t_2}\nonumber\\
&\times\Big\{ \abs{\frac{q_1}{p_1}}\left[\sinh{\frac{\Delta \Gamma_2 t_2}{2}}\cos{\Omega_3} + \sin{\Delta m_2 t_2}\sin{\Omega_3} \right]-\abs{\frac{p_1}{q_1}}\left[\sinh{\frac{\Delta \Gamma_2 t_2}{2}}\cos{\Omega_4} - \sin{\Delta m_2 t_2}\sin{\Omega_4} \right] \Big\},\label{A1-5}\end{align}
\begin{align}
S_{h,2}(t_2) =&\ r_1^2 r_2 \abs{\frac{q_1}{p_1}}\Big[\abs{g_{+,2}(t_2)}^2\cos{\pqty{\Omega_3+\Omega_5-2\Omega_6}}+\abs{g_{-,2}(t_2)}^2\cos{\pqty{\Omega_3-\Omega_5+2\Omega_6}}\Big]+r_2\abs{\frac{q_1}{p_1}}\Big[ \abs{g_{+,2}(t_2)}^2\cos{\pqty{\Omega_3-\Omega_5}}\nonumber\\
&+\abs{g_{-,2}(t_2)}^2\cos{\pqty{\Omega_3+\Omega_5}} \Big]-r_1^2r_2\abs{\frac{p_1}{q_1}}\Big[\abs{g_{+,2}(t_2)}^2\cos{\pqty{\Omega_4-\Omega_5+2\Omega_6}}+\abs{g_{-,2}(t_2)}^2\cos{\pqty{\Omega_4+\Omega_5-2\Omega_6}} \Big]\nonumber\\
&-r_2\abs{\frac{p_1}{q_1}}\Big[\abs{g_{+,2}(t_2)}^2\cos{\pqty{\Omega_4+\Omega_5}}+\abs{g_{-,2}(t_2)}^2\cos{\pqty{\Omega_4-\Omega_5}}  \Big]+r_1\cos{\pqty{\delta_1-\Omega_5+\Omega_6}}f_2(q,p)\nonumber\\
&+r_1\cos{\pqty{\delta_1+\Omega_5-\Omega_6}}f_2(p,q),\label{A1-6}
\end{align}
where 
\begin{align}
f_2(q,p) = \abs{g_{+,2}(t_2)}^2\pqty{\abs{\frac{q_1}{p_1}}-r_2^2\abs{\frac{p_1}{q_1}}}+\abs{\frac{q_2}{p_2}}^2\abs{g_{-,2}(t_2)}^2\pqty{r_2^2\abs{\frac{q_1}{p_1}}-\abs{\frac{p_1}{q_1}}}, \nonumber
\end{align}
and the function $f_2(p,q)$ is obtained by exchanging $p_{1,2} \leftrightarrow q_{1,2}$ in $f_2(q,p)$, 
\begin{align}
C_+^\prime(t_2) = &\ e^{-\Gamma_2 t_2}\Big\{r_2\sinh{\frac{\Delta \Gamma_2 t_2}{2}}\left[r_1^2\pqty{\abs{\frac{q_2}{p_2}}\cos{\Omega_3}+\abs{\frac{p_2}{q_2}}\cos{\Omega_4}}+\abs{\frac{p_2}{q_2}}\cos{\Omega_3}+\abs{\frac{q_2}{p_2}}\cos{\Omega_4} \right]-r_2\sin{\Delta m_2 t_2}\Big[r_1^2\Big(\abs{\frac{q_2}{p_2}}\sin{\Omega_3}\nonumber\\
&+\abs{\frac{p_2}{q_2}}\sin{\Omega_4}\Big)-\abs{\frac{p_2}{q_2}}\sin{\Omega_3}-\abs{\frac{q_2}{p_2}}\sin{\Omega_4} \Big]+r_1\sinh{\frac{\Delta \Gamma_2 t_2}{2}}\Big[r_2^2\pqty{\abs{\frac{q_2}{p_2}}\cos{\Omega_1}+\abs{\frac{p_2}{q_2}}\cos{\Omega_2}}+\abs{\frac{p_2}{q_2}}\cos{\Omega_1}\nonumber\\
&+\abs{\frac{q_2}{p_2}}\cos{\Omega_2} \Big]-r_1\sin{\Delta m_2 t_2}\left[r_2^2\pqty{\abs{\frac{q_2}{p_2}}\sin{\Omega_1}+\abs{\frac{p_2}{q_2}}\sin{\Omega_2}}-\abs{\frac{p_2}{q_2}}\sin{\Omega_1}-\abs{\frac{q_2}{p_2}}\sin{\Omega_2} \right]    \Big\} \nonumber\\
&+2r_1 r_2 \abs{g_{+,2}(t_2)}^2\Big[\cos{\pqty{\Omega_1-\Omega_3}}+\cos{\pqty{\Omega_2-\Omega_4}}\Big]+ 2r_1 r_2 \abs{g_{-,2}(t_2)}^2\Big[\cos{\pqty{\Omega_1+\Omega_3}}+\cos{\pqty{\Omega_2+\Omega_4}}\Big]\nonumber\\
& +\pqty{r_1^2r_2^2+1}\abs{g_{-,2}(t_2)}^2\pqty{\abs{\frac{q_2}{p_2}}^2+\abs{\frac{p_2}{q_2}}^2}+2\pqty{r_1^2+r_2^2}\abs{g_{+,2}(t_2)}^2,\label{A1-7}
\end{align}
\begin{align}
C_-^\prime(t_2) =&\ \abs{g_{-,2}(t_2)}^2\left\{r_1^2\pqty{\abs{\frac{q_1}{p_1}}^2\abs{\frac{p_2}{q_2}}^2+\abs{\frac{p_1}{q_1}}^2\abs{\frac{q_2}{p_2}}^2}+r_2^2\pqty{\abs{\frac{q_1}{p_1}}^2\abs{\frac{q_2}{p_2}}^2+\abs{\frac{p_1}{q_1}}^2\abs{\frac{p_2}{q_2}}^2} \right\}+2r_1r_2 \abs{\frac{q_1}{p_1}}^2\Big[ \abs{g_{+,2}(t_2)}^2\nonumber\\
&\times\cos{\pqty{\Omega_2+\Omega_3}}+\abs{g_{-,2}(t_2)}^2\cos{\pqty{\Omega_2-\Omega_3}}\Big]+2r_1 r_2 \abs{\frac{p_1}{q_1}}^2\Big[\abs{g_{+,2}(t_2)}^2\cos{\pqty{\Omega_1+\Omega_4}}+\abs{g_{-,2}(t_2)}^2\cos{\pqty{\Omega_1-\Omega_4}}\Big]\nonumber\\
&+e^{-\Gamma_2 t_2}\left\{r_2\abs{\frac{q_1}{p_1}}^2\abs{\frac{q_2}{p_2}}\pqty{\sinh{\frac{\Delta \Gamma_2 t_2}{2}}\cos{\Omega_3}-\sin{\Delta m_2 t_2}\sin{\Omega_3}}+r_1^2r_2\abs{\frac{q_1}{p_1}}^2\abs{\frac{p_2}{q_2}}\Big(\sinh{\frac{\Delta \Gamma_2 t_2}{2}}\cos{\Omega_3}+\sin{\Delta m_2 t_2}\right.\nonumber\\
&\left.\sin{\Omega_3}\Big)+r_2\abs{\frac{p_1}{q_1}}^2\abs{\frac{p_2}{q_2}}\pqty{\sinh{\frac{\Delta \Gamma_2 t_2}{2}}\cos{\Omega_4}-\sin{\Delta m_2 t_2}\sin{\Omega_4}}+r_1^2r_2\abs{\frac{p_1}{q_1}}^2\abs{\frac{q_2}{p_2}}\Big(\sinh{\frac{\Delta \Gamma_2 t_2}{2}}\cos{\Omega_4}+\sin{\Delta m_2 t_2}\right.\nonumber\\
&+\left.\sin{\Omega_4}\Big)r_1\sinh{\frac{\Delta \Gamma_2 t_2}{2}}\left[\abs{\frac{q_1}{p_1}}^2\cos{\Omega_2}\pqty{r_2^2\abs{\frac{q_2}{p_2}}+\abs{\frac{p_2}{q_2}}}+\abs{\frac{p_1}{q_1}}^2\cos{\Omega_1}\pqty{r_2^2\abs{\frac{p_2}{q_2}}+\abs{\frac{q_2}{p_2}}}\right] \right.\nonumber\\
&\left. + r_1\sin{\Delta m_2 t_2}\left[\abs{\frac{q_1}{p_1}}^2\sin{\Omega_2}\pqty{r_2^2\abs{\frac{q_2}{p_2}}-\abs{\frac{p_2}{q_2}}}+\abs{\frac{p_1}{q_1}}^2\sin{\Omega_1}\pqty{r_2^2\abs{\frac{p_2}{q_2}}-\abs{\frac{q_2}{p_2}}}\right]\right\}\nonumber\\
&+\pqty{r_1^2r_2^2+1}\pqty{\abs{\frac{q_1}{p_1}}^2+\abs{\frac{p_1}{q_1}}^2}\abs{g_{+,2}(t_2)}^2,\label{A1-8}
\end{align}
\begin{align}
S_{n,1}^\prime(t_2)=&\ \frac{e^{-\Gamma_2 t_2}}{2}\Big\{g(-,-,+,+)\times\sinh{\frac{\Delta \Gamma_2 t_2}{2}}\Big[r_1^2\sin{\pqty{\Omega_5-2\Omega_6}}+\sin{\Omega_5}\Big]+g(-,+,+,-)\times\sin{\Delta m_2 t_2}\Big[r_1^2\cos{\pqty{\Omega_5-2\Omega_6}}\nonumber\\
&-\cos{\Omega_5}\Big]\Big\}+r_1r_2\abs{\frac{q_2}{p_2}} \sin{\pqty{\delta_1-\Omega_5+\Omega_6}}e^{-\Gamma_2 t_2}\left\{ \abs{\frac{q_1}{p_1}}\left[\sinh{\frac{\Delta \Gamma_2 t_2}{2}}\cos{\Omega_3} - \sin{\Delta m_2 t_2}\sin{\Omega_3} \right] \right.\nonumber\\
&\left.-\abs{\frac{p_1}{q_1}}\left[\sinh{\frac{\Delta \Gamma_2 t_2}{2}}\cos{\Omega_4} + \sin{\Delta m_2 t_2}\sin{\Omega_4} \right] \right\}-r_1r_2\abs{\frac{p_2}{q_2}} \sin{\pqty{\delta_1+\Omega_5-\Omega_6}}e^{-\Gamma_2 t_2}\nonumber\\
&\times\left\{ \abs{\frac{q_1}{p_1}}\left[\sinh{\frac{\Delta \Gamma_2 t_2}{2}}\cos{\Omega_3} + \sin{\Delta m_2 t_2}\sin{\Omega_3} \right]-\abs{\frac{p_1}{q_1}}\left[\sinh{\frac{\Delta \Gamma_2 t_2}{2}}\cos{\Omega_4} - \sin{\Delta m_2 t_2}\sin{\Omega_4} \right] \right\},\label{A1-9}
\end{align}
\begin{align}
S_{n,2}^\prime(t_2)=& -r_1^2r_2\abs{\frac{q_1}{p_1}}\left\{  \abs{g_{+,2}(t_2)}^2\sin{\pqty{\Omega_3+\Omega_5-2\Omega_6}}-\abs{g_{-,2}(t_2)}^2\sin{\pqty{\Omega_3-\Omega_5+2\Omega_6}}\right\}+r_2\abs{\frac{q_1}{p_1}}\Big\{\abs{g_{+,2}(t_2)}^2\nonumber\\
&\times\sin{\pqty{\Omega_3-\Omega_5}}-\abs{g_{-,2}(t_2)}^2\sin{\pqty{\Omega_3+\Omega_5}} \Big\}-r_1^2 r_2\abs{\frac{p_1}{q_1}}\Big\{\abs{g_{+,2}(t_2)}^2\sin{\pqty{\Omega_4-\Omega_5+2\Omega_6}}-\abs{g_{-,2}(t_2)}^2\nonumber\\
&\times\sin{\pqty{\Omega_4+\Omega_5-2\Omega_6}} \Big\}+r_2\abs{\frac{p_1}{q_1}}\left\{\abs{g_{+,2}(t_2)}^2\sin{\pqty{\Omega_4+\Omega_5}}+\abs{g_{-,2}(t_2)}^2\sin{\pqty{\Omega_4-\Omega_5}}\right\}\nonumber\\
&+r_1\sin{\pqty{\delta_1-\Omega_5+\Omega_6}}f_2(q,p)-r_1\sin{\pqty{\delta_1+\Omega_5-\Omega_6}}f_2(p,q),\label{A1-10}
\end{align}
\begin{align}
S_{h,1}^\prime(t_2) =&\ \frac{e^{-\Gamma_2 t_2}}{2}\Big\{g(+,+,+,+)\sinh{\frac{\Delta \Gamma_2 t_2}{2}}\Big[r_1^2\cos{\pqty{\Omega_5-2\Omega_6}}+\cos{\Omega_5}\Big]-g(+,-,+,-)\sin{\Delta m_2 t_2}\Big[r_1^2\sin{\pqty{\Omega_5-2\Omega_6}}\nonumber\\
&-\sin{\Omega_5} \Big]\Big\}+r_1r_2\abs{\frac{q_2}{p_2}} \cos{\pqty{\delta_1-\Omega_5+\Omega_6}}e^{-\Gamma_2 t_2}\left\{ \abs{\frac{q_1}{p_1}}\left[\sinh{\frac{\Delta \Gamma_2 t_2}{2}}\cos{\Omega_3} - \sin{\Delta m_2 t_2}\sin{\Omega_3} \right] \right.\nonumber\\
&\left.+\abs{\frac{p_1}{q_1}}\left[\sinh{\frac{\Delta \Gamma_2 t_2}{2}}\cos{\Omega_4} + \sin{\Delta m_2 t_2}\sin{\Omega_4} \right] \right\}+r_1r_2\abs{\frac{p_2}{q_2}} \cos{\pqty{\delta_1+\Omega_5-\Omega_6}}e^{-\Gamma_2 t_2}\nonumber\\
&\times\left\{ \abs{\frac{q_1}{p_1}}\left[\sinh{\frac{\Delta \Gamma_2 t_2}{2}}\cos{\Omega_3} + \sin{\Delta m_2 t_2}\sin{\Omega_3} \right] +\abs{\frac{p_1}{q_1}}\left[\sinh{\frac{\Delta \Gamma_2 t_2}{2}}\cos{\Omega_4} - \sin{\Delta m_2 t_2}\sin{\Omega_4} \right] \right\},\label{A1-11}
\end{align}
\begin{align}
S_{h,2}^\prime(t_2) =&\ r_1^2 r_2 \abs{\frac{q_1}{p_1}}\Big[\abs{g_{+,2}(t_2)}^2\cos{\pqty{\Omega_3+\Omega_5-2\Omega_6}}+\abs{g_{-,2}(t_2)}^2\cos{\pqty{\Omega_3-\Omega_5+2\Omega_6}}\Big]\nonumber\\
&+r_2\abs{\frac{q_1}{p_1}}\left\{ \abs{g_{+,2}(t_2)}^2\cos{\pqty{\Omega_3-\Omega_5}}+\abs{g_{-,2}(t_2)}^2\cos{\pqty{\Omega_3+\Omega_5}} \right\}\nonumber\\
&+r_1^2r_2\abs{\frac{p_1}{q_1}}\Big[\abs{g_{+,2}(t_2)}^2\cos{\pqty{\Omega_4-\Omega_5+2\Omega_6}}+\abs{g_{-,2}(t_2)}^2\cos{\pqty{\Omega_4+\Omega_5-2\Omega_6}} \Big]\nonumber\\
&+r_2\abs{\frac{p_1}{q_1}}\left\{\abs{g_{+,2}(t_2)}^2\cos{\pqty{\Omega_4+\Omega_5}}+\abs{g_{-,2}(t_2)}^2\cos{\pqty{\Omega_4-\Omega_5}}  \right\}\nonumber\\
&+r_1\cos{\pqty{\delta_1-\Omega_5+\Omega_6}}f_1(q,p)+r_1\cos{\pqty{\delta_1+\Omega_5-\Omega_6}}f_1(p,q).\label{A1-12}
\end{align}
\end{widetext}

\end{document}